\documentclass[twoside]{article}

\usepackage{lipsum} 

\usepackage[round]{natbib}
\usepackage[sc]{mathpazo} 
\usepackage[T1]{fontenc} 
\usepackage{microtype} 

\usepackage[hmarginratio=1:1,top=32mm,columnsep=20pt]{geometry} 
\usepackage{multicol} 
\usepackage[hang, small,labelfont=bf,up,textfont=it,up]{caption} 
\usepackage{booktabs} 
\usepackage{float} 
\usepackage[draft]{hyperref} 
\usepackage{lettrine} 
\usepackage{paralist} 

\usepackage{abstract} 

\usepackage{titlesec} 
\renewcommand\thesection{\Roman{section}} 
\renewcommand\thesubsection{\Roman{subsection}} 
\titleformat{\section}[block]{\large\scshape\centering}{\thesection.}{1em}{} 
\titleformat{\subsection}[block]{\large}{\thesubsection.}{1em}{} 

\usepackage{caption}
\usepackage{subcaption}
\usepackage{amsmath}

\usepackage{fancyhdr} 
\usepackage{graphicx}
\DeclareGraphicsExtensions{.jpg}
\usepackage{verbatim}

\title{\vspace{-15mm}\fontsize{23pt}{10pt}\selectfont\textbf{ Structure and Dynamics of Brain Lobe's Functional Networks at the Onset of Anesthesia-Induced Loss of Consciousness }} 

\author{
\large
\textsc{Eduardo C. Padovani}
\thanks{Email: \texttt{eduardo.padovani@alumni.usp.br}}\\[2mm] 
\normalsize 
\vspace{-5mm}
}
\date{}

\begin{document}

\maketitle 

\thispagestyle{fancy} 


\begin{abstract}

\noindent \ Anesthetic agents are neurotropic drugs capable of inducing significant alterations in the thalamocortical system, promoting a profound decrease in awareness and level of consciousness. There is experimental evidence that general anesthesia affects large-scale functional networks, leading to alterations in the brain's state. However, the specific impact on the network structure assumed by functional connectivity locally in different cortical regions has not yet been reported. Within this context, the present study has sought to characterize the functional brain networks relative to the frontal, parietal, temporal, and occipital lobes. In this study, electrophysiological neural activity was recorded using a dense ECoG-electrode array placed directly on the cortical surface of an old-world monkey of the species Macaca fuscata. Networks were estimated serially over time every five seconds, while the animal model was under controlled experimental conditions of a Ketamine-Medetomidine anesthetic induction. In each of the four cortical brain lobes, noticeable alterations in distinct properties of the networks evidenced a transition in the network’s architecture, which occurred within approximately one and a half minutes after administering the anesthetics. The characterization of functional brain networks performed in this study provides relevant experimental evidence that expands our understanding of the neural correlates of consciousness in terms of the structure and properties of the functional brain networks.

\end{abstract}


\begin{multicols}{2} 

\section{Introduction}
\lettrine[nindent=0em,lines=3]{O}ur understanding of how the brain works has deepened as neuroscience has evolved and consolidated. In general, our knowledge of the brain has been gradually built through the integration of several discoveries and reports by many researchers, accomplished mainly over the past 150 years. Most of the advances in neuroscience have been associated with the development of new technologies and methods that have provided novel information about the brain and neural elements, permitting scientists to formulate new hypotheses and theories.
In the early years of the $20^{Th}$ century, the German neurologist Korbinian Brodmann conducted a cornerstone research \citep{brodmann1908beitrage} which changed the view we had about the brain. Studying the cortex's cytoarchitectural organization, he discovered that distinct anatomical areas were composed of populations of cells with different morphologies and densities. This experimental evidence enabled him to make the remarkable assumption that distinctions in the cytoarchitecture across specific cortical regions could have functional implications.
Since then, a significant amount of research has been conducted to classify the cortex into sets of specialized anatomical areas and to comprehend the brain's functioning in terms of the activation and deactivation of these specialized regions. This approach has been fruitful, and numerous experiments demonstrate that various cortical regions are responsible for specific functions and activities. 
 
 \newpage
 
Over the last few decades, advancements in recording techniques have led to higher accuracy and resolution, allowing for precise and simultaneous recording of multiple cortical regions and brain structures. Experiments involving these techniques revealed a novel aspect of the brain's functioning. They have provided evidence that many distinct specialized areas functionally interact while the brain performs its activities \citep{padovani2016characterization2}. Based on this evidence, many neuroscientists are arriving at a consensus that, in essence, specialized cortical areas do not work in isolation. Instead, they consider that neural activities are accounted for by networks that involve interactions among several distinct cortical areas. These networks are considered to represent the precise manner in which the brain processes and integrates information \citep{stam2007graph,sporns2011networks}. It is also believed that these networks serve as a substrate that supports various dynamics and neural processes \citep{bassett2006small}. Until now, a complete understanding of how cognitive capacities or distinct physiological states of the organism are translated into networks with specific properties has yet not been achieved, which has motivated research and experiments in this field. Among many relevant topics, the present study aims to investigate the impacts of general anesthesia on the functional brain networks that are established locally in distinct brain lobes.

General anesthesia is a physiologically stable and reversible state induced by drugs characterized by analgesia, immobility, amnesia, and loss of consciousness \citep{schwartz2010general}. In addition to their undeniable importance and usefulness for clinical medicine, anesthetic agents can also serve as tools of great importance for neuroscience. Once their pharmacological effects involve a drastic and fast reduction in awareness and levels of consciousness, they offer means for the study of consciousness and neural correlates of consciousness \citep{uhrig2014cerebral}, providing possibilities for the investigation of fundamental processes and phenomena that happen in the brain \citep{hameroff1998toward}.
 
The characterization of the large-scale functional brain networks during general anesthesia has already been performed and reported \citep{padovani2016characterization1}. However, once different cortical areas are functionally specialized, the effects of general anesthesia may also be diverse, motivating further investigation of the structure assumed by functional interactions over distinct specialized cortical regions. Within this context, this study has sought to characterize the structure and properties of functional networks in each of the cortical brain lobes (frontal, parietal, temporal, and occipital)\footnote{Networks of interactions that involve cortical areas of each brain lobe are considered without taking into account interactions that occur across areas of distinct brain lobes. }.

\textbf{Note}:
\textit{The networks presented in \citep{padovani2016characterization1} are not ``the sum'' of the networks of this manuscript, once the large-scale functional brain networks \citep{padovani2016characterization1} also involve interactions established among areas of distinct brain lobes. There is experimental evidence that segregated and highly functionally integrated neural processes that involve large cortical areas and transcend brain lobes anatomical divisions occur most of the time \citep{padovani2016characterization2}.}

\section{Methods}

This study involved the analysis of a database provided by the Laboratory of Adaptive Intelligence at the \texttt{RIKEN- Brain Science Institute}, Saitama, Japan. The electrophysiological data is respective to a \textit{Ketamine} and \textit{Medetomidine} anesthetic induction experiment conducted in non-human primate subject of the species \textit{Macaca fuscata}. A cutting-edge technique, the \textit{MDR-ECoG}, is based on a high-density ECoG electrode array implanted chronically over the cortical surface, beneath the dura mater \citep{yanagawa2013large,fukushima2014electrocorticographic}, was used to record cortical electrophysiological activity. This technique offered extensive coverage of the left cortical lateral surface and medial walls in conjunction with high temporal (1Khz) and spatial resolution (5mm). The characteristics of the data records permitted us to characterize and evaluate the functional neural networks established individually at the frontal, parietal, temporal, and occipital lobes along the anesthetic induction.

\subsubsection*{Anesthetic Agents}

The experiment was respective to a Ketamine-Medetomidine anesthetic induction.
 
 Ketamine is a non-competitive antagonist of the receptor N-methyl-D-asparthate \citep{green2011clinical}. It induces an anesthetic state characterized by the dissociation between the thalamocortical and limbic systems \citep{bergman1999ketamine}. Medetomidine (an agonist of the Alpha-2 Adrenergic Receptor) has been administered combined with Ketamine to promote muscle relaxation during the anesthetic induction procedure \citep{young1999short}.

\subsubsection*{Neural Connectivity Estimator}

As a neural connectivity estimator, a method based on Granger causality in the frequency domain \citep{granger1969investigating,seth2010matlab} was used to infer statistical dependencies between the time series of the electrodes on five physiological frequency bands. When applied in neuroscience, Granger causality provides an estimate of the information exchange from one cortical area to another \citep{seth2007distinguishing,seth2010matlab}.

\subsubsection*{Experimental Procedures - Summary of Steps}

The modeling of functional neural activities was performed according to the following procedures:

\begin{enumerate}

\item The database comprised MDR-ECoG cortical electrophysiological activity records. The matrix of ECoG electrodes continuously covered the left brain's lateral cortical surface and parts of the cortical medial walls.
\item Each one of the electrodes in the array was considered a vertex of the network and represented the cortical area in which it was positioned.
\item Granger causality in the frequency domain was used to compute functional connectivity association values between the electrode's records (time series).
\item For each brain lobe, an adjacency matrix was assembled, containing the pairwise association values between the corresponding nodes of the respective brain lobe.
\item Complex Networks measures were employed to study and characterize the topology of the obtained networks.

\end{enumerate}

For each brain lobe, networks were estimated serially in time intervals of five seconds in five physiological frequency bands throughout the experiment. For each frequency band, all the networks in the sequence were computed using exactly the same procedures and parameters\footnote{For each frequency band, the same threshold was applied on the networks of different cortical lobes.}. Therefore, the alterations observed between the measures of distinct time-resolved networks came from changes in the records of neural activity.\\

\subsubsection*{Database - Anesthetic Induction Experimental Procedures}{

The monkey was accommodated in a properly fitted chair, with its head and arms restrained. The recording of cortical electrophysiological activity began with the macaque awake with its open eyes. Then, the eyes were covered with a patch to prevent evoked visual responses. About 10 minutes later, a cocktail of Ketamine and Medetomidine (5.6mg/Kg ketamine + 0.01mg/Kg medetomidine) was injected intramuscularly to induce anesthesia. The point of loss of consciousness (LOC) was experimentally determined at the moment when the monkey no longer reacted to external stimuli (such as touching the nostrils or opening the hands). After establishing the LOC, neural activity was recorded for about 25–30 minutes. Vital signs were monitored throughout the entire experiment.

All experimental and surgical procedures were idealized and conducted by researchers from the Laboratory of Adaptive Intelligence at the \textit{Riken - Brain Science Institute} following experimental protocols (No. H24-2-203(4)) approved by the RIKEN ethics committee and the recommendations of the Weatherall report, "The use of non-human primates in research". For additional information regarding the neural record methodology and experimental protocols, see \citep{nagasaka2011multidimensional} and (\texttt{http://neurotycho.org}).
}

\subsection{Signal Processing and Granger Causality in the Frequency Domain}

\subsubsection*{Data Processing}

\begin{enumerate}

\item A reject-band \texttt{IIR-notch} filter was used to attenuate components of the signal at 50Hz.

\item The signal was down-sampled from 1KHz to 200Hz.

\item The signal was separated into windows of 1000 points (equivalent to a five-second recording of neural activity).

\item For each of the 128-time series, the trend was removed, and the average was subtracted.

\item To verify the stationary condition of the time series, the tests KPSS \citep{kwiatkowski1992testing} and ADF [Augmented Dickey Fuller]\citep{hamilton1989new} were applied.

\end{enumerate}

\subsubsection*{Libraries Used}

For the computation of association values using Granger causality in the frequency domain, with some adaptations, the following libraries were used: \texttt{MVGC GRANGER TOOLBOX}, developed by Ph.D. Anil Seth (Sussex University, UK), described in \mbox{\citep{seth2010matlab}}, available at \texttt{www.anilseth.com}, and the library \texttt{BSMART toolbox} (\textit{\textbf{B}rain-\textbf{S}ystem for \textbf{M}ultivariate \textbf{A}uto\textbf{R}egressive \textbf{T}imeseries} \texttt{toolbox}) described in \citep{cui2008bsmart} and available at \texttt{www.brain-smart.org}.

\subsubsection*{Computation of Causal Interactions}

\begin{enumerate}

\item {Model Order}

To find the model order (number of observations to be used in the regression model), the criteria for the selection of models from Akaike (AIC) and Bayes/Schwartz (BIC) were used.
Both methods returned the order of the model equal to seven.
\vspace*{2mm}

\item {Causal Interactions}

At each window of 1000 points, Granger causality in the frequency domain interactions were pair-wise computed among the time series respective to each brain lobe by the function \texttt{cca\_pwcausal()} (\texttt{MVGC GRANGER TOOLBOX}).

\item {Frequency Bands}

 Granger causality interactions were computed in five physiological frequency bands: Delta (0-4Hz), Theta (4-8Hz), Alpha (8-12Hz), Beta (13-30Hz), and Gamma (25-100Hz). 
The interaction values obtained were saved into adjacency matrices.

\end{enumerate}

\subsubsection*{Graphs and Networks}

\begin{enumerate}

\item Assemble Networks

For each sequence of graphs relative to a frequency band, a threshold was chosen, and only the interactions with a magnitude value higher than this threshold were considered as edges of the graphs. The same threshold was applied to the distinct brain lobes analyzed.
\begin{itemize}
\item Delta (0-4Hz), threshold = $0.6$
\item Theta (4-8Hz), threshold = $0.4$
\item Alpha (8-12Hz), threshold = $0.3$
\item Beta (13-30Hz), threshold = $0.6$
\item Gamma (25-100Hz), threshold = $2.0$
\end{itemize}
\end{enumerate}
As discussed in \citep{bullmore2009complex,sporns2011networks}, scientists can use different criteria to determine the threshold parameter. In the present study, due to experimental conditions, each sequence of networks contained graphs with distinct connectivity. Thresholds were chosen in such a way as to prevent graphs with lower connectivity in each sequence from presenting many disconnected parts or vertices, which might introduce distortions in the analysis.\\
Networks contained only vertices respective to electrodes positioned over the corresponding brain lobe analysed\footnote{A few electrodes positioned over sulci divisions were included on both brain lobes.}.
\\
\\

\vspace{+0.5\baselineskip}
The number of vertices of the networks of each brain lobe was:

\begin{itemize}
\item Frontal lobe = 47 vertices
\item Parietal lobe = 23 vertices 
\item Temporal lobe = 26 vertices 
\item Occipital lobe = 34 vertices

\end{itemize}

After obtaining non-weighted graphs, the directions of the edges were removed, resulting in undirected and non-weighted networks.

\subsubsection*{Analysis of the Topology}

Network measures were used to
characterize the topology and properties
of the graphs.

\end{multicols}

\newpage

\section{Results}

\subsection{Measures Related to Connectivity}

\subsubsection{Average Degree}

\subsubsection*{Delta 0-4Hz}

\begin{figure}[!h]
\begin{subfigure}{.5\textwidth}
  \centering
  \includegraphics[width=1\linewidth]{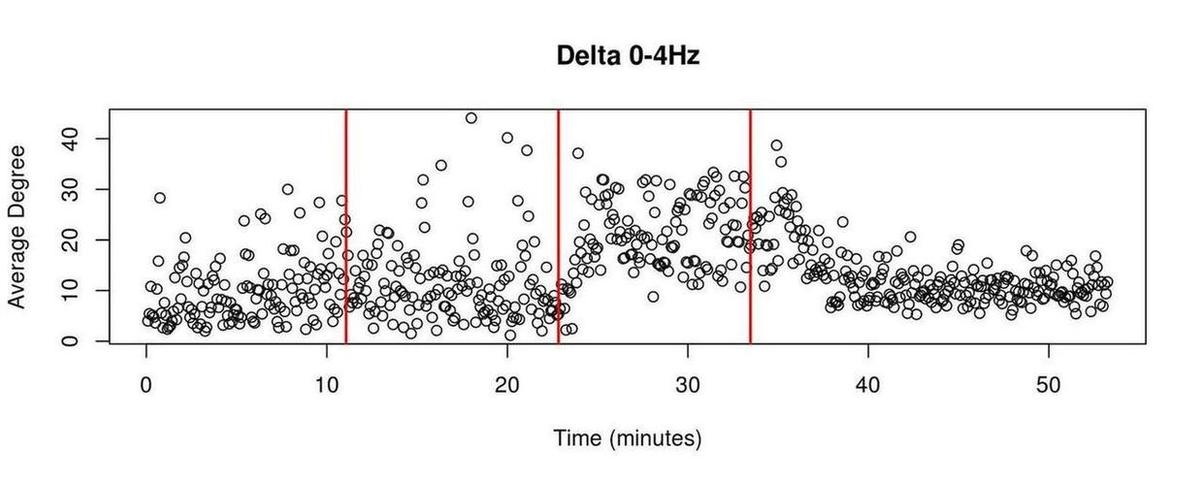}
  \caption{Frontal Lobe}
  \label{fig:sfig1}
\end{subfigure}%
\begin{subfigure}{.5\textwidth}
  \centering
  \includegraphics[width=1\linewidth]{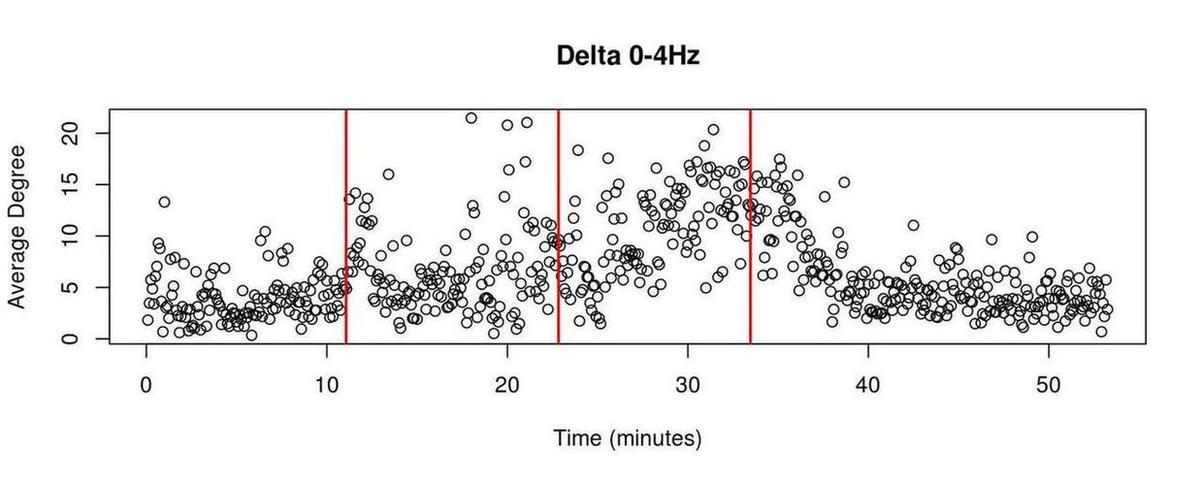}
 \caption{Parietal Lobe}
  \label{fig:sfig2}
\end{subfigure}\\
\centering
\begin{subfigure}{.5\textwidth}
\includegraphics[width=1\linewidth]{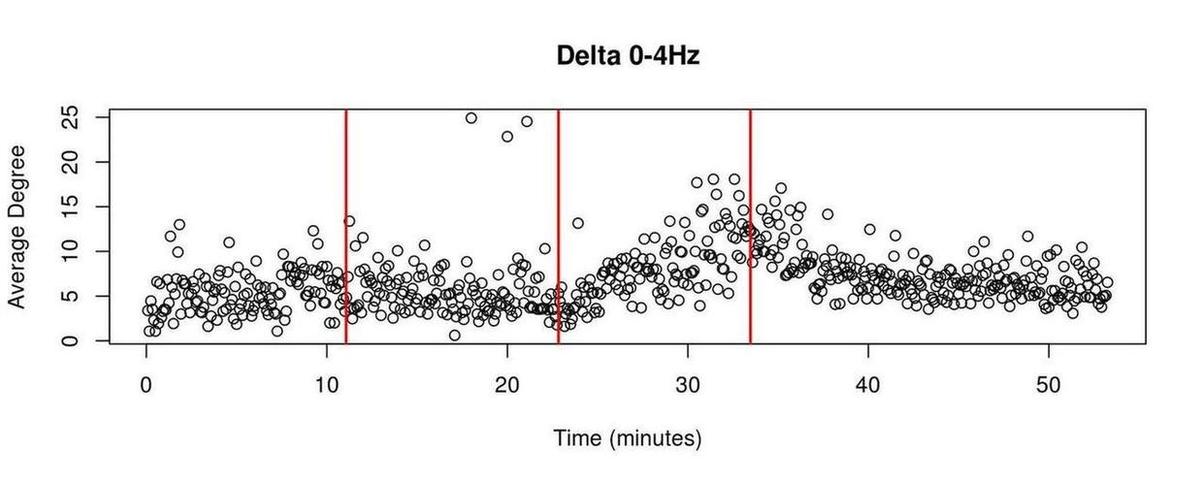}
  \caption{Temporal Lobe}
  \label{fig:sfig3}
\end{subfigure}%
\begin{subfigure}{.5\textwidth}
  \centering
  \includegraphics[width=1\linewidth]{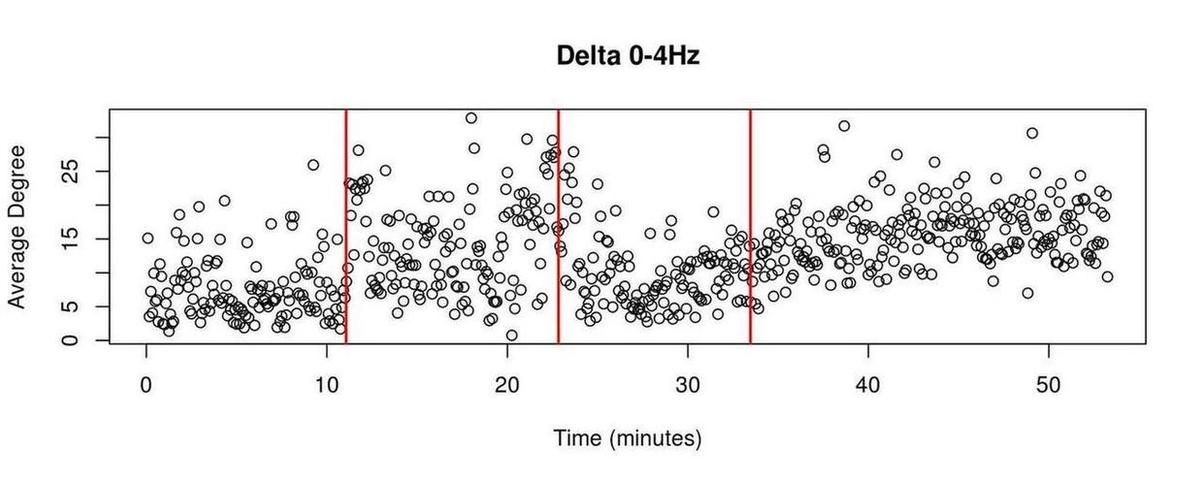}
  \caption{Occipital Lobe}
  \label{fig:sfig4}
\end{subfigure}\\
\caption{{\normalfont \textbf{Average Degree.}} Average degree vertex respective to the networks of the Delta frequency band (0-4Hz). Vertical axis average degree; Horizontal axis time (minutes). At t=11 minutes, the monkey was blindfolded. The first red line in each sub-figure represents the moment when a patch was placed over the eyes. At t=23 minutes, the Ketamine-Medetomidine cocktail was injected, represented by the second red line. The third red line indicates the point of loss of consciousness (LOC) that occurred at t=33 minutes. Sub-figures: (\textbf{a}) Frontal Lobe; (\textbf{b}) Parietal Lobe; (\textbf{c}) Temporal Lobe; (\textbf{d}) Occipital Lobe.}
\label{fig:fig}
\end{figure}

\begin{table}[!h]
\centering
\caption{{\normalfont \textbf{Average Degree.}} Mean, variance (Var), and standard deviation (SD) of the average degree vertex of the networks respective to each one of the four cortical lobes on the three different conditions in which the monkey was exposed during the experiment: awake with eyes open, awake with eyes closed and anesthesia (eyes closed). Frequency band Delta (0-4Hz). }
\vspace{0.5cm}
\begin{tabular}{l|lcr|lcr|lcr}
\hline 
\textbf{Delta Band (0-4Hz)} & \multicolumn{3}{c}{Eyes Open} \vline &\multicolumn{3}{c}{Eyes Closed} \vline &\multicolumn{3}{c}{Anesthesia}\\
\hline
Corresponding Graph & Mean   & Var  & SD  & Mean  & Var  & SD & Mean   & Var  & SD\\ 
\hline                             

Frontal Lobe    & 9.62  & 37.5 &6.12       & 11.2 & 63.4  & 7.96     &  14.7  & 48.0   &  6.93 \\

Parietal Lobe   &3.91  &5.5  & 2.35     &6.25  &16.7   &4.08      &7.23    &20.7  &4.55   \\

Temporal Lobe   & 5.45 & 5.80 & 2.41     & 5.55 & 13.0  & 3.61     &  7.86  & 8.87   &  2.98  \\

Occipital Lobe  & 7.68 & 22.0 & 4.69     & 13.9 & 47.4  &  6.89    &  14.3  &  25.5  &  5.05  \\   

\end{tabular}
\end{table}

\clearpage

\subsubsection*{Theta 4-8Hz}

\begin{figure}[!h]
\begin{subfigure}{.5\textwidth}
  \centering
  \includegraphics[width=1\linewidth]{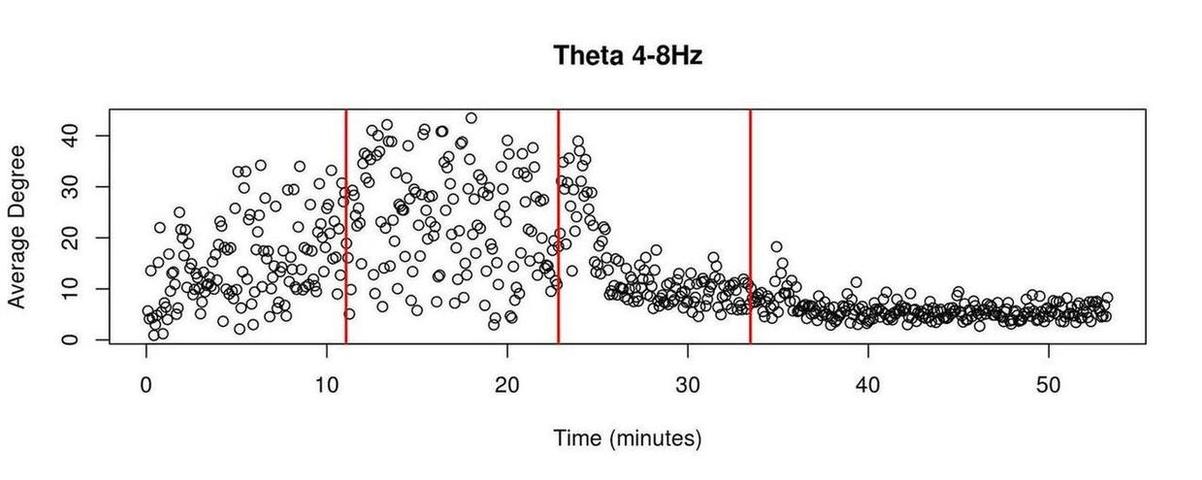}
  \caption{Frontal Lobe}
  \label{fig:sfig1}
\end{subfigure}%
\begin{subfigure}{.5\textwidth}
  \centering
  \includegraphics[width=1\linewidth]{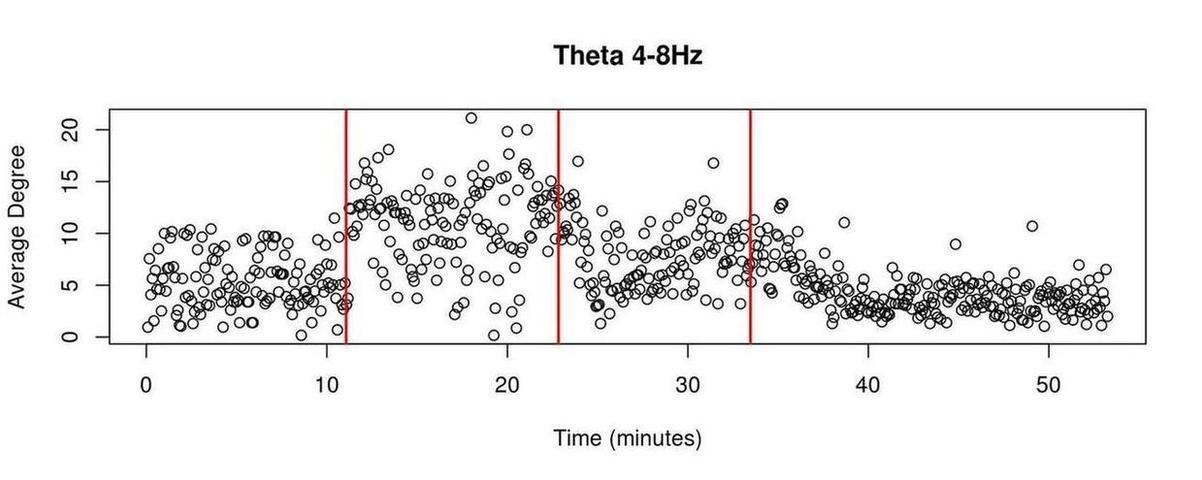}
 \caption{Parietal Lobe}
  \label{fig:sfig2}
\end{subfigure}\\
\centering
\begin{subfigure}{.5\textwidth}
\includegraphics[width=1\linewidth]{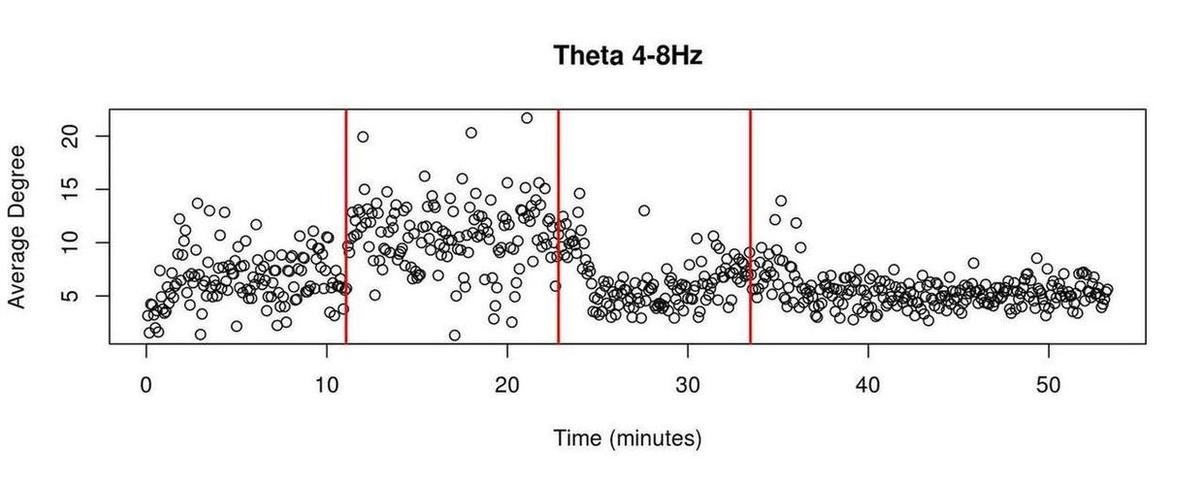}
  \caption{Temporal Lobe}
  \label{fig:sfig3}
\end{subfigure}%
\begin{subfigure}{.5\textwidth}
  \centering
  \includegraphics[width=1\linewidth]{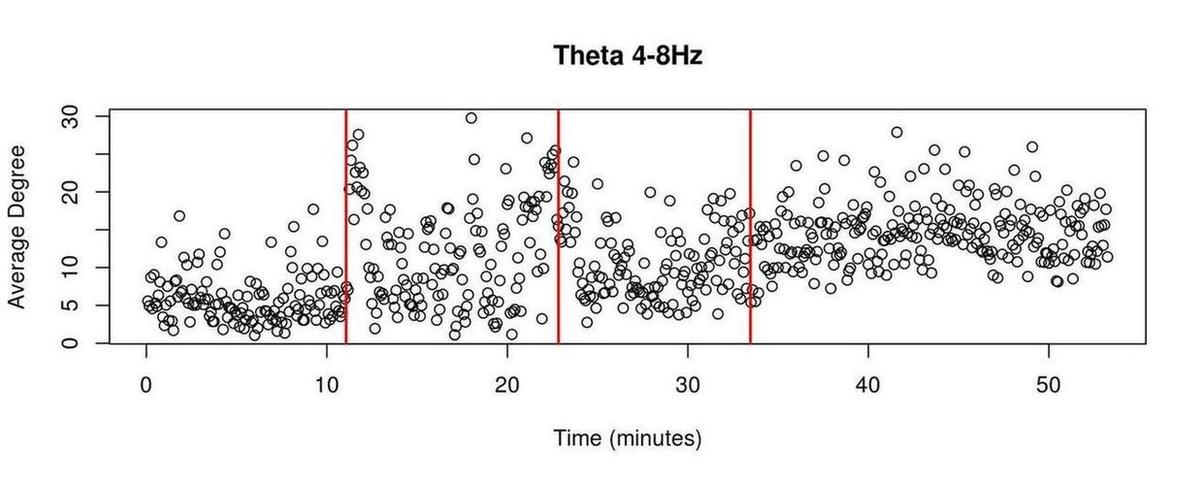}
  \caption{Occipital Lobe}
  \label{fig:sfig4}
\end{subfigure}\\
\caption{{\normalfont \textbf{Average Degree.}} Average degree vertex respective to the networks of the  Theta frequency band (4-8Hz). Vertical axis average degree; Horizontal axis time (minutes). At t=11 minutes, the monkey was blindfolded, the first red line in each sub-figure represents the moment when a patch was placed over the eyes. At t=23 minutes the Ketamine-Medetomidine cocktail was injected, being represented by the second red line. The point of loss of consciousness (LOC) was registered at t=33 minutes, and is indicated by the third red line. Sub-figures: (\textbf{a}) Frontal Lobe; (\textbf{b}) Parietal Lobe; (\textbf{c}) Temporal Lobe; (\textbf{d}) Occipital Lobe.}
\label{fig:fig}
\end{figure}

\begin{table}[!h]
\centering
\caption{{\normalfont \textbf{Average Degree.}} Mean, variance (Var), and standard deviation (SD) of the average degree vertex of the networks respective to each one of the four cortical lobes, on the three different conditions in which the monkey was exposed during the experiment: awake with eyes open, awake with eyes closed and anesthesia (eyes closed). Frequency band Theta (4-8Hz). }
\vspace{0.5cm}
\begin{tabular}{l|lcr|lcr|lcr}
\hline 
\textbf{Theta Band (4-8Hz)} & \multicolumn{3}{c}{Eyes Open} \vline &\multicolumn{3}{c}{Eyes Closed} \vline &\multicolumn{3}{c}{Anesthesia}\\
\hline
Corresponding Graph & Mean   & Var  & SD  & Mean  & Var  & SD & Mean   & Var  & SD\\ 
\hline                             

Frontal Lobe    &14.4  & 60.7 & 7.79      &22.7  &106   & 10.3     &6.7    & 6.7   &   2.59\\

Parietal Lobe   &5.33  &6.88  &2.62      &10.8  &16.8   &4.10      &5.18    &7.95  &2.82   \\

Temporal Lobe   & 6.62 & 6.27 &2.50      &10.6  & 10.0  & 3.21     & 5.61   &  2.57  & 1.60   \\

Occipital Lobe  & 5.96 & 11.3 & 3.35     & 11.5 & 44.1  &  6.64    &  13.58  &  19.1  & 4.37   \\   

\end{tabular}
\end{table}

\clearpage

\subsubsection*{Alpha 8-12Hz}

\begin{figure}[!h]
\begin{subfigure}{.5\textwidth}
  \centering
  \includegraphics[width=1\linewidth]{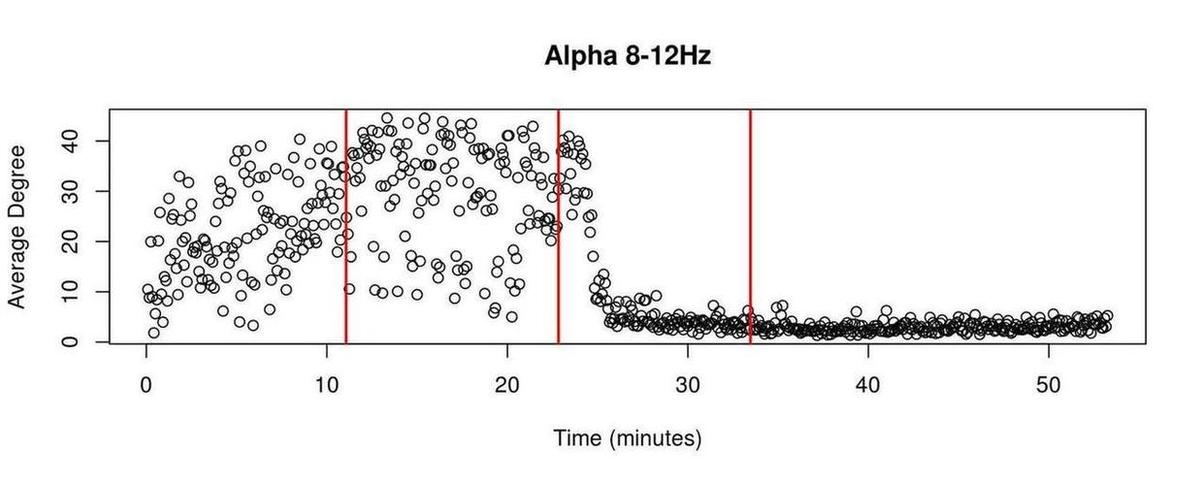}
  \caption{Frontal Lobe}
  \label{fig:sfig1}
\end{subfigure}%
\begin{subfigure}{.5\textwidth}
  \centering
  \includegraphics[width=1\linewidth]{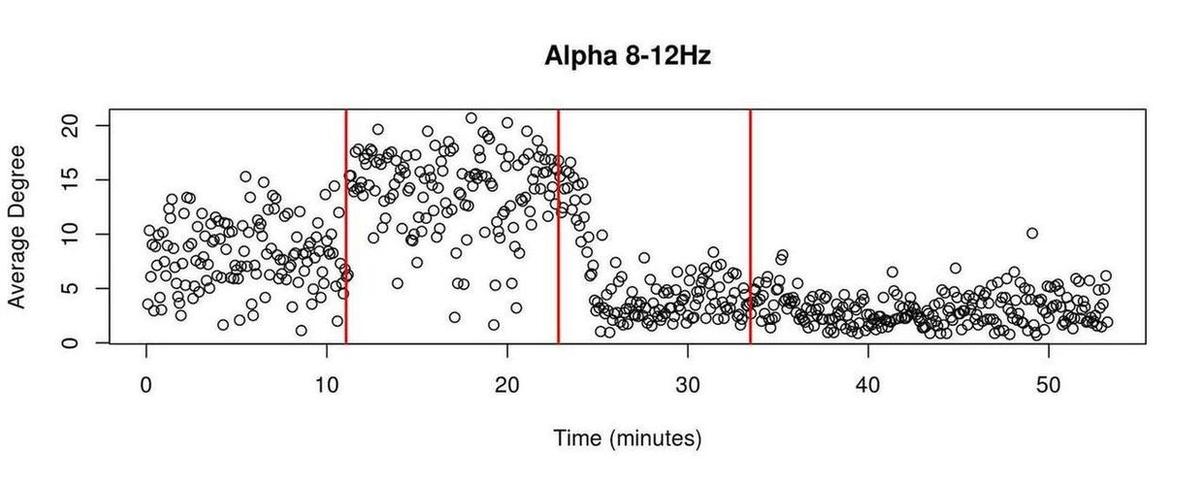}
 \caption{Parietal Lobe}
  \label{fig:sfig2}
\end{subfigure}\\
\centering
\begin{subfigure}{.5\textwidth}
\includegraphics[width=1\linewidth]{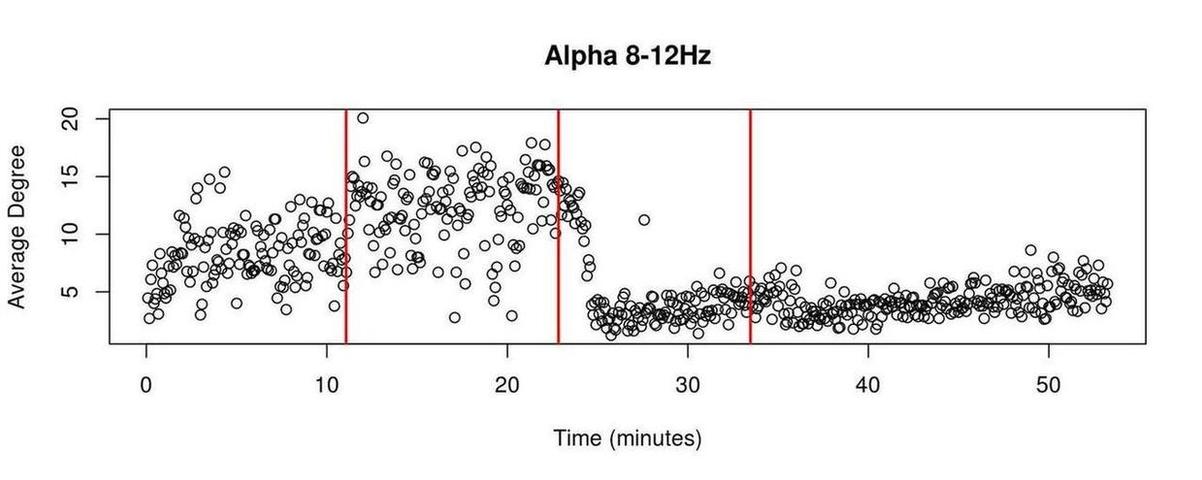}
  \caption{Temporal Lobe}
  \label{fig:sfig3}
\end{subfigure}%
\begin{subfigure}{.5\textwidth}
  \centering
  \includegraphics[width=1\linewidth]{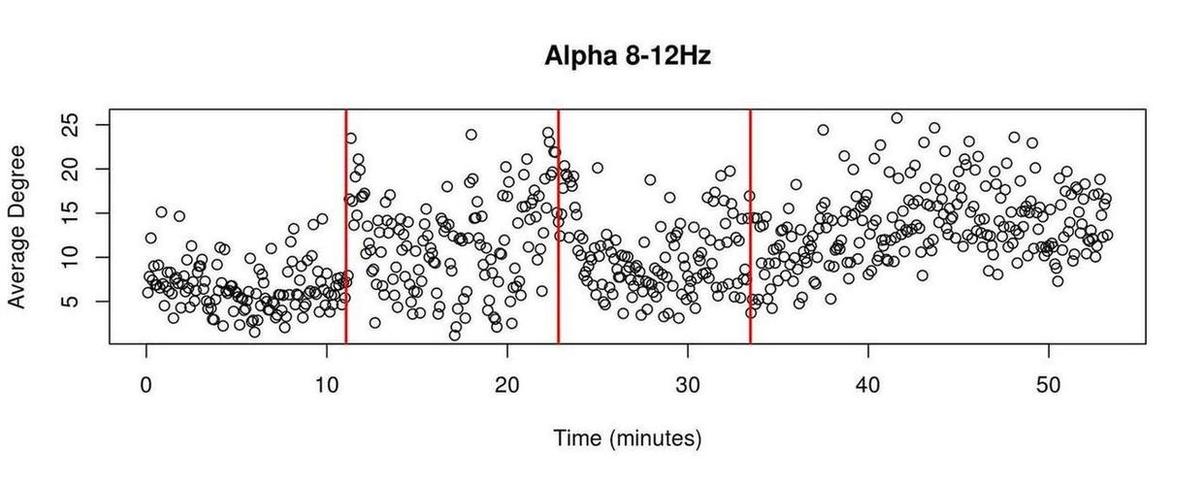}
  \caption{Occipital Lobe}
  \label{fig:sfig4}
\end{subfigure}\\
\caption{{\normalfont \textbf{Average Degree.}} Average degree vertex respective to the networks of the  Alpha frequency band (8-12Hz). Vertical axis average degree; Horizontal axis time (minutes). At t=11 minutes, the monkey was blindfolded, the first red line in each sub-figure represents the moment when a patch was placed over the eyes. At t=23 minutes the Ketamine-Medetomidine cocktail was injected, being represented by the second red line. The point of loss of consciousness (LOC) was registered at t=33 minutes, and is indicated by the third red line. Sub-figures: (\textbf{a}) Frontal Lobe; (\textbf{b}) Parietal Lobe; (\textbf{c}) Temporal Lobe; (\textbf{d}) Occipital Lobe.}
\label{fig:fig}
\end{figure}

\begin{table}[!h]
\centering
\caption{{\normalfont \textbf{Average Degree.}}  Mean, variance (Var), and standard deviation (SD) of the average degree vertex of the networks respective to each one of the four cortical lobes, on the three different conditions in which the monkey was exposed during the experiment: awake with eyes open, awake with eyes closed and anesthesia (eyes closed). Frequency band Alpha (8-12Hz).}
\vspace{0.5cm}
\begin{tabular}{l|lcr|lcr|lcr}
\hline 
\textbf{Alpha Band (8-12Hz)} & \multicolumn{3}{c}{Eyes Open} \vline &\multicolumn{3}{c}{Eyes Closed} \vline &\multicolumn{3}{c}{Anesthesia}\\
\hline
Corresponding Graph & Mean   & Var  & SD  & Mean  & Var  & SD & Mean   & Var  & SD\\ 
\hline                             

Frontal Lobe  &21.5   &85.6  &9.25       &29.7  &118   &10.9      &3.14   &1.30   &1.14   \\

Parietal Lobe   &8.10  &9.57  &3.09      &13.8  &15.0   &3.87      &3.30    &2.46  &1.57   \\

Temporal Lobe  &8.22   &6.87  &2.62       &12.3  &10.7   &3.28      &4.12   &1.54   &1.24     \\

Occipital Lobe  &6.60   &7.55  &2.75       &11.4  &28.5   &5.34      &12.8   &19.5   & 4.42   \\   

\end{tabular}
\end{table}

\clearpage

\subsubsection*{Beta 13-30Hz}

\begin{figure}[!h]
\begin{subfigure}{.5\textwidth}
  \centering
  \includegraphics[width=1\linewidth]{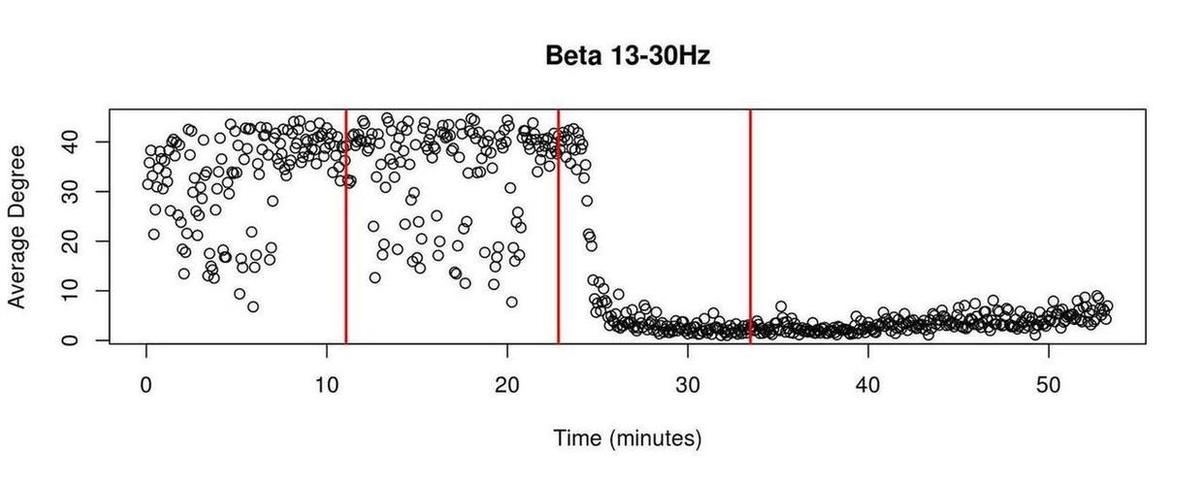}
  \caption{Frontal Lobe}
  \label{fig:sfig1}
\end{subfigure}%
\begin{subfigure}{.5\textwidth}
  \centering
  \includegraphics[width=1\linewidth]{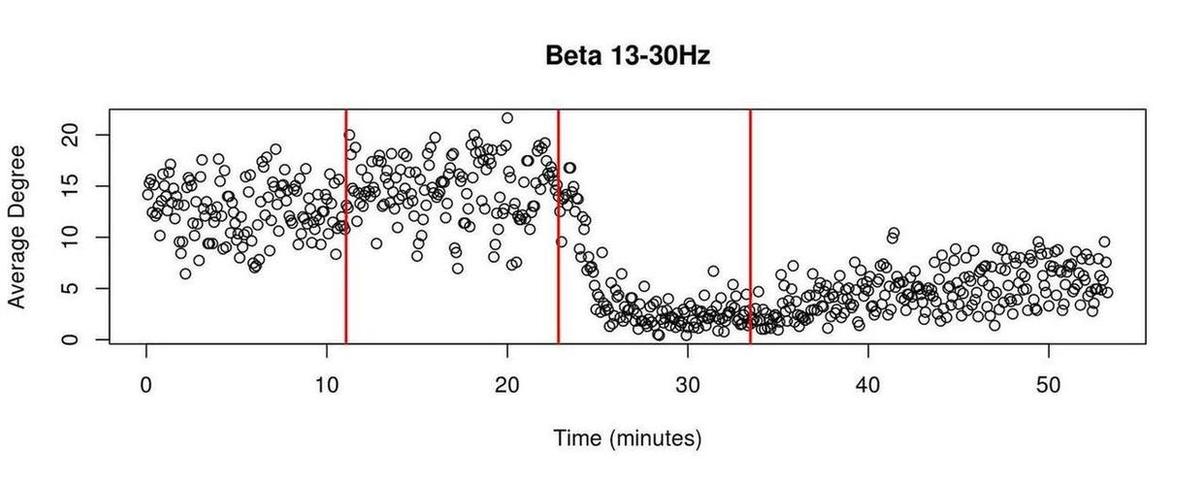}
 \caption{Parietal Lobe}
  \label{fig:sfig2}
\end{subfigure}\\
\centering
\begin{subfigure}{.5\textwidth}
\includegraphics[width=1\linewidth]{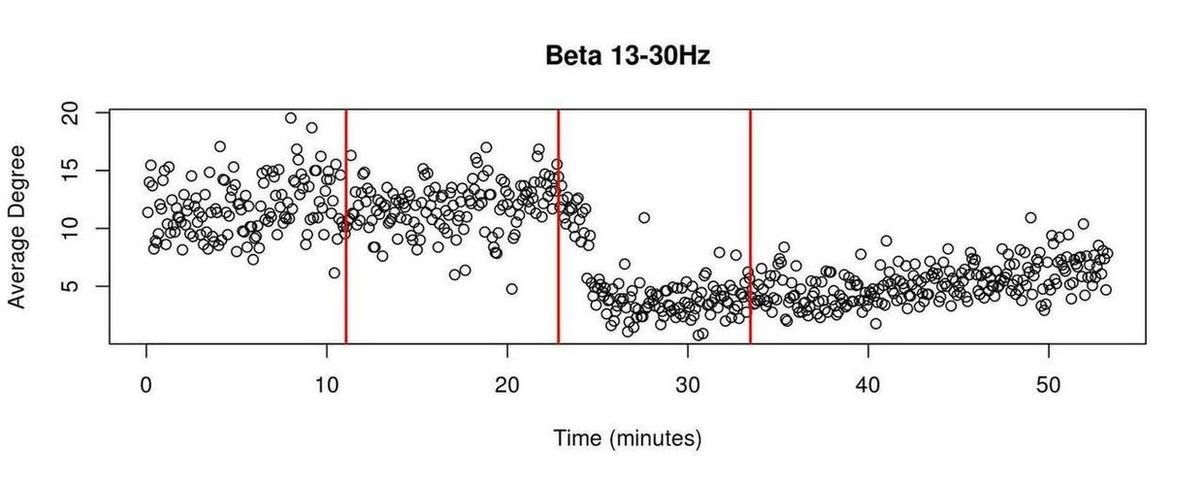}
  \caption{Temporal Lobe}
  \label{fig:sfig3}
\end{subfigure}%
\begin{subfigure}{.5\textwidth}
  \centering
  \includegraphics[width=1\linewidth]{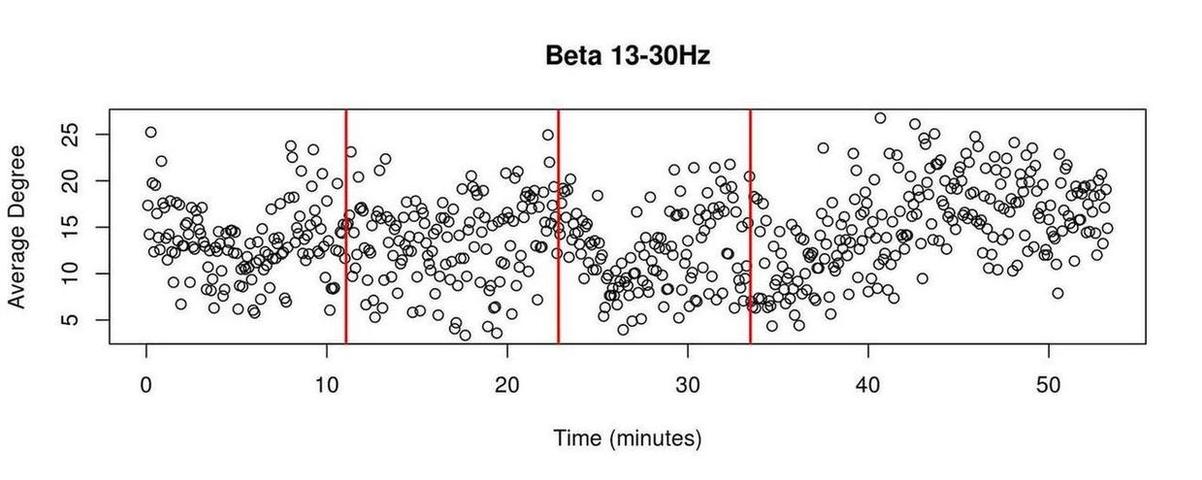}
  \caption{Occipital Lobe}
  \label{fig:sfig4}
\end{subfigure}\\
\caption{{\normalfont \textbf{Average Degree.}} Average degree vertex respective to the networks of the Beta frequency band (13-30Hz). Vertical axis average degree; Horizontal axis time (minutes). At t=11 minutes, the monkey was blindfolded, the first red line in each sub-figure represents the moment when a patch was placed over the eyes. At t=23 minutes the Ketamine-Medetomidine cocktail was injected, being represented by the second red line. The point of loss of consciousness (LOC) was registered at t=33 minutes, and is indicated by the third red line. Sub-figures: (\textbf{a}) Frontal Lobe; (\textbf{b}) Parietal Lobe; (\textbf{c}) Temporal Lobe; (\textbf{d}) Occipital Lobe.}
\label{fig:fig}
\end{figure}

\begin{table}[!h]
\centering
\caption{{\normalfont \textbf{Average Degree.}}  Mean, variance (Var), and standard deviation (SD) of the average degree vertex of the networks respective to each one of the four cortical lobes, on the three different conditions in which the monkey was exposed during the experiment: awake with eyes open, awake with eyes closed and anesthesia (eyes closed). Frequency band Beta (13-30Hz). }
\vspace{0.5cm}
\begin{tabular}{l|lcr|lcr|lcr}
\hline 
\textbf{Beta Band (13-30Hz)} & \multicolumn{3}{c}{Eyes Open} \vline &\multicolumn{3}{c}{Eyes Closed} \vline &\multicolumn{3}{c}{Anesthesia}\\
\hline
Corresponding Graph & Mean   & Var  & SD  & Mean  & Var  & SD & Mean   & Var  & SD\\ 
\hline                             

Frontal Lobe    &32.9   &89.3  &9.45       &34.1  &99.2   &9.96      &3.29    &2.48    &1.57   \\

Parietal Lobe   &12.7  &7.27  &2.70      &14.7  &9.39   &3.06      &4.24    &4.77  &2.18   \\

Temporal Lobe   &11.8  &5.90  &2.43      &12.0  &4.84   &2.20     &4.97    & 2.90   &    1.70\\

Occipital Lobe  &13.5  & 15.1 &3.88      &13.6  &20.1   & 4.48     &14.8    &23.2    &4.82    \\   

\end{tabular}
\end{table}

\clearpage

\subsubsection*{Gamma 25-100Hz}
\begin{figure}[!h]
\begin{subfigure}{.5\textwidth}
  \centering
  \includegraphics[width=1\linewidth]{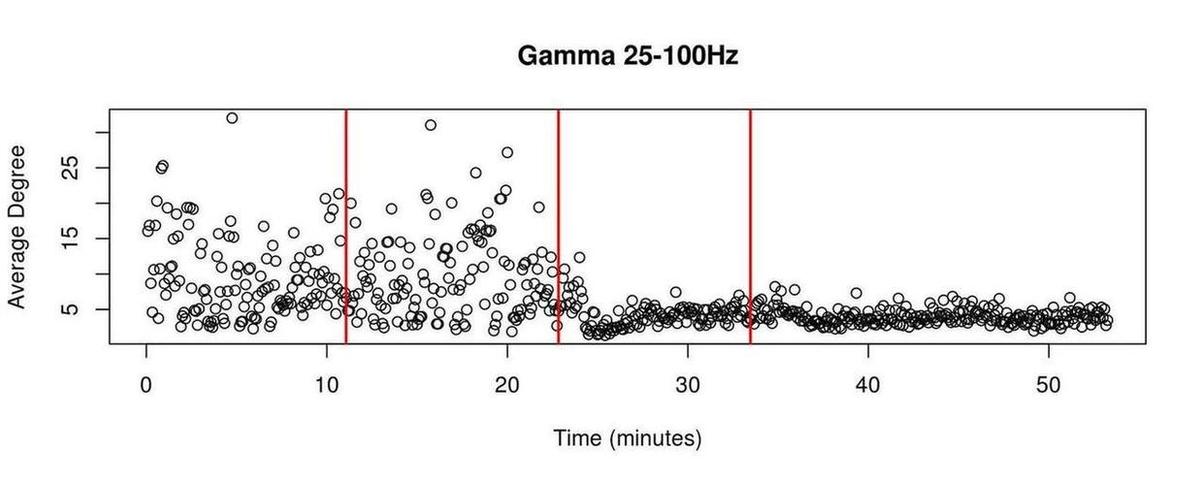}
  \caption{Frontal Lobe}
  \label{fig:sfig1}
\end{subfigure}%
\begin{subfigure}{.5\textwidth}
  \centering
  \includegraphics[width=1\linewidth]{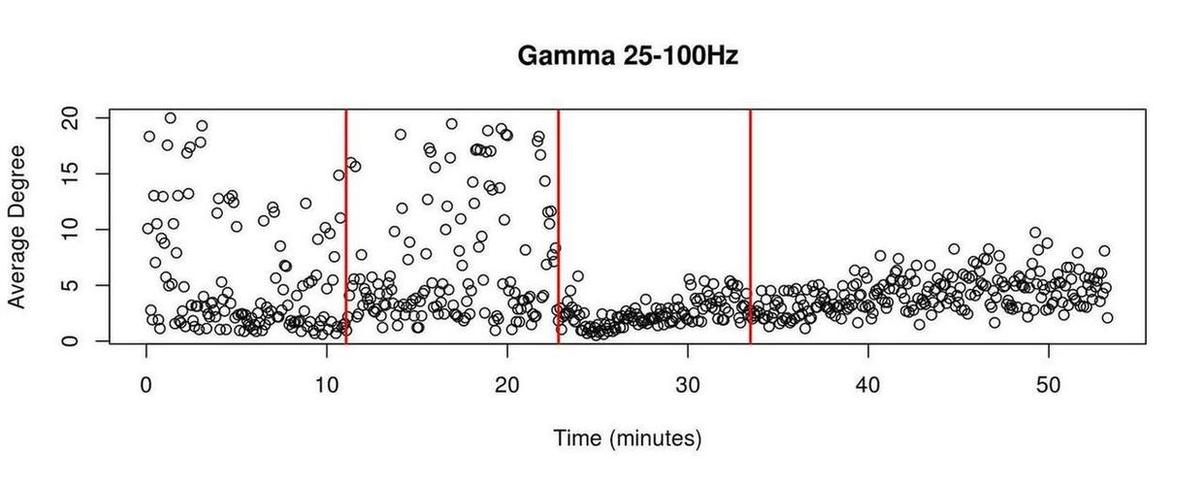}
 \caption{Parietal Lobe}
  \label{fig:sfig2}
\end{subfigure}\\
\centering
\begin{subfigure}{.5\textwidth}
\includegraphics[width=1\linewidth]{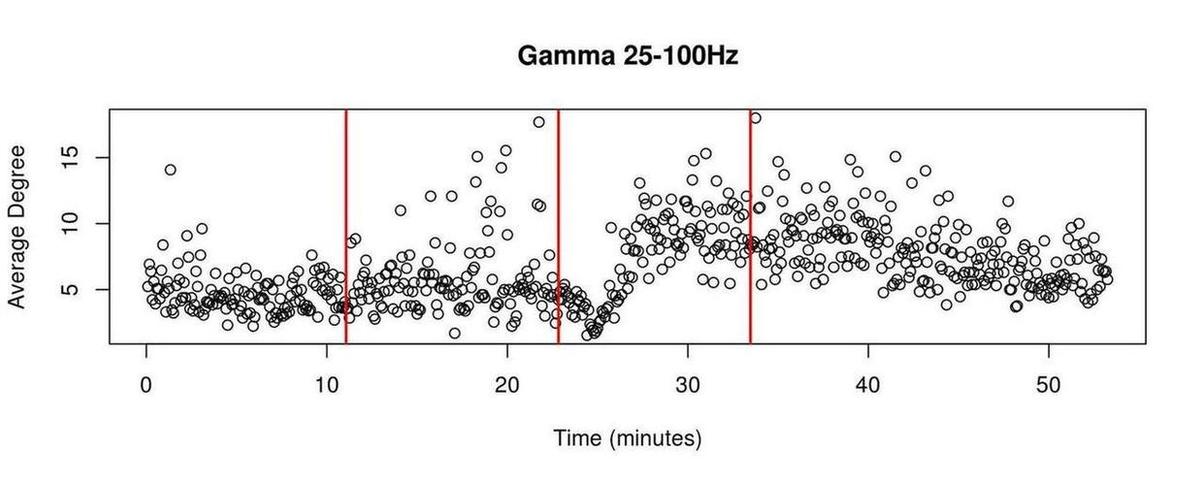}
  \caption{Temporal Lobe}
  \label{fig:sfig3}
\end{subfigure}%
\begin{subfigure}{.5\textwidth}
  \centering
  \includegraphics[width=1\linewidth]{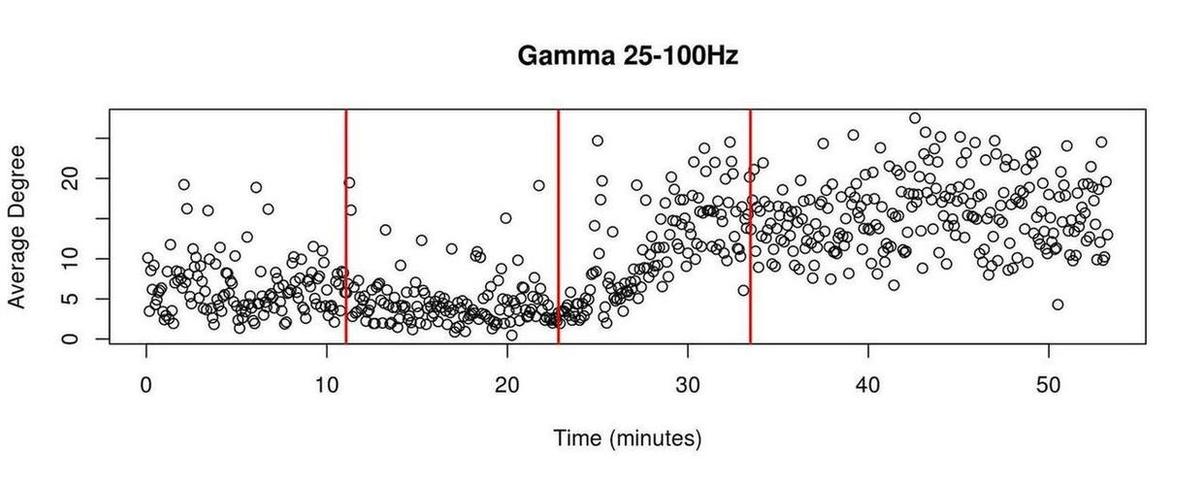}
  \caption{Occipital Lobe}
  \label{fig:sfig4}
\end{subfigure}\\
\caption{{\normalfont \textbf{Average Degree.}} Average degree vertex respective to the networks of the  Gamma frequency band (25-30Hz). Vertical axis average degree; Horizontal axis time (minutes). At t=11 minutes, the monkey was blindfolded, the first red line in each sub-figure represents the moment when a patch was placed over the eyes. At t=23 minutes the Ketamine-Medetomidine cocktail was injected, being represented by the second red line. The point of loss of consciousness (LOC) was registered at t=33 minutes, and is indicated by the third red line. Sub-figures: (\textbf{a}) Frontal Lobe; (\textbf{b}) Parietal Lobe; (\textbf{c}) Temporal Lobe; (\textbf{d}) Occipital Lobe.}
\label{fig:fig}
\end{figure}

\begin{table}[!h]
\centering
\caption{{\normalfont \textbf{Average Degree.}} Mean, variance (Var), and standard deviation (SD) of the average degree vertex of the networks respective to each one of the four cortical lobes, on the three different conditions in which the monkey was exposed during the experiment: awake with eyes open, awake with eyes closed and anesthesia (eyes closed). Frequency band Gamma (25-100Hz). }
\vspace{0.5cm}
\begin{tabular}{l|lcr|lcr|lcr}
\hline 
\textbf{Gamma Band (25-100Hz)} & \multicolumn{3}{c}{Eyes Open} \vline &\multicolumn{3}{c}{Eyes Closed} \vline &\multicolumn{3}{c}{Anesthesia}\\
\hline
Corresponding Graph & Mean   & Var  & SD  & Mean  & Var  & SD & Mean   & Var  & SD\\ 
\hline                             

Frontal Lobe    & 9.28  & 31.2 &5.59       &9.64  & 36.1  &6.01      & 4.08   &  1.17  & 1.08  \\

Parietal Lobe   &5.18  &24.1  &4.91      &7.03  &31.5   &5.62      &3.94    &2.53  &1.59   \\

Temporal Lobe   & 4.75 & 2.65 & 1.63     &5.92  & 8.43  &  2.90    & 8.16   &   5.85 &  2.42  \\

Occipital Lobe  & 6.47 & 11.9 &  3.43    & 4.37 &  8.65 &  2.94    &  15.2  & 19.4  &    4.40\\   

\end{tabular}
\end{table}

\clearpage

\subsubsection{Assortativity}

\subsubsection*{Delta 0-4Hz}
\begin{figure}[!h]
\begin{subfigure}{.5\textwidth}
  \centering
  \includegraphics[width=1\linewidth]{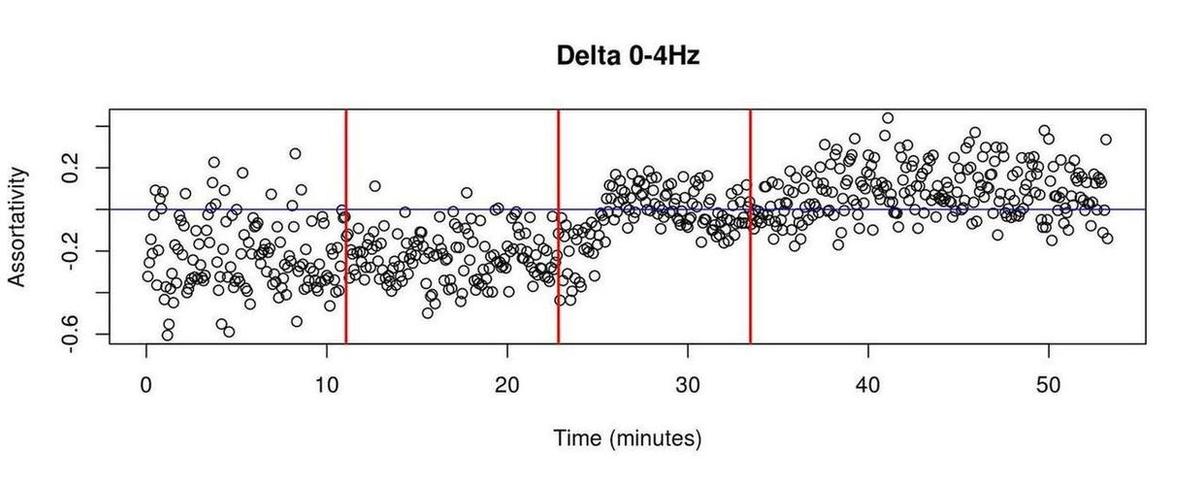}
  \caption{Frontal Lobe}
  \label{fig:sfig1}
\end{subfigure}%
\begin{subfigure}{.5\textwidth}
  \centering
  \includegraphics[width=1\linewidth]{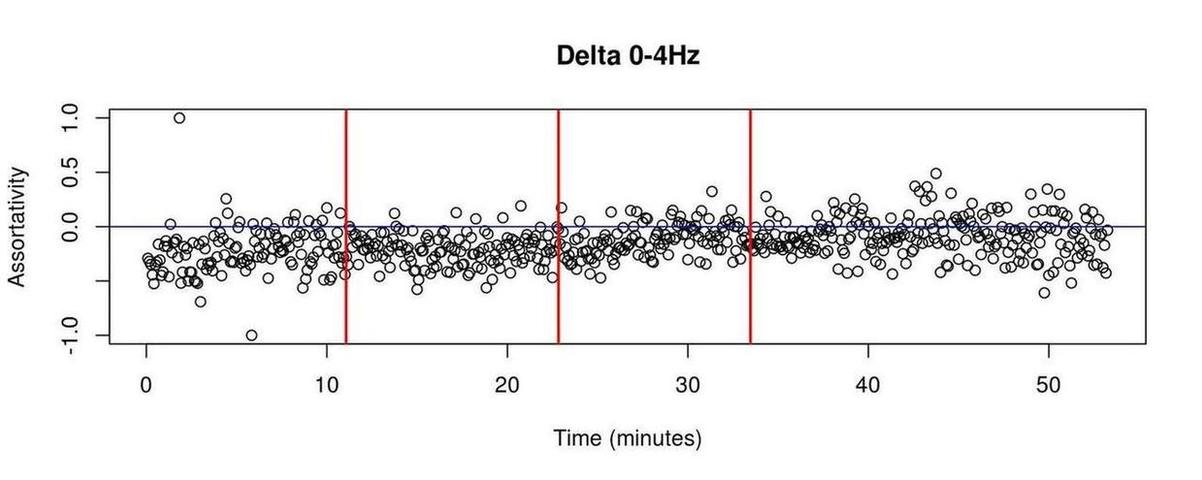}
 \caption{Parietal Lobe}
  \label{fig:sfig2}
\end{subfigure}\\
\centering
\begin{subfigure}{.5\textwidth}
\includegraphics[width=1\linewidth]{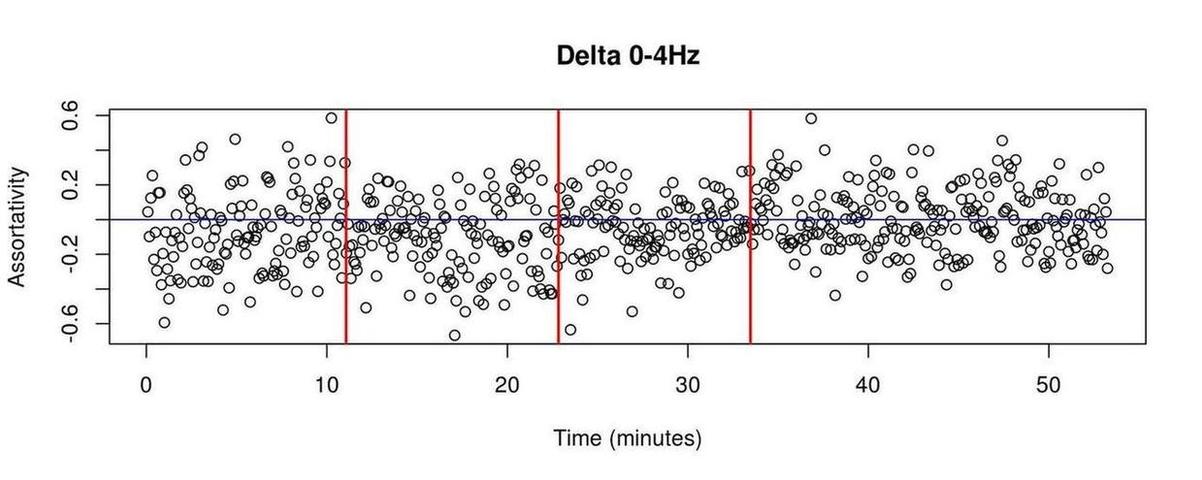}
  \caption{Temporal Lobe}
  \label{fig:sfig3}
\end{subfigure}%
\begin{subfigure}{.5\textwidth}
  \centering
  \includegraphics[width=1\linewidth]{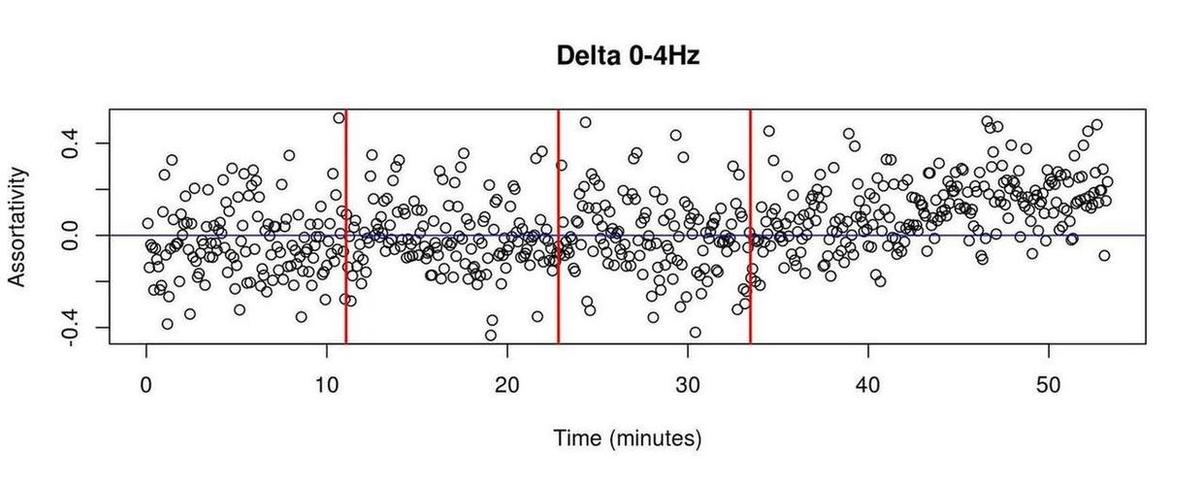}
  \caption{Occipital Lobe}
  \label{fig:sfig4}
\end{subfigure}\\
\caption{{\normalfont \textbf{Assortativity.} } Assortativity respective to the networks of the  Delta frequency band (0-4Hz). Vertical axis assortativity; Horizontal axis time (minutes). At t=11 minutes, the monkey was blindfolded, the first red line in each sub-figure represents the moment when a patch was placed over the eyes. At t=23 minutes the Ketamine-Medetomidine cocktail was injected, being represented by the second red line. The point of loss of consciousness (LOC) was registered at t=33 minutes, and is indicated by the third red line. Sub-figures: (\textbf{a}) Frontal Lobe; (\textbf{b}) Parietal Lobe; (\textbf{c}) Temporal Lobe; (\textbf{d}) Occipital Lobe.}
\label{fig:fig}
\end{figure}

\begin{table}[!h]
\centering
\caption{{\normalfont \textbf{Assortativity.}}  Mean, variance (Var), and standard deviation (SD) of the assortativity of the networks respective to each one of the four cortical lobes, on the three different conditions in which the monkey was exposed during the experiment: awake with eyes open, awake with eyes closed and anesthesia (eyes closed). Frequency band Delta (0-4Hz). }
\vspace{0.5cm}
\begin{tabular}{l|lcr|lcr|lcr}
\hline 
\textbf{Delta Band (0-4Hz)} & \multicolumn{3}{c}{Eyes Open} \vline &\multicolumn{3}{c}{Eyes Closed} \vline &\multicolumn{3}{c}{Anesthesia}\\
\hline
Corresponding Graph & Mean   & Var  & SD  & Mean  & Var  & SD & Mean   & Var  & SD\\ 
\hline                             

Frontal Lobe    & -0.22  & 0.03 &0.18       & -0.23 & 0.01  & 0.12     & 0.06   &   0.01 & 0.12  \\

Parietal Lobe   &-0.22  & 0.05 &0.22      &-0.22  &0.02   &0.15      &-0.09    &0.03  &0.17   \\

Temporal Lobe   & -0.07 & 0.05 & 0.22     & -0.11 &  0.05 &  0.22    & -0.01   & 0.03   &  0.17  \\

Occipital Lobe  & -0.04 & 0.02 & 0.15     & -0.01 &  0.02 & 0.14     & 0.08   &  0.03  &  0.16  \\   

\end{tabular}
\end{table}

\clearpage

\subsubsection*{Theta 4-8Hz}

\begin{figure}[!h]
\begin{subfigure}{.5\textwidth}
  \centering
  \includegraphics[width=1\linewidth]{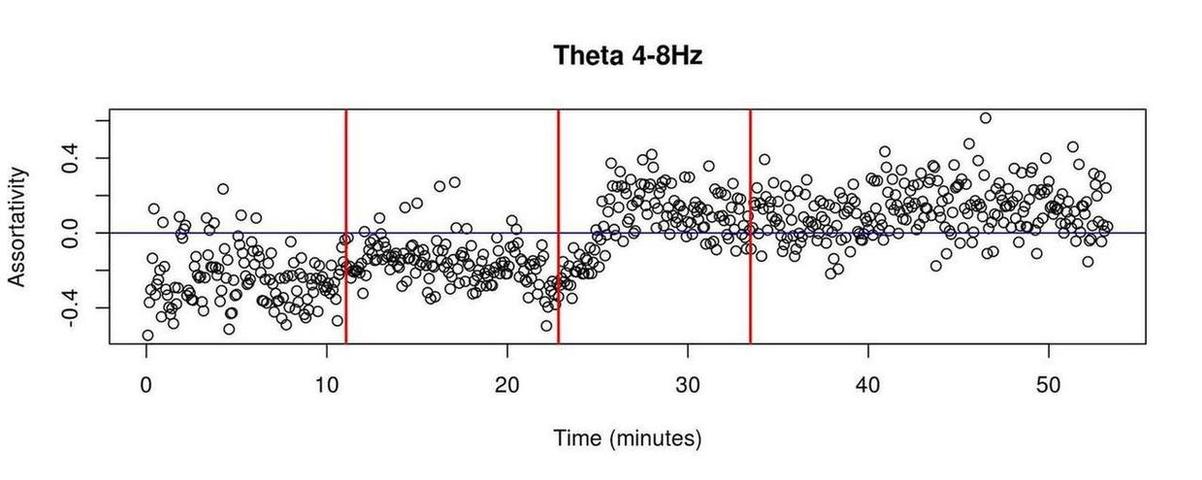}
  \caption{Frontal Lobe}
  \label{fig:sfig1}
\end{subfigure}%
\begin{subfigure}{.5\textwidth}
  \centering
  \includegraphics[width=1\linewidth]{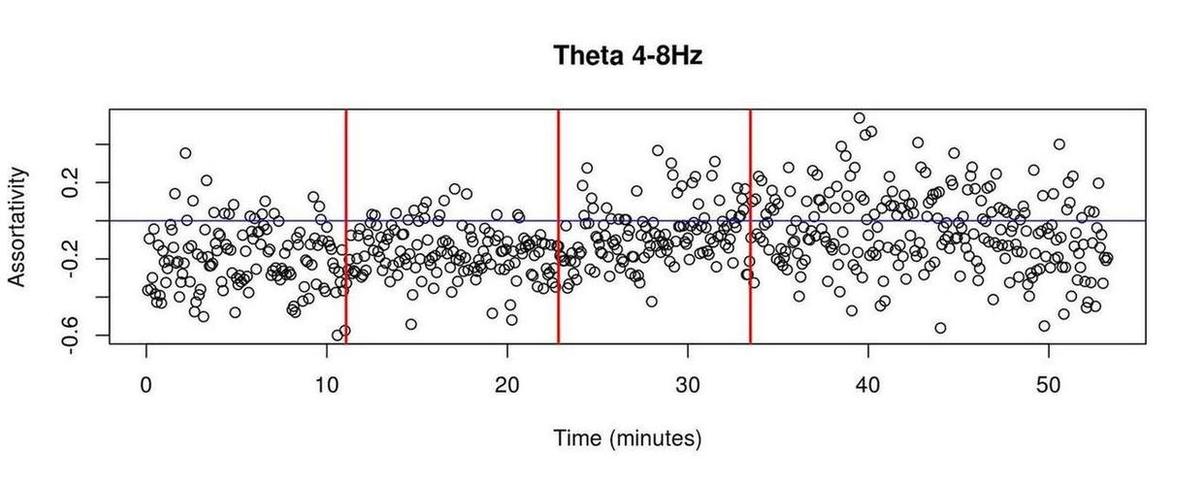}
 \caption{Parietal Lobe}
  \label{fig:sfig2}
\end{subfigure}\\
\centering
\begin{subfigure}{.5\textwidth}
\includegraphics[width=1\linewidth]{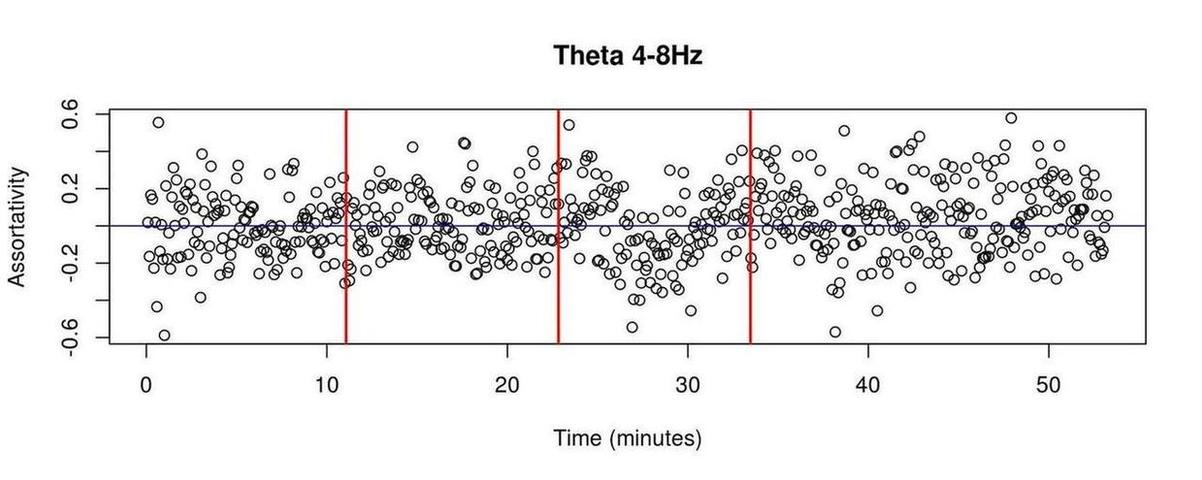}
  \caption{Temporal Lobe}
  \label{fig:sfig3}
\end{subfigure}%
\begin{subfigure}{.5\textwidth}
  \centering
  \includegraphics[width=1\linewidth]{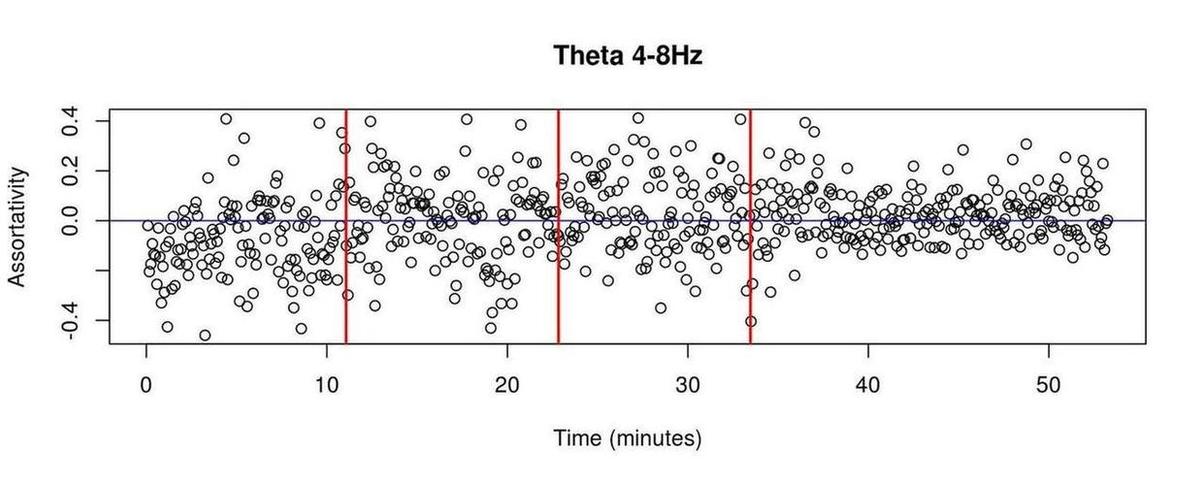}
  \caption{Occipital Lobe}
  \label{fig:sfig4}
\end{subfigure}\\
\caption{{\normalfont \textbf{Assortativity.}} Assortativity respective to the networks of the  Theta frequency band (4-8Hz). Vertical axis  assortativity; Horizontal axis time (minutes). At t=11 minutes, the monkey was blindfolded, the first red line in each sub-figure represents the moment when a patch was placed over the eyes. At t=23 minutes the Ketamine-Medetomidine cocktail was injected, being represented by the second red line. The point of loss of consciousness (LOC) was registered at t=33 minutes, and is indicated by the third red line. Sub-figures: (\textbf{a}) Frontal Lobe; (\textbf{b}) Parietal Lobe; (\textbf{c}) Temporal Lobe; (\textbf{d}) Occipital Lobe.}
\label{fig:fig}
\end{figure}


\begin{table}[!h]
\centering
\caption{{\normalfont \textbf{Assortativity.}}  Mean, variance (Var), and standard deviation (SD) of the assortativity of the networks respective to each one of the four cortical lobes, on the three different conditions in which the monkey was exposed during the experiment: awake with eyes open, awake with eyes closed and anesthesia (eyes closed). Frequency band Theta (4-8Hz). }
\vspace{0.5cm}
\begin{tabular}{l|lcr|lcr|lcr}
\hline 
\textbf{Theta Band (4-8Hz)} & \multicolumn{3}{c}{Eyes Open} \vline &\multicolumn{3}{c}{Eyes Closed} \vline &\multicolumn{3}{c}{Anesthesia}\\
\hline
Corresponding Graph & Mean   & Var  & SD  & Mean  & Var  & SD & Mean   & Var  & SD\\ 
\hline                             

Frontal Lobe    &-0.24   & 0.02 &0.15       &-0.17  &0.01   &0.12      & 0.12   &   0.02 & 0.13  \\

Parietal Lobe   &-0.19  &0.03  &0.17      &-0.17  &0.02   &0.13      &-0.05    &0.04  &0.19   \\

Temporal Lobe   &-0.01  & 0.03 &0.18      & 0.02 & 0.03  &  0.17    &  0.03  & 0.04   &  0.19  \\

Occipital Lobe  & -0.08 & 0.02 & 0.15     & 0.0 & 0.02  & 0.15    &  0.02  & 0.01   &    0.12\\   

\end{tabular}
\end{table}

\clearpage

\subsubsection*{Alpha 8-12Hz}

\begin{figure}[!h]
\begin{subfigure}{.5\textwidth}
  \centering
  \includegraphics[width=1\linewidth]{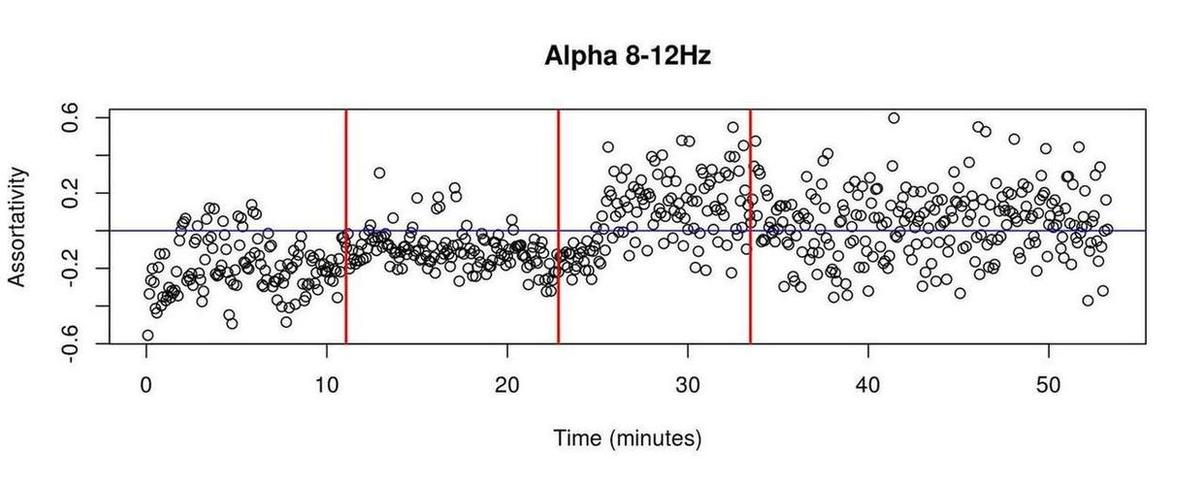}
  \caption{Frontal Lobe}
  \label{fig:sfig1}
\end{subfigure}%
\begin{subfigure}{.5\textwidth}
  \centering
  \includegraphics[width=1\linewidth]{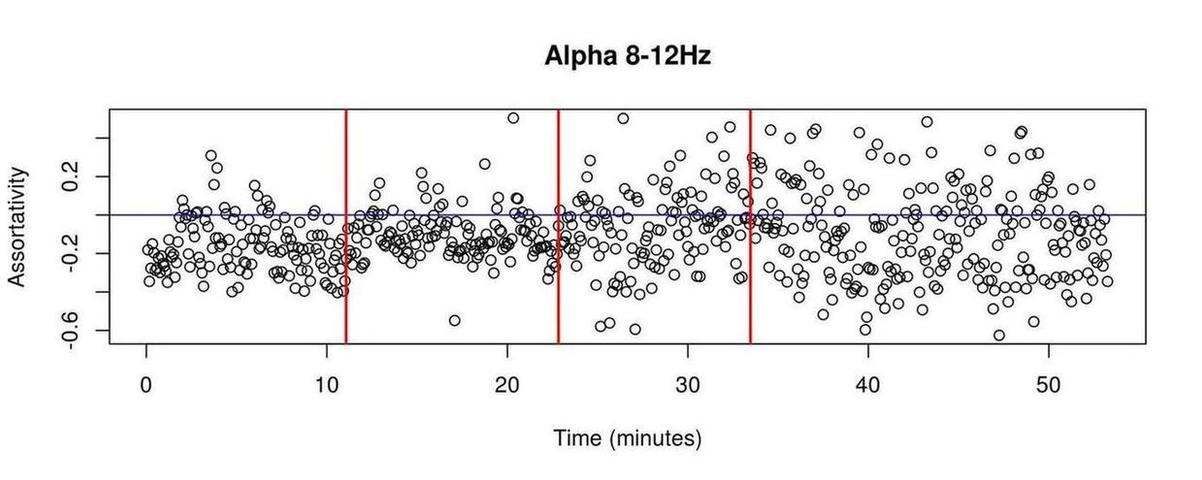}
 \caption{Parietal Lobe}
  \label{fig:sfig2}
\end{subfigure}\\
\centering
\begin{subfigure}{.5\textwidth}
\includegraphics[width=1\linewidth]{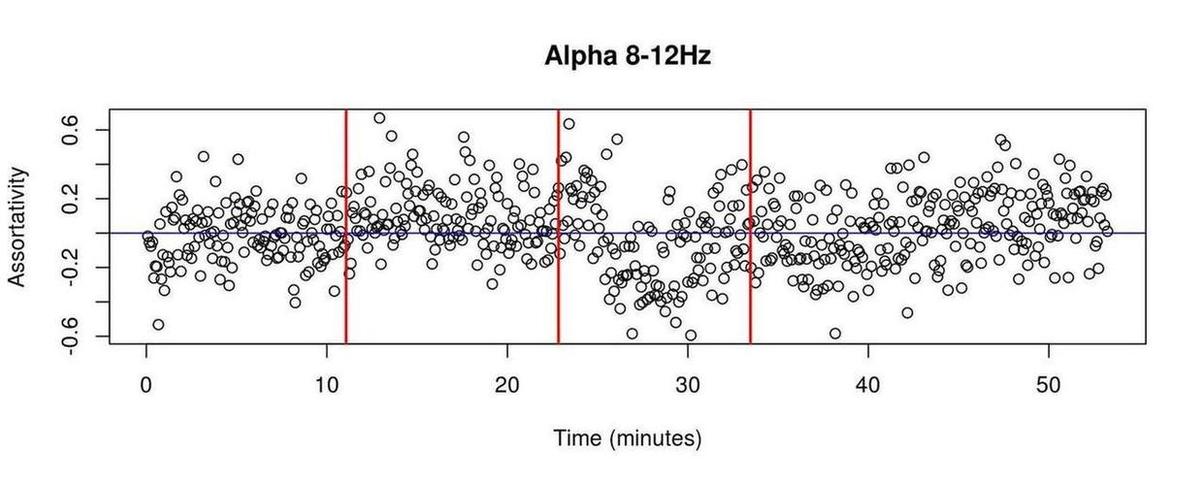}
  \caption{Temporal Lobe}
  \label{fig:sfig3}
\end{subfigure}%
\begin{subfigure}{.5\textwidth}
  \centering
  \includegraphics[width=1\linewidth]{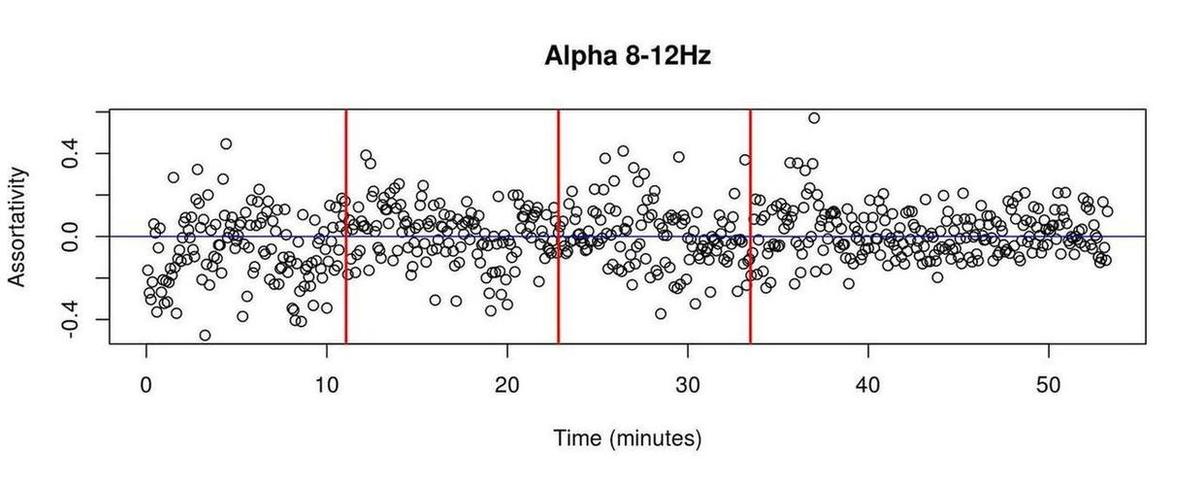}
  \caption{Occipital Lobe}
  \label{fig:sfig4}
\end{subfigure}\\
\caption{{\normalfont \textbf{Assortativity.}} Assortativity respective to the networks of the  Alpha frequency band (8-12Hz). Vertical axis  assortativity; Horizontal axis time (minutes). At t=11 minutes, the monkey was blindfolded, the first red line in each sub-figure represents the moment when a patch was placed over the eyes. At t=23 minutes the Ketamine-Medetomidine cocktail was injected, being represented by the second red line. The point of loss of consciousness (LOC) was registered at t=33 minutes, and is indicated by the third red line. Sub-figures: (\textbf{a}) Frontal Lobe; (\textbf{b}) Parietal Lobe; (\textbf{c}) Temporal Lobe; (\textbf{d}) Occipital Lobe.}
\label{fig:fig}
\end{figure}


\begin{table}[!h]
\centering
\caption{{\normalfont \textbf{Assortativity.}}  Mean, variance (Var), and standard deviation (SD) of the assortativity of the networks respective to each one of the four cortical lobes, on the three different conditions in which the monkey was exposed during the experiment: awake with eyes open, awake with eyes closed and anesthesia (eyes closed). Frequency band Alpha (8-12Hz). }
\vspace{0.5cm}
\begin{tabular}{l|lcr|lcr|lcr}
\hline 
\textbf{Alpha Band (8-12Hz)} & \multicolumn{3}{c}{Eyes Open} \vline &\multicolumn{3}{c}{Eyes Closed} \vline &\multicolumn{3}{c}{Anesthesia}\\
\hline
Corresponding Graph & Mean   & Var  & SD  & Mean  & Var  & SD & Mean   & Var  & SD\\ 
\hline                             

Frontal Lobe  &-0.20   &0.02  &0.15       &-0.11  &0.01   &0.10      &0.06   &0.03   &0.18   \\

Parietal Lobe   &-0.16  &0.02  &0.14      &-0.11  &0.02   &0.13      &-0.08    &0.05  &0.23   \\

Temporal Lobe  &-0.02   &0.03  &0.17       &0.10  &0.03   &0.17      &0.00   &0.04   &0.21      \\

Occipital Lobe  &-0.07   &0.03  &0.17       &0.01  &0.02   &0.13      &0.01   &0.02   &0.12     \\   

\end{tabular}
\end{table}

\clearpage

\subsubsection*{Beta 13-30Hz}

\begin{figure}[!h]
\begin{subfigure}{.5\textwidth}
  \centering
  \includegraphics[width=1\linewidth]{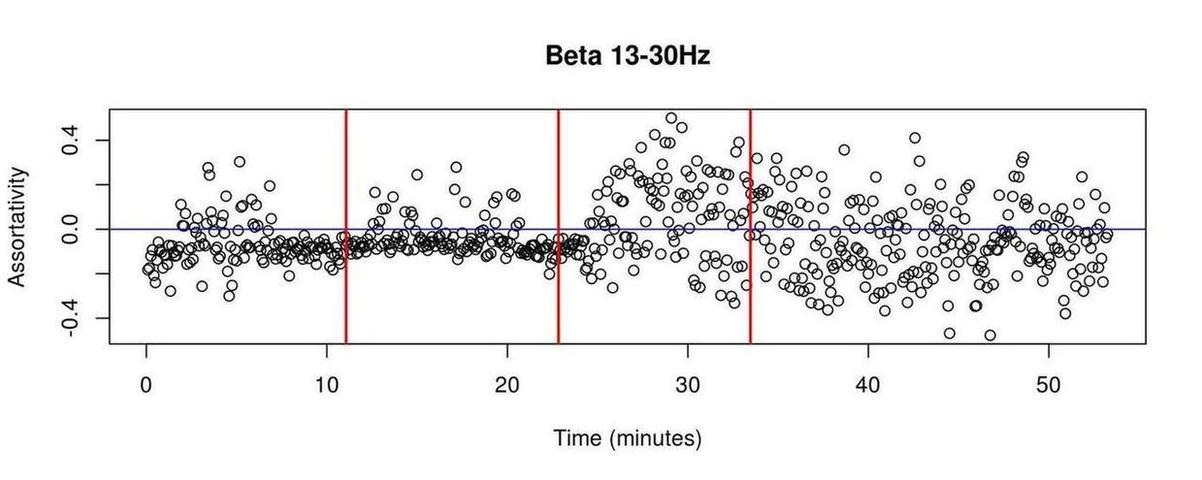}
  \caption{Frontal Lobe}
  \label{fig:sfig1}
\end{subfigure}%
\begin{subfigure}{.5\textwidth}
  \centering
  \includegraphics[width=1\linewidth]{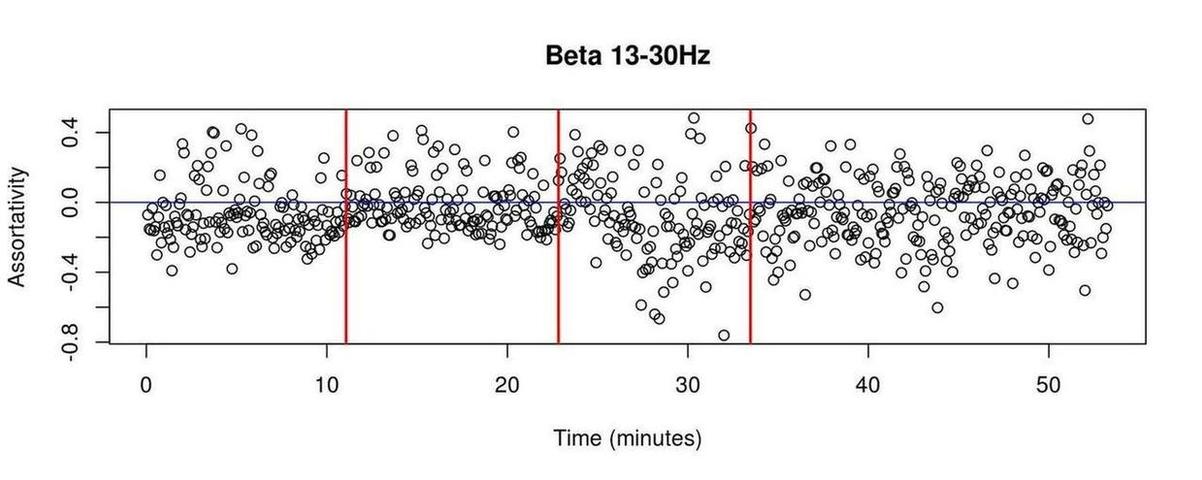}
 \caption{Parietal Lobe}
  \label{fig:sfig2}
\end{subfigure}\\
\centering
\begin{subfigure}{.5\textwidth}
\includegraphics[width=1\linewidth]{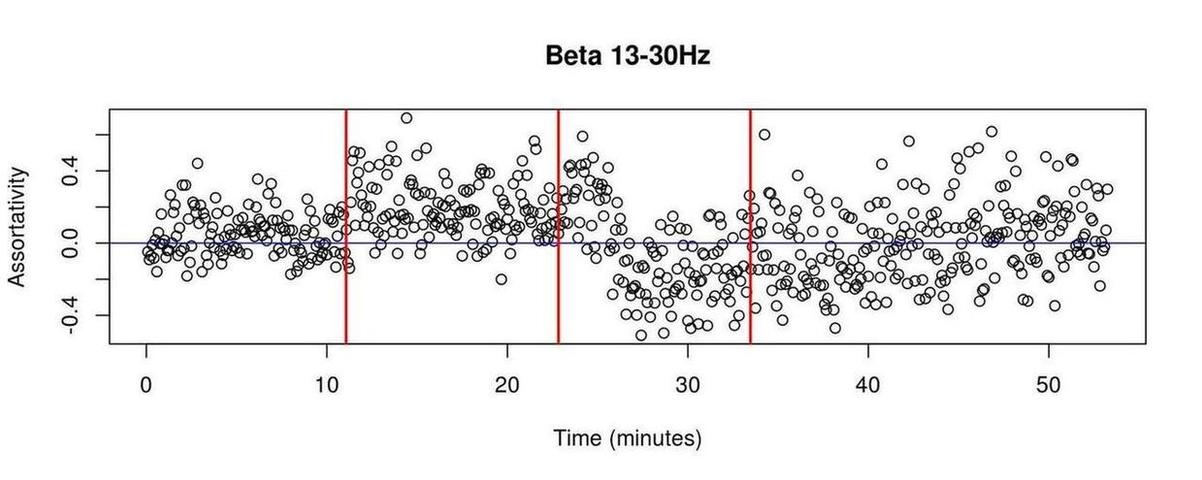}
  \caption{Temporal Lobe}
  \label{fig:sfig3}
\end{subfigure}%
\begin{subfigure}{.5\textwidth}
  \centering
  \includegraphics[width=1\linewidth]{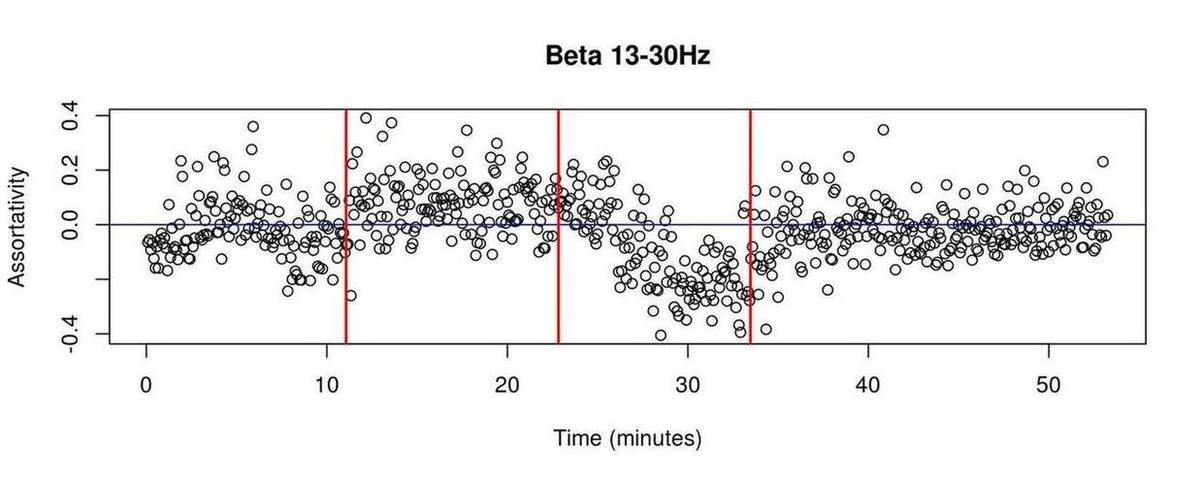}
  \caption{Occipital Lobe}
  \label{fig:sfig4}
\end{subfigure}\\
\caption{{\normalfont \textbf{Assortativity.}} Assortativity respective to the networks of the  Beta frequency band (13-30Hz). Vertical axis  assortativity; Horizontal axis time (minutes). At t=11 minutes, the monkey was blindfolded, the first red line in each sub-figure represents the moment when a patch was placed over the eyes. At t=23 minutes the Ketamine-Medetomidine cocktail was injected, being represented by the second red line. The point of loss of consciousness (LOC) was registered at t=33 minutes, and is indicated by the third red line. Sub-figures: (\textbf{a}) Frontal Lobe; (\textbf{b}) Parietal Lobe; (\textbf{c}) Temporal Lobe; (\textbf{d}) Occipital Lobe.}
\label{fig:fig}
\end{figure}


\begin{table}[!h]
\centering
\caption{{\normalfont \textbf{Assortativity.}}  Mean, variance (Var), and standard deviation (SD) of the assortativity of the networks respective to each one of the four cortical lobes, on the three different conditions in which the monkey was exposed during the experiment: awake with eyes open, awake with eyes closed and anesthesia (eyes closed). Frequency band Beta (13-30Hz). }
\vspace{0.5cm}
\begin{tabular}{l|lcr|lcr|lcr}
\hline 
\textbf{Beta Band (13-30Hz)} & \multicolumn{3}{c}{Eyes Open} \vline &\multicolumn{3}{c}{Eyes Closed} \vline &\multicolumn{3}{c}{Anesthesia}\\
\hline
Corresponding Graph & Mean   & Var  & SD  & Mean  & Var  & SD & Mean   & Var  & SD\\ 
\hline                             

Frontal Lobe & -0.07  &0.01  &0.10       & -0.04 &0.01   & 0.08     &-0.02 &0.03    & 0.18  \\

Parietal Lobe   &-0.07  &0.03  &0.17      &-0.02  &0.02   &0.15      &-0.07    &0.04  &0.20   \\

Temporal Lobe   &0.05  &0.02  &0.12      & 0.18 &0.02   & 0.15          &-0.03    & 0.05   & 0.22   \\

Occipital Lobe  &-0.02  &0.01  &0.11      & 0.08 & 0.01  & 0.10         &-0.06    & 0.01   & 0.12   \\   

\end{tabular}
\end{table}

\clearpage

\subsubsection*{Gamma 25-100Hz}

\begin{figure}[!h]
\begin{subfigure}{.5\textwidth}
  \centering
  \includegraphics[width=1\linewidth]{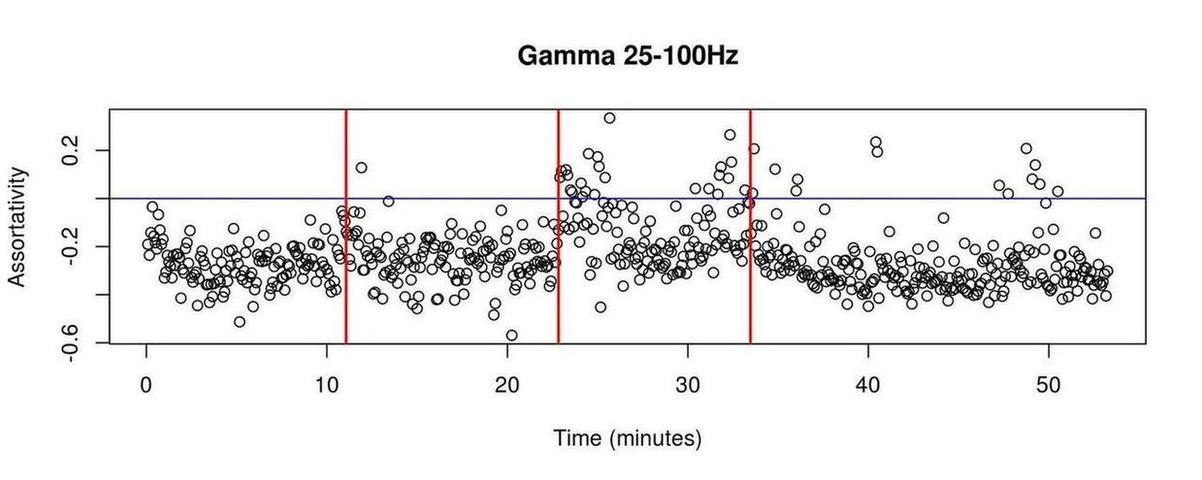}
  \caption{Frontal Lobe}
  \label{fig:sfig1}
\end{subfigure}%
\begin{subfigure}{.5\textwidth}
  \centering
  \includegraphics[width=1\linewidth]{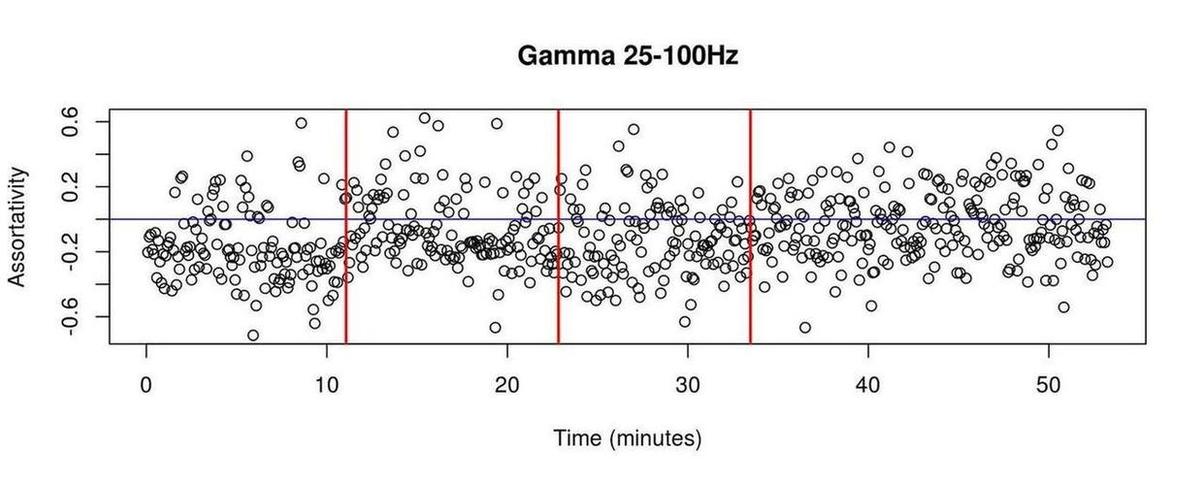}
 \caption{Parietal Lobe}
  \label{fig:sfig2}
\end{subfigure}\\
\centering
\begin{subfigure}{.5\textwidth}
\includegraphics[width=1\linewidth]{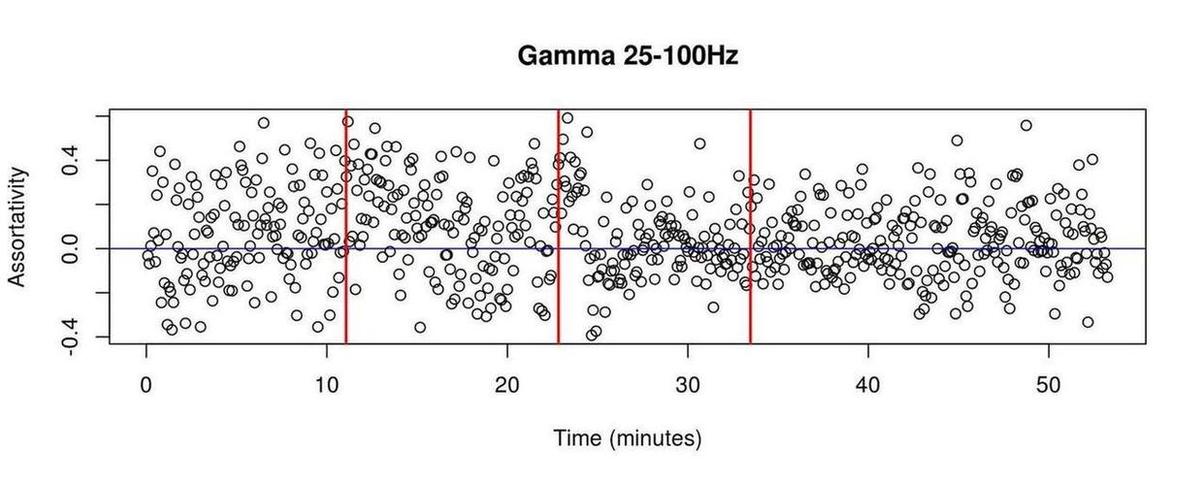}
  \caption{Temporal Lobe}
  \label{fig:sfig3}
\end{subfigure}%
\begin{subfigure}{.5\textwidth}
  \centering
  \includegraphics[width=1\linewidth]{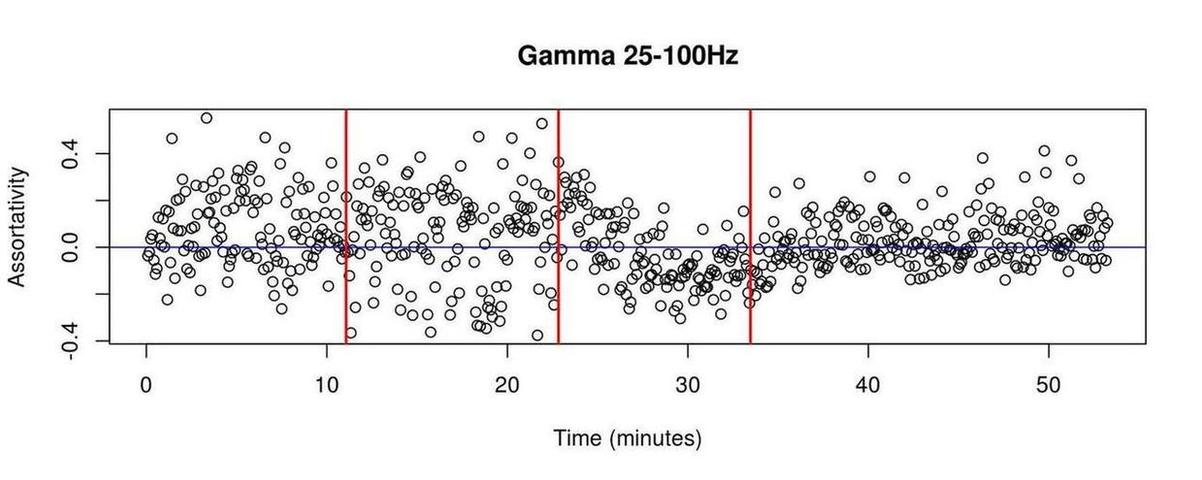}
  \caption{Occipital Lobe}
  \label{fig:sfig4}
\end{subfigure}\\
\caption{{\normalfont \textbf{Assortativity.}} Assortativity respective to the networks of the  Gamma frequency band (25-100Hz). Vertical axis  assortativity; Horizontal axis time (minutes). At t=11 minutes, the monkey was blindfolded, the first red line in each sub-figure represents the moment when a patch was placed over the eyes. At t=23 minutes the Ketamine-Medetomidine cocktail was injected, being represented by the second red line. The point of loss of consciousness (LOC) was registered at t=33 minutes, and is indicated by the third red line. Sub-figures: (\textbf{a}) Frontal Lobe; (\textbf{b}) Parietal Lobe; (\textbf{c}) Temporal Lobe; (\textbf{d}) Occipital Lobe.}
\label{fig:fig}
\end{figure}


\begin{table}[!h]
\centering
\caption{{\normalfont \textbf{Assortativity.}}  Mean, variance (Var), and standard deviation (SD) of the assortativity of the networks respective to each one of the four cortical lobes, on the three different conditions in which the monkey was exposed during the experiment: awake with eyes open, awake with eyes closed and anesthesia (eyes closed). Frequency band Gamma (25-100Hz). }
\vspace{0.5cm}
\begin{tabular}{l|lcr|lcr|lcr}
\hline 
\textbf{Gamma Band (25-100Hz)} & \multicolumn{3}{c}{Eyes Open} \vline &\multicolumn{3}{c}{Eyes Closed} \vline &\multicolumn{3}{c}{Anesthesia}\\
\hline
Corresponding Graph & Mean   & Var  & SD  & Mean  & Var  & SD & Mean   & Var  & SD\\ 
\hline                             

Frontal Lobe    &-0.27   &0.01  & 0.08      &-0.26 & 0.01  & 0.09     & -0.25   &   0.002 & 0.13  \\

Parietal Lobe   &-0.17  &0.05  &0.22      &0.07  &0.05   &0.22      &-0.06    &0.04  &0.20   \\

Temporal Lobe   &0.07  & 0.04 & 0.21     & 0.08 & 0.04  & 0.21     & 0.03   & 0.02   &  0.15  \\

Occipital Lobe  & 0.08 & 0.03 & 0.16     & 0.05 & 0.04  & 0.21     & 0.00   &  0.01  &  0.12  \\   

\end{tabular}
\end{table}

\clearpage

\subsection{Measures Related to Global Integration}

\subsubsection{Average Path Length}

\subsubsection*{Delta 0-4Hz}

\begin{figure}[!h]
\begin{subfigure}{.5\textwidth}
  \centering
  \includegraphics[width=1\linewidth]{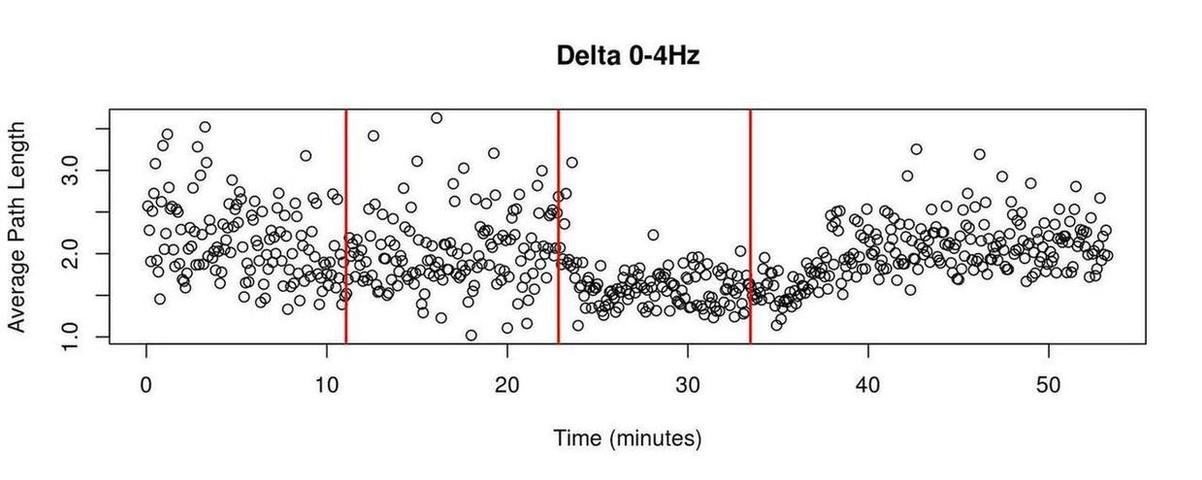}
  \caption{Frontal Lobe}
  \label{fig:sfig1}
\end{subfigure}%
\begin{subfigure}{.5\textwidth}
  \centering
  \includegraphics[width=1\linewidth]{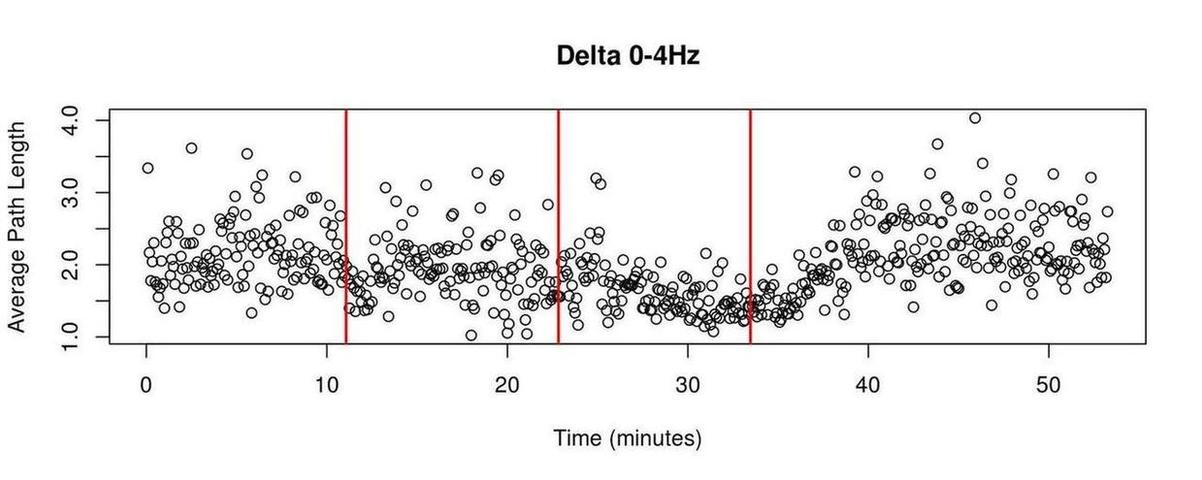}
 \caption{Parietal Lobe}
  \label{fig:sfig2}
\end{subfigure}\\
\centering
\begin{subfigure}{.5\textwidth}
\includegraphics[width=1\linewidth]{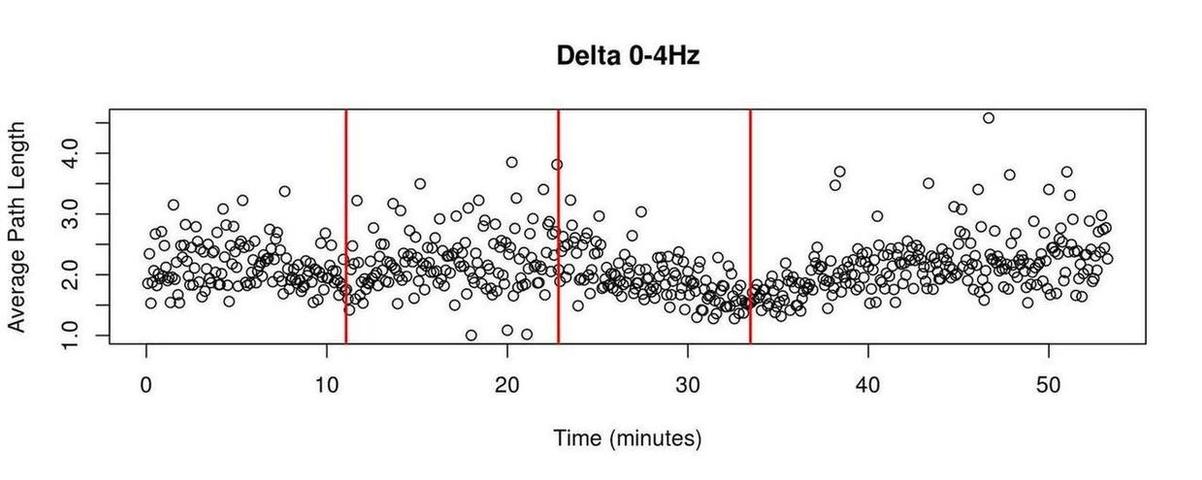}
  \caption{Temporal Lobe}
  \label{fig:sfig3}
\end{subfigure}%
\begin{subfigure}{.5\textwidth}
  \centering
  \includegraphics[width=1\linewidth]{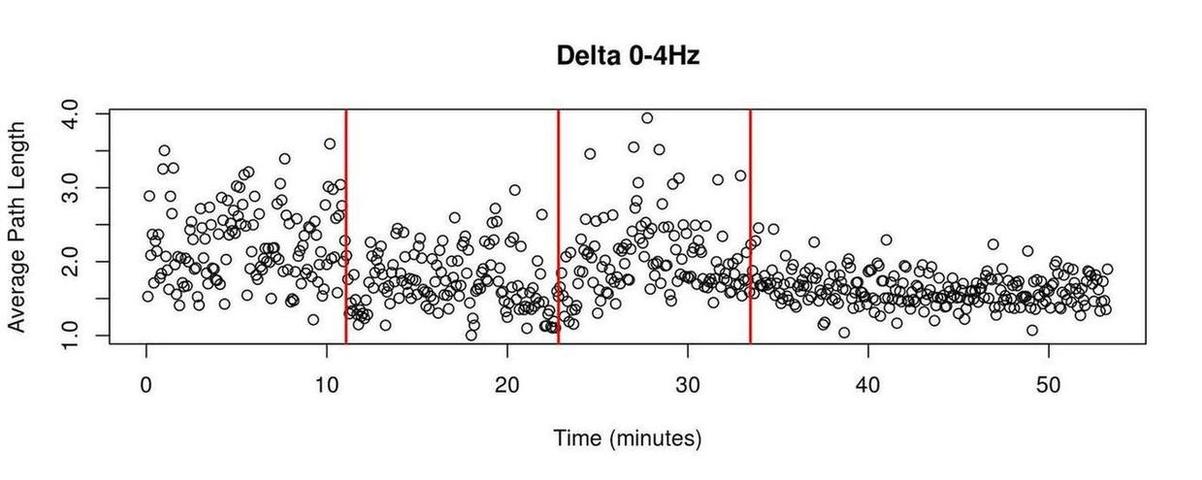}
  \caption{Occipital Lobe}
  \label{fig:sfig4}
\end{subfigure}\\
\caption{{\normalfont \textbf{Average Path Length.}} Average path length respective to the networks of the Delta frequency band (0-4Hz). Vertical axis average path length; Horizontal axis time (minutes). At t=11 minutes, the monkey was blindfolded, the first red line in each sub-figure represents the moment when a patch was placed over the eyes. At t=23 minutes the Ketamine-Medetomidine cocktail was injected, being represented by the second red line. The point of loss of consciousness (LOC) was registered at t=33 minutes, and is indicated by the third red line. Sub-figures: (\textbf{a}) Frontal Lobe; (\textbf{b}) Parietal Lobe; (\textbf{c}) Temporal Lobe; (\textbf{d}) Occipital Lobe.}
\label{fig:fig}
\end{figure}


\begin{table}[!h]
\centering
\caption{{\normalfont \textbf{Average Path Length.}}  Mean, variance (Var), and standard deviation (SD) of the average path length of the networks respective to each one of the four cortical lobes, on the three different conditions in which the monkey was exposed during the experiment: awake with eyes open, awake with eyes closed and anesthesia (eyes closed). Frequency band Delta (0-4Hz). }
\vspace{0.5cm}
\begin{tabular}{l|lcr|lcr|lcr}
\hline 
\textbf{Delta Band (0-4Hz)} & \multicolumn{3}{c}{Eyes Open} \vline &\multicolumn{3}{c}{Eyes Closed} \vline &\multicolumn{3}{c}{Anesthesia}\\
\hline
Corresponding Graph & Mean   & Var  & SD  & Mean  & Var  & SD & Mean   & Var  & SD\\ 
\hline                             

Frontal Lobe    & 2.16  & 0.22 & 0.46      & 2.03 & 0.20  & 0.45     & 1.91   & 0.13   &  0.37 \\

Parietal Lobe   &2.16  &0.20  &0.45      &1.97  &0.21   &0.45      &1.99    &0.28  &0.53   \\

Temporal Lobe   & 2.16 & 0.14 & 0.38     & 2.25 & 0.25  & 0.5     &  2.06  & 0.21   &    0.45\\

Occipital Lobe  & 2.19 & 0.24 & 0.49     & 1.74 & 0.16  &  0.4    &  1.72  &  0.11  &    0.33\\   

\end{tabular}
\end{table}

\clearpage

\subsubsection*{Theta 4-8Hz}

\begin{figure}[!h]
\begin{subfigure}{.5\textwidth}
  \centering
  \includegraphics[width=1\linewidth]{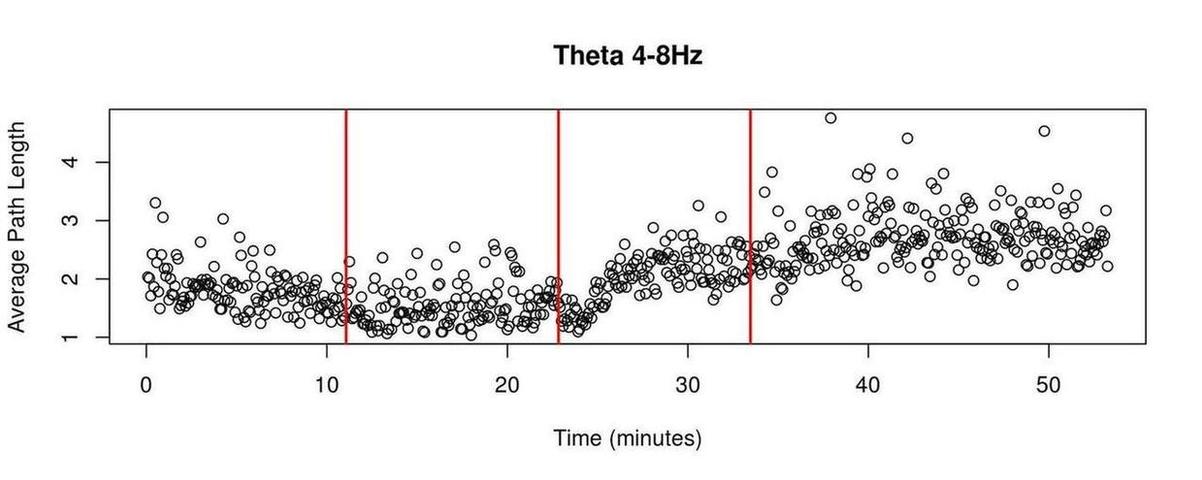}
  \caption{Frontal Lobe}
  \label{fig:sfig1}
\end{subfigure}%
\begin{subfigure}{.5\textwidth}
  \centering
  \includegraphics[width=1\linewidth]{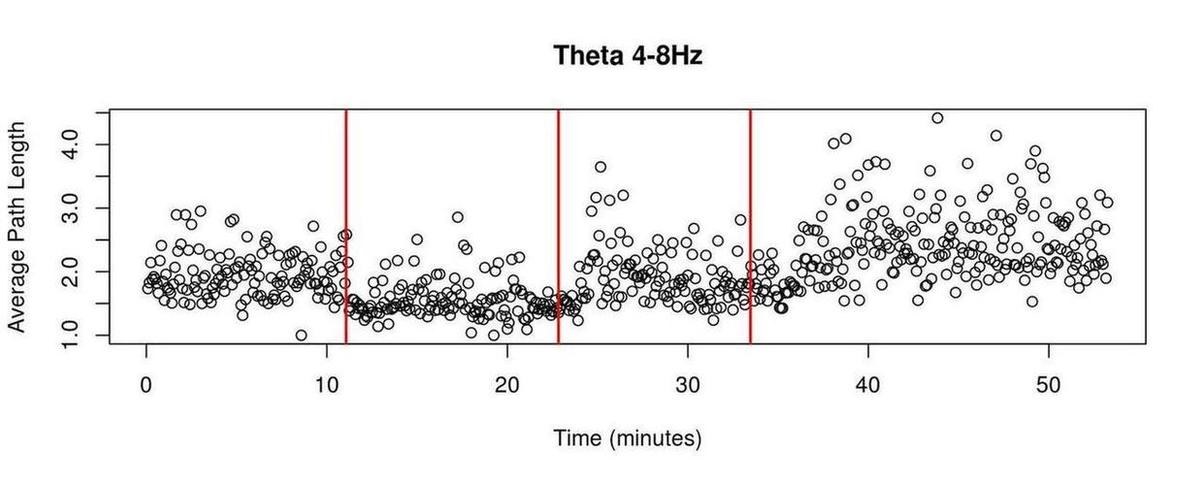}
 \caption{Parietal Lobe}
  \label{fig:sfig2}
\end{subfigure}\\
\centering
\begin{subfigure}{.5\textwidth}
\includegraphics[width=1\linewidth]{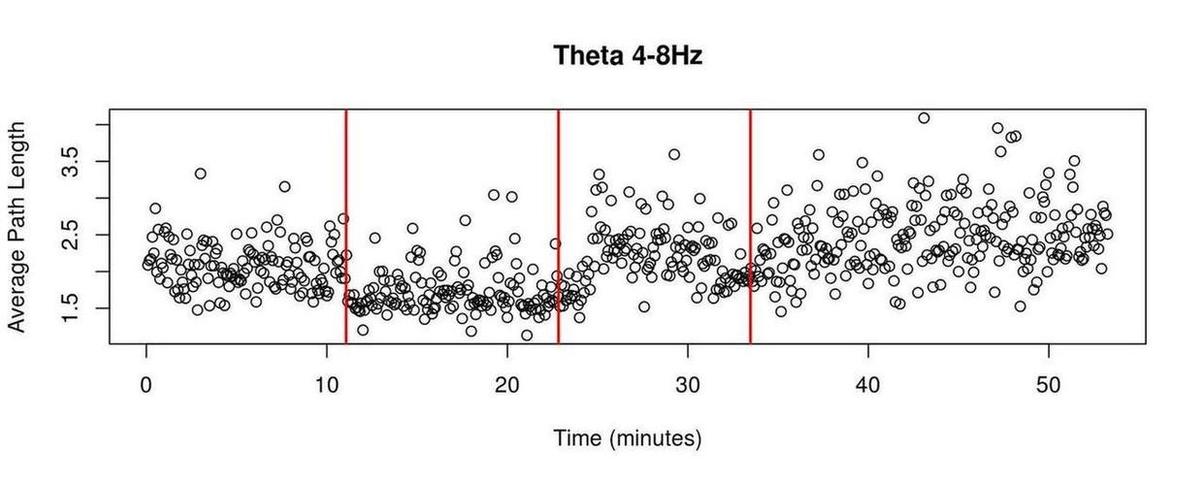}
  \caption{Temporal Lobe}
  \label{fig:sfig3}
\end{subfigure}%
\begin{subfigure}{.5\textwidth}
  \centering
  \includegraphics[width=1\linewidth]{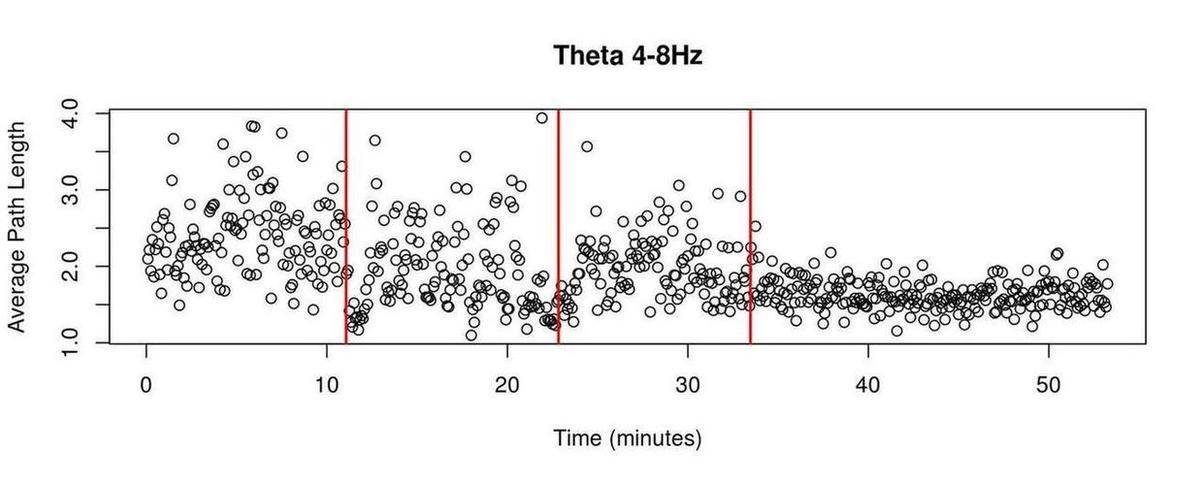}
  \caption{Occipital Lobe}
  \label{fig:sfig4}
\end{subfigure}\\
\caption{{\normalfont \textbf{Average Path Length.}} Average path length respective to the networks of the Theta frequency band (4-8Hz). Vertical axis average path length; Horizontal axis time (minutes). At t=11 minutes, the monkey was blindfolded, the first red line in each sub-figure represents the moment when a patch was placed over the eyes. At t=23 minutes the Ketamine-Medetomidine cocktail was injected, being represented by the second red line. The point of loss of consciousness (LOC) was registered at t=33 minutes, and is indicated by the third red line. Sub-figures: (\textbf{a}) Frontal Lobe; (\textbf{b}) Parietal Lobe; (\textbf{c}) Temporal Lobe; (\textbf{d}) Occipital Lobe.}
\label{fig:fig}
\end{figure}


\begin{table}[!h]
\centering
\caption{ {\normalfont \textbf{Average Path Length.}} Mean, variance (Var), and standard deviation (SD) of the average path length of the networks respective to each one of the four cortical lobes, on the three different conditions in which the monkey was exposed during the experiment: awake with eyes open, awake with eyes closed and anesthesia (eyes closed). Frequency band Theta (4-8Hz). }
\vspace{0.5cm}
\begin{tabular}{l|lcr|lcr|lcr}
\hline 
\textbf{Theta Band (4-8Hz)} & \multicolumn{3}{c}{Eyes Open} \vline &\multicolumn{3}{c}{Eyes Closed} \vline &\multicolumn{3}{c}{Anesthesia}\\
\hline
Corresponding Graph & Mean   & Var  & SD  & Mean  & Var  & SD & Mean   & Var  & SD\\ 
\hline                             

Frontal Lobe    &1.83  & 0.13 & 0.37      &1.57  & 0.13  & 0.36     & 2.60   & 0.28   & 0.47  \\

Parietal Lobe   &1.95  &0.13  &0.35      &1.57  &0.09   &0.31      &2.26    &0.33  &0.57   \\

Temporal Lobe   & 2.07 &0.10  & 0.32     & 1.74 & 0.11  &   0.32   & 2.40   & 0.21  &    0.46\\

Occipital Lobe  & 2.39 &0.27  & 0.52     & 1.95 & 0.29  &   0.54   & 1.71   &  0.09  & 0.30   \\   

\end{tabular}
\end{table}

\clearpage

\subsubsection*{Alpha 8-12Hz}

\begin{figure}[!h]
\begin{subfigure}{.5\textwidth}
  \centering
  \includegraphics[width=1\linewidth]{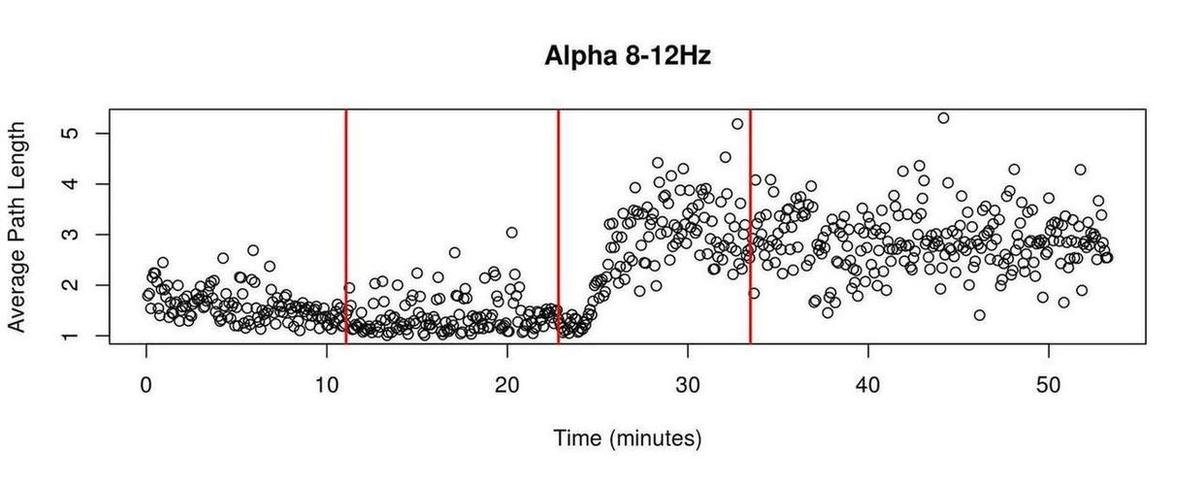}
  \caption{Frontal Lobe}
  \label{fig:sfig1}
\end{subfigure}%
\begin{subfigure}{.5\textwidth}
  \centering
  \includegraphics[width=1\linewidth]{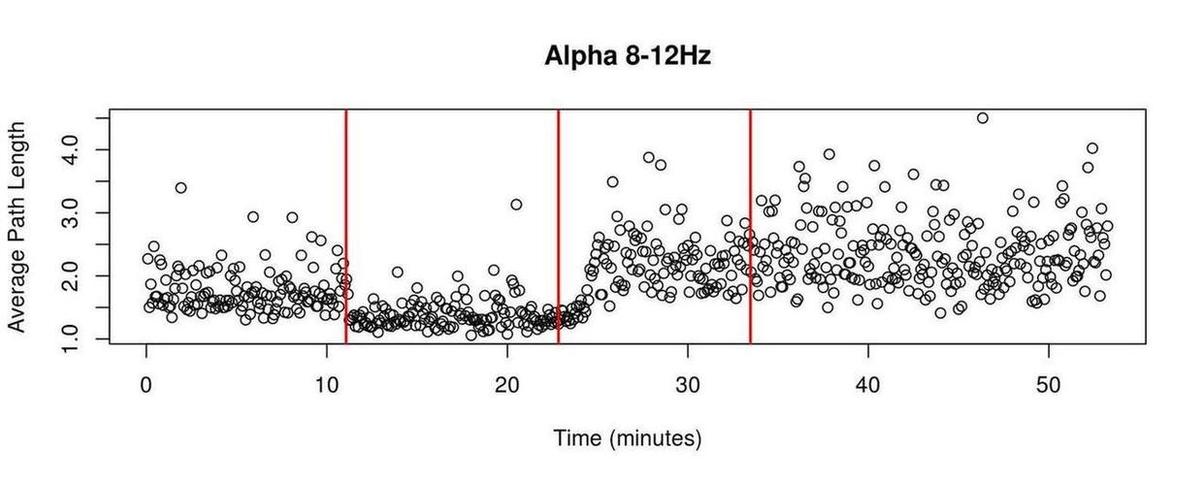}
 \caption{Parietal Lobe}
  \label{fig:sfig2}
\end{subfigure}\\
\centering
\begin{subfigure}{.5\textwidth}
\includegraphics[width=1\linewidth]{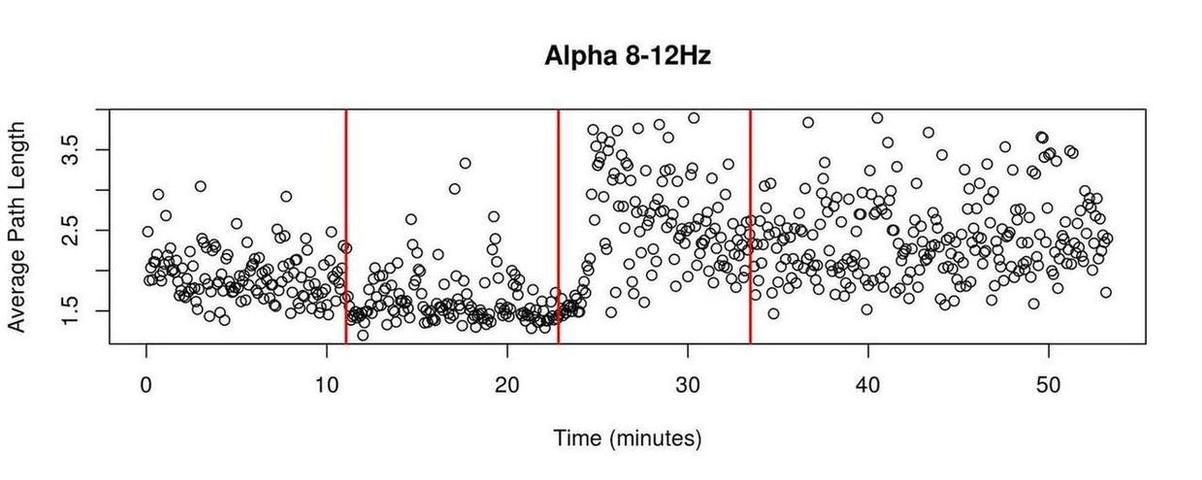}
  \caption{Temporal Lobe}
  \label{fig:sfig3}
\end{subfigure}%
\begin{subfigure}{.5\textwidth}
  \centering
  \includegraphics[width=1\linewidth]{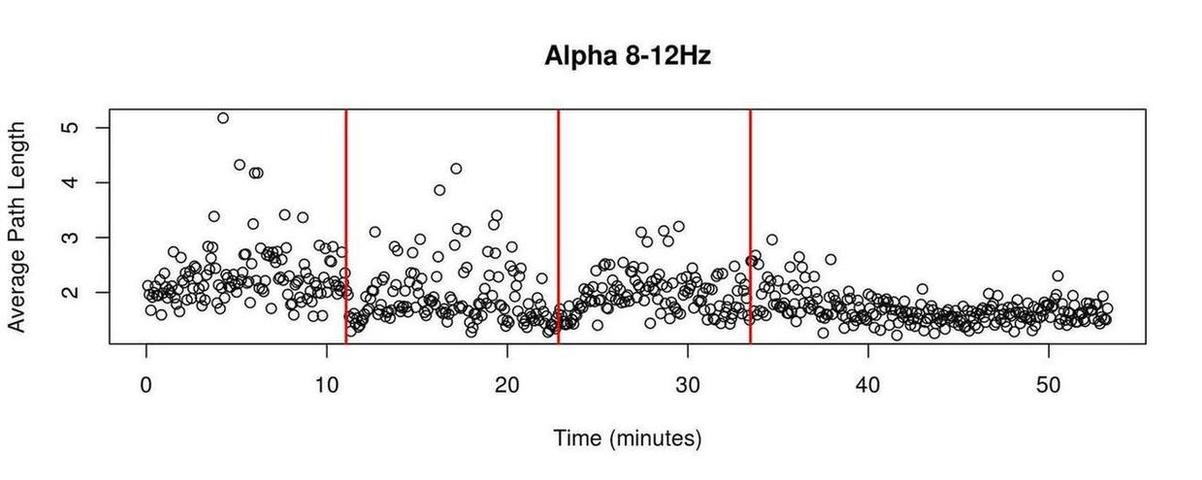}
  \caption{Occipital Lobe}
  \label{fig:sfig4}
\end{subfigure}\\
\caption{{\normalfont \textbf{Average Path Length.}} Average path length respective to the networks of the Alpha frequency band (8-12Hz). Vertical axis average path length; Horizontal axis time (minutes). At t=11 minutes, the monkey was blindfolded, the first red line in each sub-figure represents the moment when a patch was placed over the eyes. At t=23 minutes the Ketamine-Medetomidine cocktail was injected, being represented by the second red line. The point of loss of consciousness (LOC) was registered at t=33 minutes, and is indicated by the third red line. Sub-figures: (\textbf{a}) Frontal Lobe; (\textbf{b}) Parietal Lobe; (\textbf{c}) Temporal Lobe; (\textbf{d}) Occipital Lobe.}
\label{fig:fig}
\end{figure}


\begin{table}[!h]
\centering
\caption{{\normalfont \textbf{Average Path Length.}} Mean, variance (Var), and standard deviation (SD) of the average path length of the networks respective to each one of the four cortical lobes, on the three different conditions in which the monkey was exposed during the experiment: awake with eyes open, awake with eyes closed and anesthesia (eyes closed). Frequency band Alpha (8-12Hz). }
\vspace{0.5cm}
\begin{tabular}{l|lcr|lcr|lcr}
\hline 
\textbf{Alpha Band (8-12Hz)} & \multicolumn{3}{c}{Eyes Open} \vline &\multicolumn{3}{c}{Eyes Closed} \vline &\multicolumn{3}{c}{Anesthesia}\\
\hline
Corresponding Graph & Mean   & Var  & SD  & Mean  & Var  & SD & Mean   & Var  & SD\\ 
\hline                             

Frontal Lobe  &1.61   &0.11  &0.33       &1.41  &0.12   &0.35      &2.95   &0.33   &0.58      \\

Parietal Lobe   &1.77  &0.12  &0.35      &1.40  &0.07   &0.26      &2.32    &0.27  &0.52   \\

Temporal Lobe  &1.95   &0.10  &0.31       &1.64  &0.12   &0.34      &2.43   &0.24   &0.49     \\

Occipital Lobe  &2.30   &0.31  &0.56       &1.93  &0.29   &0.54      &1.75   &0.10   &0.32     \\   

\end{tabular}
\end{table}

\clearpage

\subsubsection*{Beta 13-30Hz}

\begin{figure}[!h]
\begin{subfigure}{.5\textwidth}
  \centering
  \includegraphics[width=1\linewidth]{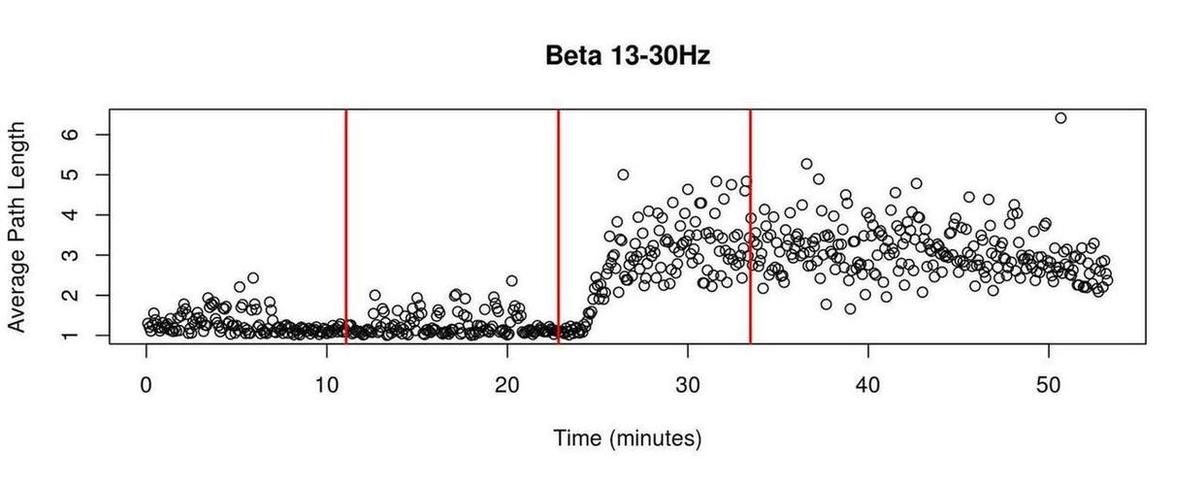}
  \caption{Frontal Lobe}
  \label{fig:sfig1}
\end{subfigure}%
\begin{subfigure}{.5\textwidth}
  \centering
  \includegraphics[width=1\linewidth]{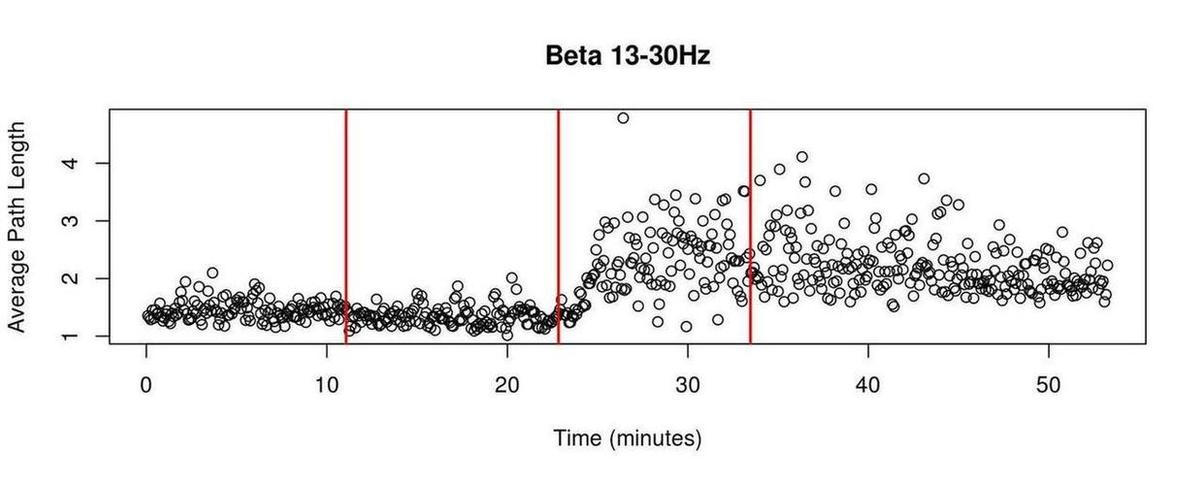}
 \caption{Parietal Lobe}
  \label{fig:sfig2}
\end{subfigure}\\
\centering
\begin{subfigure}{.5\textwidth}
\includegraphics[width=1\linewidth]{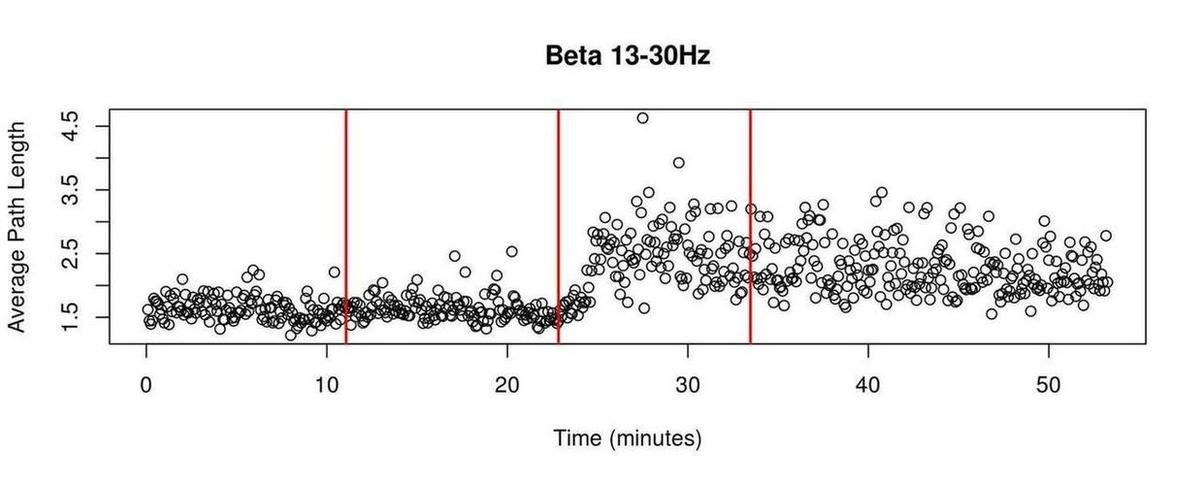}
  \caption{Temporal Lobe}
  \label{fig:sfig3}
\end{subfigure}%
\begin{subfigure}{.5\textwidth}
  \centering
  \includegraphics[width=1\linewidth]{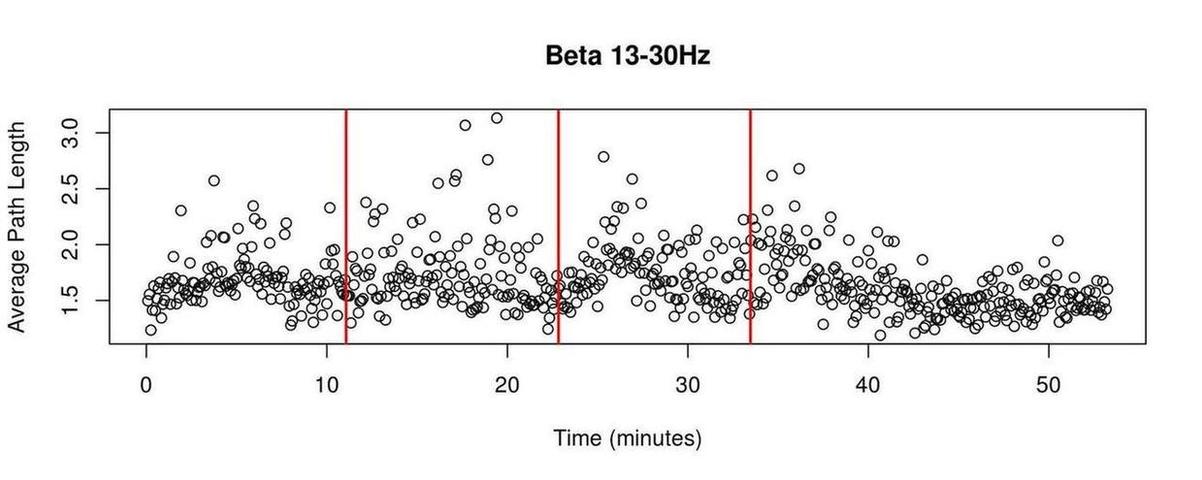}
  \caption{Occipital Lobe}
  \label{fig:sfig4}
\end{subfigure}\\
\caption{{\normalfont \textbf{Average Path Length.}} Average path length respective to the networks of the Beta frequency band (13-30Hz). Vertical axis average path length; Horizontal axis time (minutes). At t=11 minutes, the monkey was blindfolded, the first red line in each sub-figure represents the moment when a patch was placed over the eyes. At t=23 minutes the Ketamine-Medetomidine cocktail was injected, being represented by the second red line. The point of loss of consciousness (LOC) was registered at t=33 minutes, and is indicated by the third red line. Sub-figures: (\textbf{a}) Frontal Lobe; (\textbf{b}) Parietal Lobe; (\textbf{c}) Temporal Lobe; (\textbf{d}) Occipital Lobe.}
\label{fig:fig}
\end{figure}


\begin{table}[!h]
\centering
\caption{{\normalfont \textbf{Average Path Length.}} Mean, variance (Var), and standard deviation (SD) of the average path length of the networks respective to each one of the four cortical lobes, on the three different conditions in which the monkey was exposed during the experiment: awake with eyes open, awake with eyes closed and anesthesia (eyes closed). Frequency band Beta (13-30Hz). }
\vspace{0.5cm}
\begin{tabular}{l|lcr|lcr|lcr}
\hline 
\textbf{Beta Band (13-30Hz)} & \multicolumn{3}{c}{Eyes Open} \vline &\multicolumn{3}{c}{Eyes Closed} \vline &\multicolumn{3}{c}{Anesthesia}\\
\hline
Corresponding Graph & Mean   & Var  & SD  & Mean  & Var  & SD & Mean   & Var  & SD\\ 
\hline                             

Frontal Lobe    &1.29   &0.07  &0.26       &1.26  &0.07   & 0.27     & 3.13   & 0.42   & 0.65  \\

Parietal Lobe   &1.45  &0.03  &0.17      &1.35  &0.03   &0.17      &2.25    &0.25  &0.50   \\

Temporal Lobe   &1.63  &0.04  &0.19      &1.63  & 0.04  & 0.20     & 2.33   &  0.18  &  0.42  \\

Occipital Lobe  &1.68  &0.05  &0.22      &1.73  & 0.11  &  0.34    & 1.62   & 0.06   & 0.24   \\   

\end{tabular}
\end{table}

\clearpage

\subsubsection*{Gamma 25-100Hz}

\begin{figure}[!h]
\begin{subfigure}{.5\textwidth}
  \centering
  \includegraphics[width=1\linewidth]{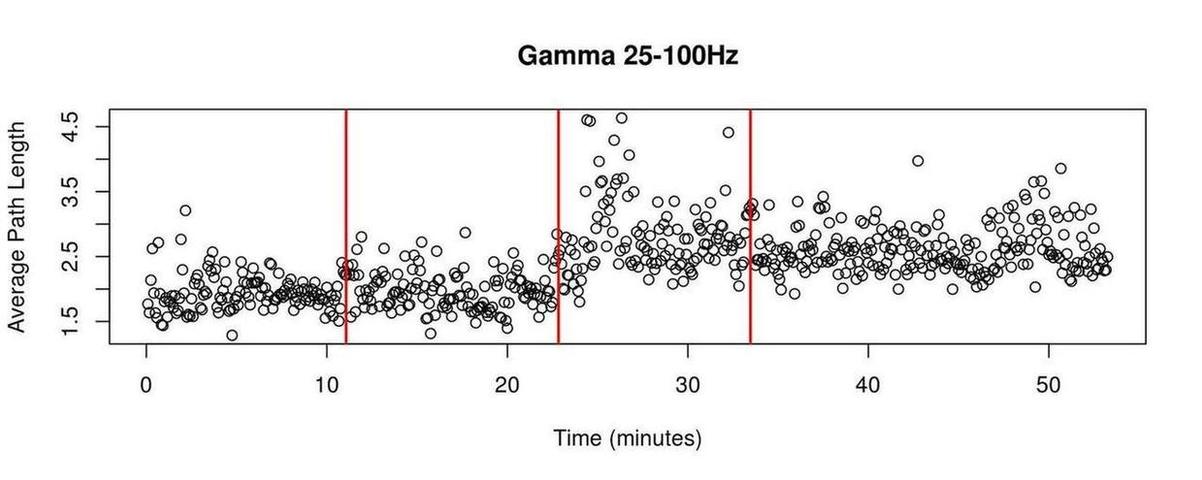}
  \caption{Frontal Lobe}
  \label{fig:sfig1}
\end{subfigure}%
\begin{subfigure}{.5\textwidth}
  \centering
  \includegraphics[width=1\linewidth]{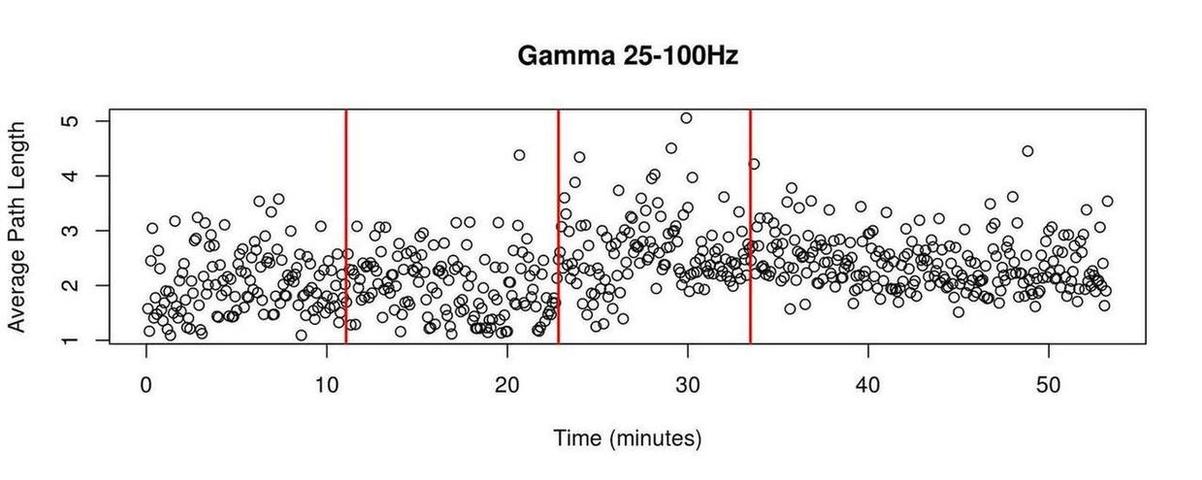}
 \caption{Parietal Lobe}
  \label{fig:sfig2}
\end{subfigure}\\
\centering
\begin{subfigure}{.5\textwidth}
\includegraphics[width=1\linewidth]{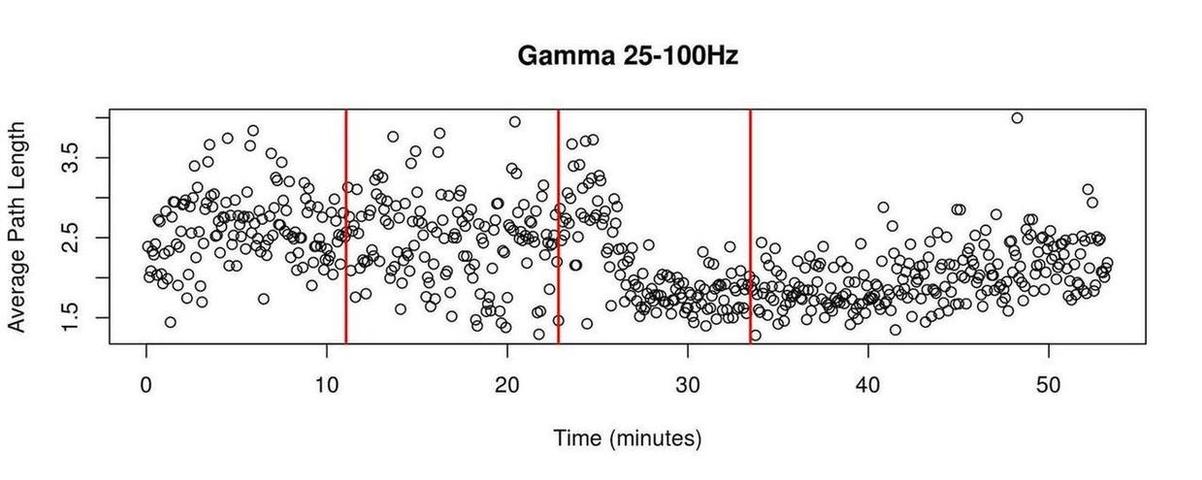}
  \caption{Temporal Lobe}
  \label{fig:sfig3}
\end{subfigure}%
\begin{subfigure}{.5\textwidth}
  \centering
  \includegraphics[width=1\linewidth]{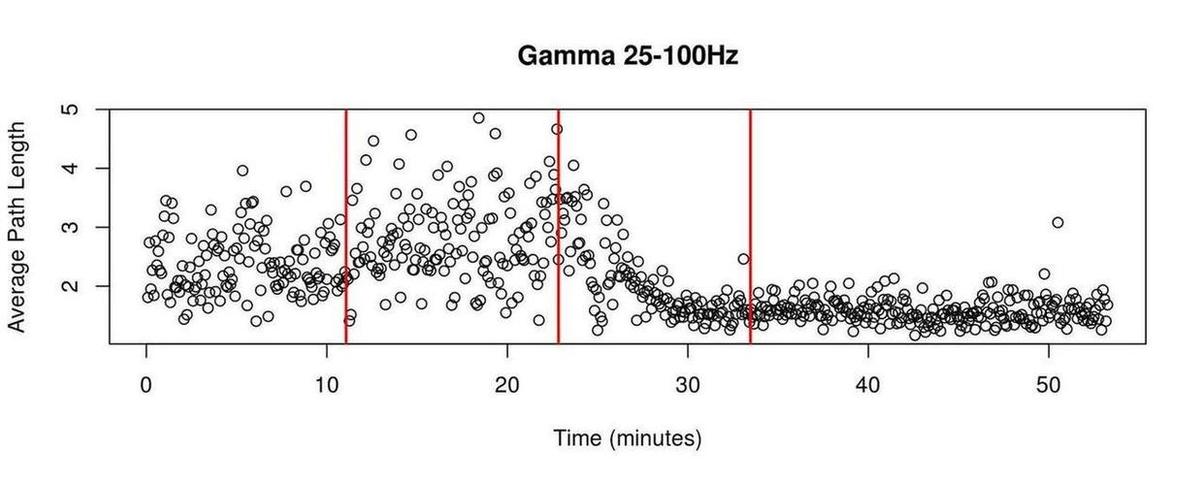}
  \caption{Occipital Lobe}
  \label{fig:sfig4}
\end{subfigure}\\
\caption{{\normalfont \textbf{Average Path Length.}} Average path length respective to the networks of the Gamma frequency band (25-100Hz). Vertical axis average path length; Horizontal axis time (minutes). At t=11 minutes, the monkey was blindfolded, the first red line in each sub-figure represents the moment when a patch was placed over the eyes. At t=23 minutes the Ketamine-Medetomidine cocktail was injected, being represented by the second red line. The point of loss of consciousness (LOC) was registered at t=33 minutes, and is indicated by the third red line. Sub-figures: (\textbf{a}) Frontal Lobe; (\textbf{b}) Parietal Lobe; (\textbf{c}) Temporal Lobe; (\textbf{d}) Occipital Lobe.}
\label{fig:fig}
\end{figure}


\begin{table}[!h]
\centering
\caption{{\normalfont \textbf{Average Path Length.}} Mean, variance (Var), and standard deviation (SD) of the average path length of the networks respective to each one of the four cortical lobes, on the three different conditions in which the monkey was exposed during the experiment: awake with eyes open, awake with eyes closed and anesthesia (eyes closed). Frequency band Gamma (25-100Hz). }
\vspace{0.5cm}
\begin{tabular}{l|lcr|lcr|lcr}
\hline 
\textbf{Gamma Band (25-100Hz)} & \multicolumn{3}{c}{Eyes Open} \vline &\multicolumn{3}{c}{Eyes Closed} \vline &\multicolumn{3}{c}{Anesthesia}\\
\hline
Corresponding Graph & Mean   & Var  & SD  & Mean  & Var  & SD & Mean   & Var  & SD\\ 
\hline                             

Frontal Lobe    & 1.95  & 0.08 & 0.29      & 1.97 & 0.09  & 0.31     & 2.63   & 0.14   &  0.37 \\

Parietal Lobe   &2.04  &0.33  &0.57      &1.96  &0.35   &0.59      &2.46    &0.28  &0.53   \\

Temporal Lobe   & 2.61 & 0.20 & 0.45     & 2.45 & 0.31  &  0.56    & 1.97   &  0.12  &  0.34  \\

Occipital Lobe  &2.36  & 0.28 & 0.53     & 2.82 & 0.51  &  0.71    & 1.60   & 0.05   &  0.22  \\   

\end{tabular}
\end{table}

\clearpage

\subsubsection{Diameter}

\subsubsection*{Delta 0-4Hz}

\begin{figure}[!h]
\begin{subfigure}{.5\textwidth}
  \centering
  \includegraphics[width=1\linewidth]{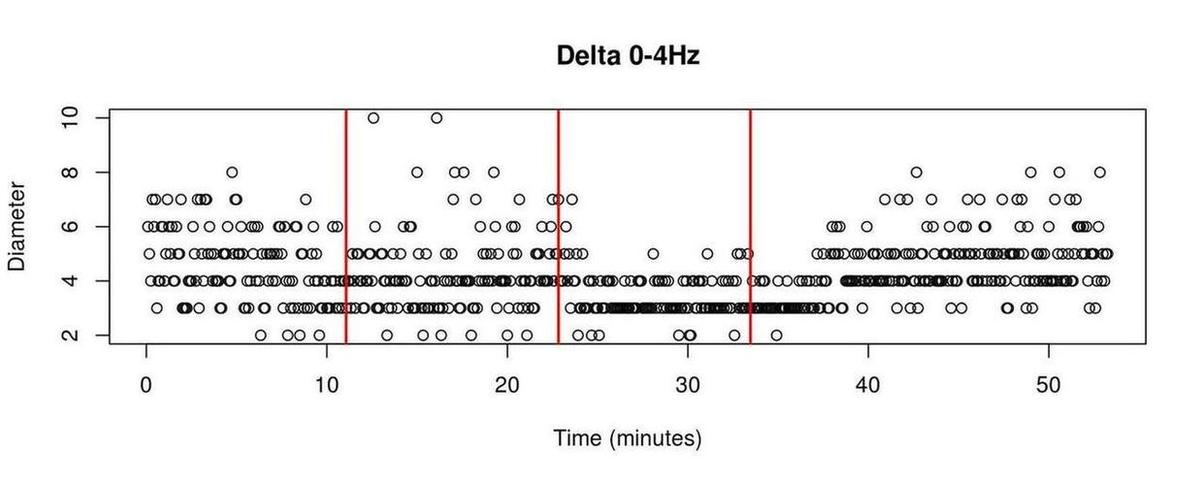}
  \caption{Frontal Lobe}
  \label{fig:sfig1}
\end{subfigure}%
\begin{subfigure}{.5\textwidth}
  \centering
  \includegraphics[width=1\linewidth]{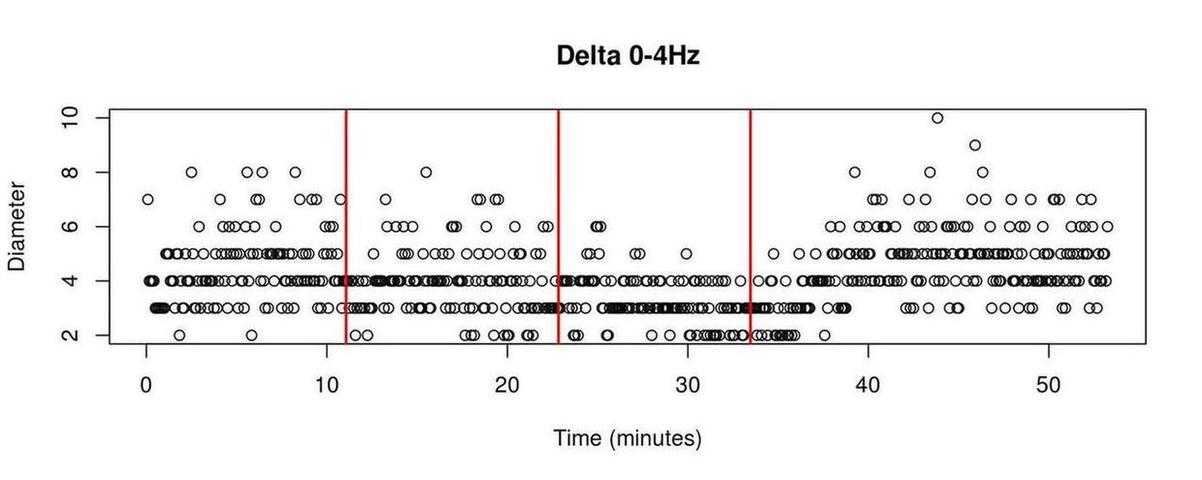}
 \caption{Parietal Lobe}
  \label{fig:sfig2}
\end{subfigure}\\
\centering
\begin{subfigure}{.5\textwidth}
\includegraphics[width=1\linewidth]{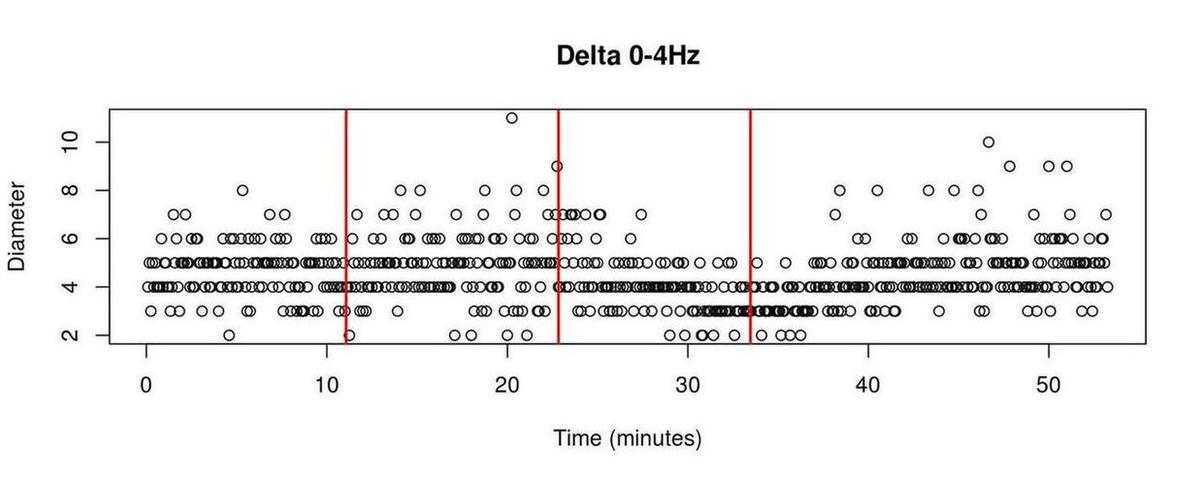}
  \caption{Temporal Lobe}
  \label{fig:sfig3}
\end{subfigure}%
\begin{subfigure}{.5\textwidth}
  \centering
  \includegraphics[width=1\linewidth]{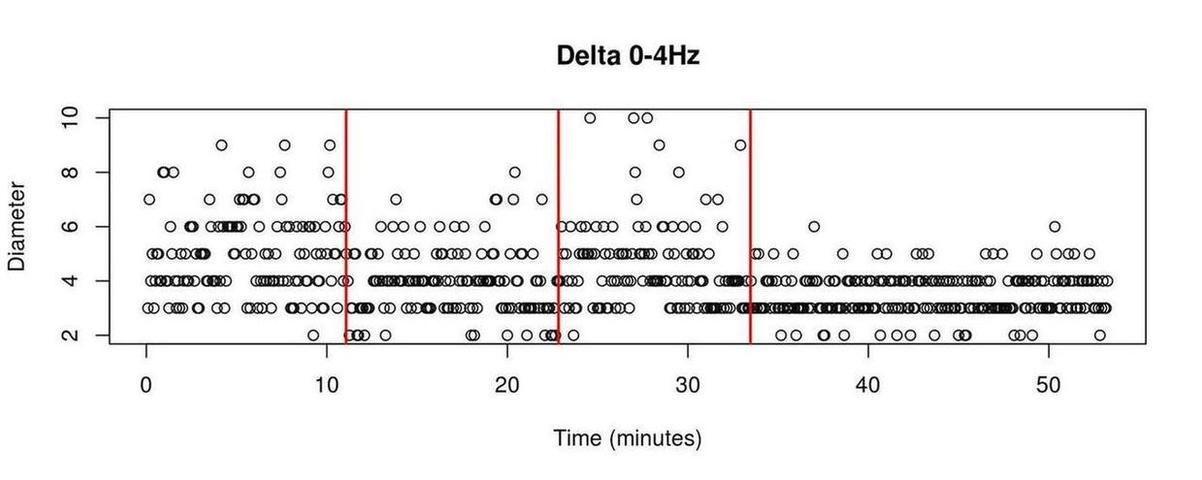}
  \caption{Occipital Lobe}
  \label{fig:sfig4}
\end{subfigure}\\
\caption{{\normalfont \textbf{Diameter.}} Diameter respective to the networks of the Delta frequency band (0-4Hz). Vertical axis diameter; Horizontal axis time (minutes). At t=11 minutes, the monkey was blindfolded, the first red line in each sub-figure represents the moment when a patch was placed over the eyes. At t=23 minutes the Ketamine-Medetomidine cocktail was injected, being represented by the second red line. The point of loss of consciousness (LOC) was registered at t=33 minutes, and is indicated by the third red line. Sub-figures: (\textbf{a}) Frontal Lobe; (\textbf{b}) Parietal Lobe; (\textbf{c}) Temporal Lobe; (\textbf{d}) Occipital Lobe.}
\label{fig:fig}
\end{figure}


\begin{table}[!h]
\centering
\caption{{\normalfont \textbf{Diameter.}}  Mean, variance (Var), and standard deviation (SD) of the diameter of the networks respective to each one of the four cortical lobes, on the three different conditions in which the monkey was exposed during the experiment: awake with eyes open, awake with eyes closed and anesthesia (eyes closed). Frequency band Delta (0-4Hz). }
\vspace{0.5cm}
\begin{tabular}{l|lcr|lcr|lcr}
\hline 
\textbf{Delta Band (0-4Hz)} & \multicolumn{3}{c}{Eyes Open} \vline &\multicolumn{3}{c}{Eyes Closed} \vline &\multicolumn{3}{c}{Anesthesia}\\
\hline
Corresponding Graph & Mean   & Var  & SD  & Mean  & Var  & SD & Mean   & Var  & SD\\ 
\hline                             

Frontal Lobe    &4.62   &1.82  &1.35       &4.29  &2.06   &1.43      &4.19    &1.37    & 1.17  \\

Parietal Lobe   &4.47  &1.68  &1.30      &4.03  &1.65   &1.29      &4.16    &1.97  &1.40   \\

Temporal Lobe   &4.67  &1.13  & 1.06     &4.98  &2.19   &1.48      &4.31    &    1.60& 1.27   \\

Occipital Lobe  &4.84  &2.20  & 1.48     &3.97  &1.53   & 1.24     & 3.65   &    1.01& 1.00   \\   

\end{tabular}
\end{table}

\clearpage

\subsubsection*{Theta 4-8Hz}

\begin{figure}[!h]
\begin{subfigure}{.5\textwidth}
  \centering
  \includegraphics[width=1\linewidth]{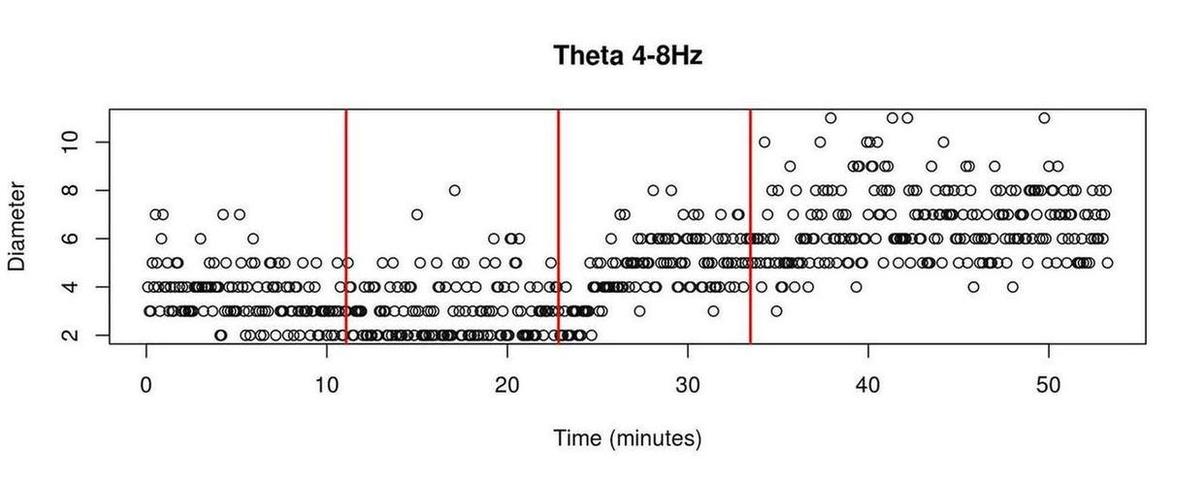}
  \caption{Frontal Lobe}
  \label{fig:sfig1}
\end{subfigure}%
\begin{subfigure}{.5\textwidth}
  \centering
  \includegraphics[width=1\linewidth]{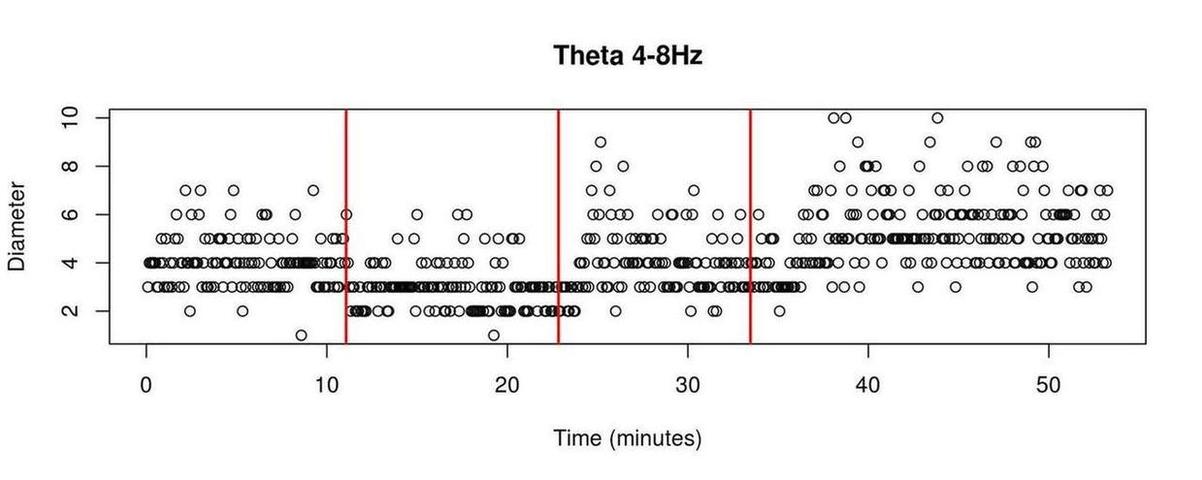}
 \caption{Parietal Lobe}
  \label{fig:sfig2}
\end{subfigure}\\
\centering
\begin{subfigure}{.5\textwidth}
\includegraphics[width=1\linewidth]{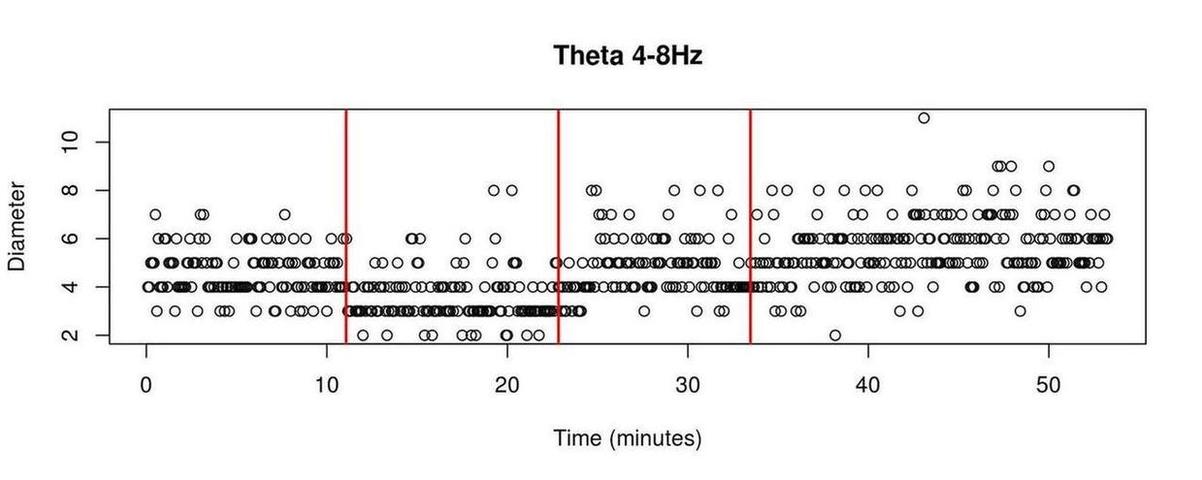}
  \caption{Temporal Lobe}
  \label{fig:sfig3}
\end{subfigure}%
\begin{subfigure}{.5\textwidth}
  \centering
  \includegraphics[width=1\linewidth]{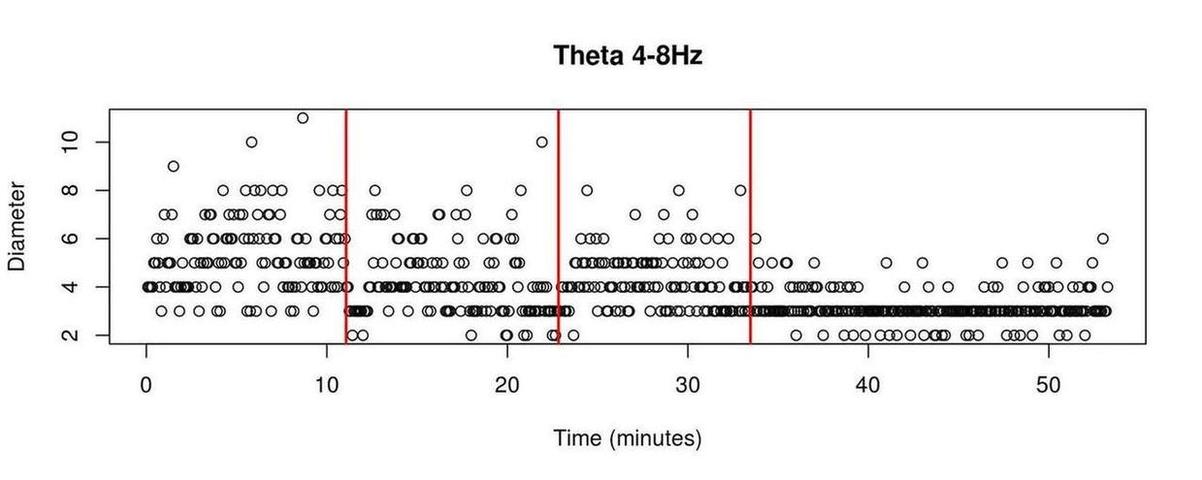}
  \caption{Occipital Lobe}
  \label{fig:sfig4}
\end{subfigure}\\
\caption{{\normalfont \textbf{Diameter.}} Diameter respective to the networks of the Theta frequency band (4-8Hz). Vertical axis diameter; Horizontal axis time (minutes). At t=11 minutes, the monkey was blindfolded, the first red line in each sub-figure represents the moment when a patch was placed over the eyes. At t=23 minutes the Ketamine-Medetomidine cocktail was injected, being represented by the second red line. The point of loss of consciousness (LOC) was registered at t=33 minutes, and is indicated by the third red line. Sub-figures: (\textbf{a}) Frontal Lobe; (\textbf{b}) Parietal Lobe; (\textbf{c}) Temporal Lobe; (\textbf{d}) Occipital Lobe.}
\label{fig:fig}
\end{figure}


\begin{table}[!h]
\centering
\caption{{\normalfont \textbf{Diameter.}} Mean, variance (Var), and standard deviation (SD) of the diameter of the networks respective to each one of the four cortical lobes, on the three different conditions in which the monkey was exposed during the experiment: awake with eyes open, awake with eyes closed and anesthesia (eyes closed). Frequency band Theta (4-8Hz). }
\vspace{0.5cm}
\begin{tabular}{l|lcr|lcr|lcr}
\hline 
\textbf{Theta Band (4-8Hz)} & \multicolumn{3}{c}{Eyes Open} \vline &\multicolumn{3}{c}{Eyes Closed} \vline &\multicolumn{3}{c}{Anesthesia}\\
\hline
Corresponding Graph & Mean   & Var  & SD  & Mean  & Var  & SD & Mean   & Var  & SD\\ 
\hline                             

Frontal Lobe    &3.69   &1.29  &1.14       &3.11  &1.61   &1.27      &6.31    &2.27   & 1.51  \\

Parietal Lobe   &4.03  &1.16  &1.08      &2.99  &0.94   &0.97      &4.90    &2.39  &1.55   \\

Temporal Lobe   &4.55  &0.96  &0.98      &3.58  &1.21   &1.10      &5.42    &    1.72& 1.31   \\

Occipital Lobe  &5.28  &2.39  &1.54      &4.24  &2.22   &1.49      &3.40    &    0.93&  0.96  \\   

\end{tabular}
\end{table}

\clearpage

\subsubsection*{Alpha 8-12Hz}

\begin{figure}[!h]
\begin{subfigure}{.5\textwidth}
  \centering
  \includegraphics[width=1\linewidth]{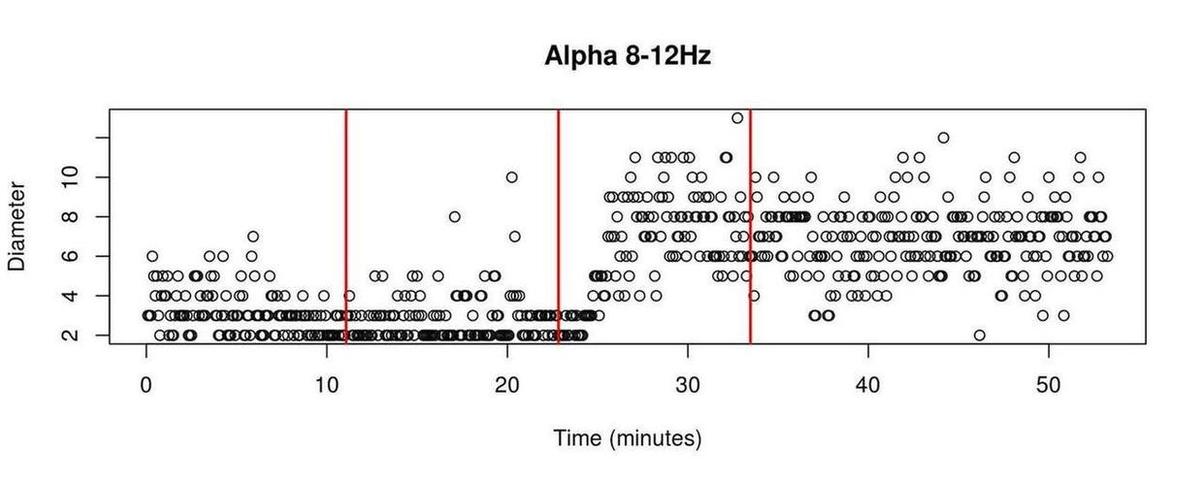}
  \caption{Frontal Lobe}
  \label{fig:sfig1}
\end{subfigure}%
\begin{subfigure}{.5\textwidth}
  \centering
  \includegraphics[width=1\linewidth]{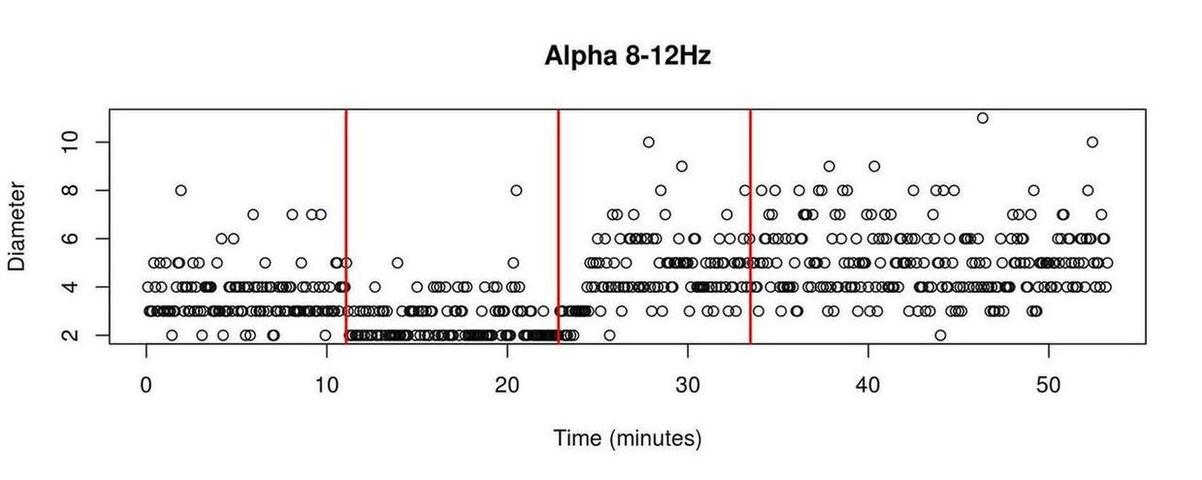}
 \caption{Parietal Lobe}
  \label{fig:sfig2}
\end{subfigure}\\
\centering
\begin{subfigure}{.5\textwidth}
\includegraphics[width=1\linewidth]{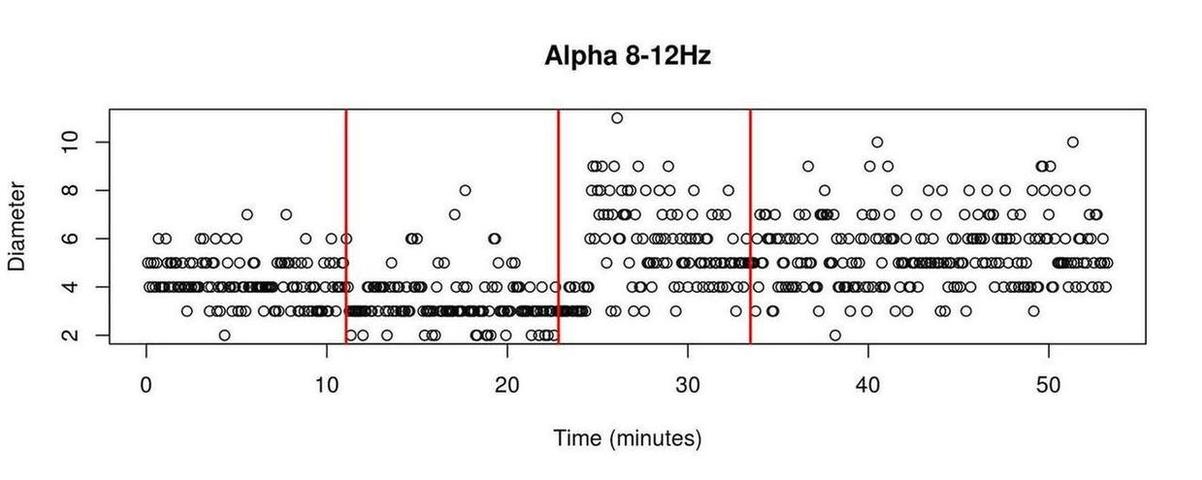}
  \caption{Temporal Lobe}
  \label{fig:sfig3}
\end{subfigure}%
\begin{subfigure}{.5\textwidth}
  \centering
  \includegraphics[width=1\linewidth]{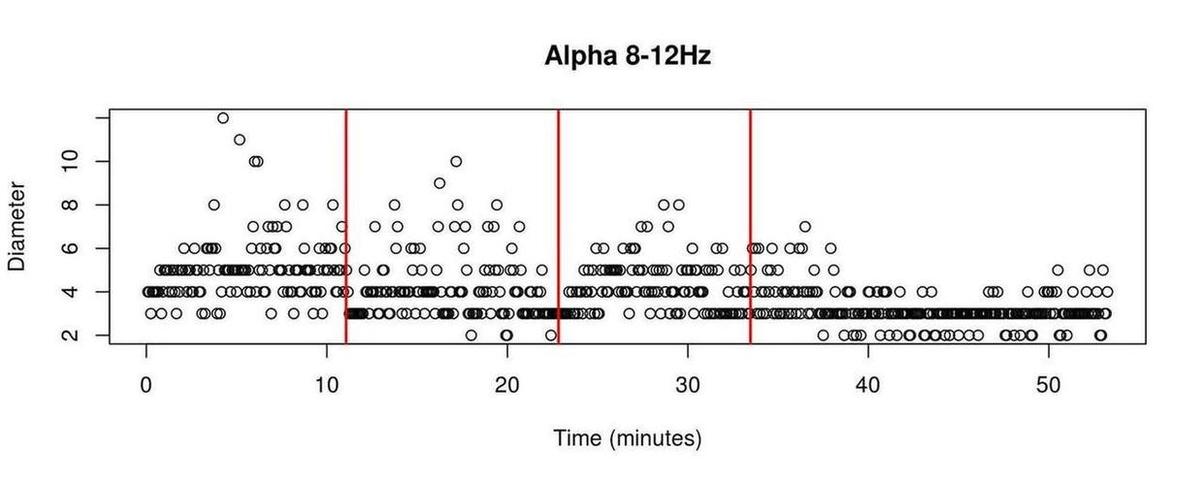}
  \caption{Occipital Lobe}
  \label{fig:sfig4}
\end{subfigure}\\
\caption{{\normalfont \textbf{Diameter.}} Diameter respective to the networks of the Alpha frequency band (8-12Hz). Vertical axis diameter; Horizontal axis time (minutes). At t=11 minutes, the monkey was blindfolded, the first red line in each sub-figure represents the moment when a patch was placed over the eyes. At t=23 minutes the Ketamine-Medetomidine cocktail was injected, being represented by the second red line. The point of loss of consciousness (LOC) was registered at t=33 minutes, and is indicated by the third red line. Sub-figures: (\textbf{a}) Frontal Lobe; (\textbf{b}) Parietal Lobe; (\textbf{c}) Temporal Lobe; (\textbf{d}) Occipital Lobe.}
\label{fig:fig}
\end{figure}


\begin{table}[!h]
\centering
\caption{{\normalfont \textbf{Diameter.}} Mean, variance (Var), and standard deviation (SD) of the diameter of the networks respective to each one of the four cortical lobes, on the three different conditions in which the monkey was exposed during the experiment: awake with eyes open, awake with eyes closed and anesthesia (eyes closed). Frequency band Alpha (8-12Hz). }
\vspace{0.5cm}
\begin{tabular}{l|lcr|lcr|lcr}
\hline 
\textbf{Alpha Band (8-12Hz)} & \multicolumn{3}{c}{Eyes Open} \vline &\multicolumn{3}{c}{Eyes Closed} \vline &\multicolumn{3}{c}{Anesthesia}\\
\hline
Corresponding Graph & Mean   & Var  & SD  & Mean  & Var  & SD & Mean   & Var  & SD\\ 
\hline                             

Frontal Lobe  &3.24   &1.39  &1.18       &2.82  &1.40   &1.18      &7.02   &3.01   &1.74       \\

Parietal Lobe   &3.65  &1.20  &1.10      &2.60  &0.84   &0.91      &5.02    &2.09  &1.44   \\

Temporal Lobe  &4.22   &0.86  &0.93       &3.40  &1.12   &1.06      &5.42   &1.96   & 1.40    \\

Occipital Lobe   &5.00   &2.32  &1.52       &4.15  &2.27   &1.51      &3.44   &1.04   &1.02    \\   

\end{tabular}
\end{table}

\clearpage

\subsubsection*{Beta 13-30Hz}

\begin{figure}[!h]
\begin{subfigure}{.5\textwidth}
  \centering
  \includegraphics[width=1\linewidth]{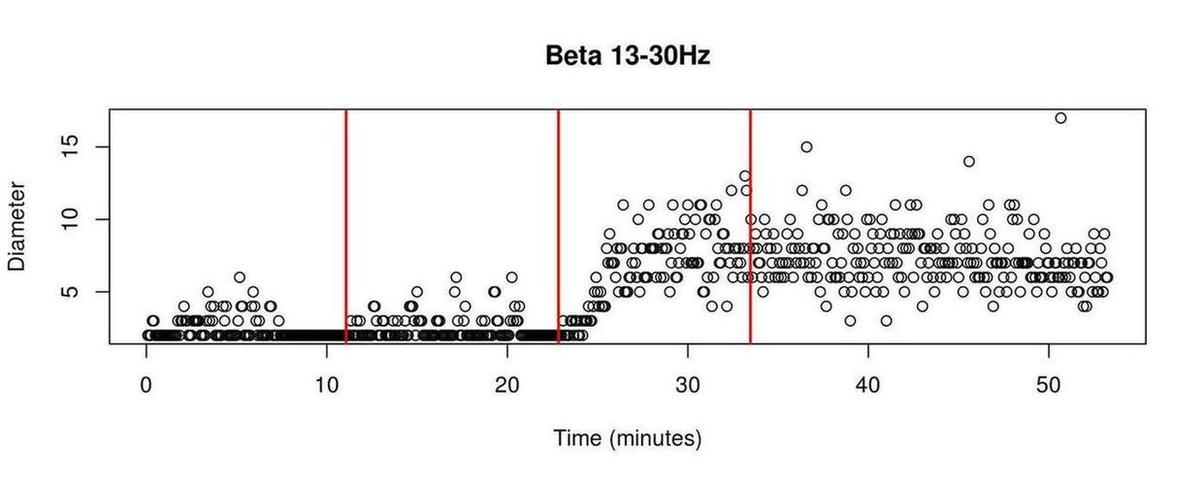}
  \caption{Frontal Lobe}
  \label{fig:sfig1}
\end{subfigure}%
\begin{subfigure}{.5\textwidth}
  \centering
  \includegraphics[width=1\linewidth]{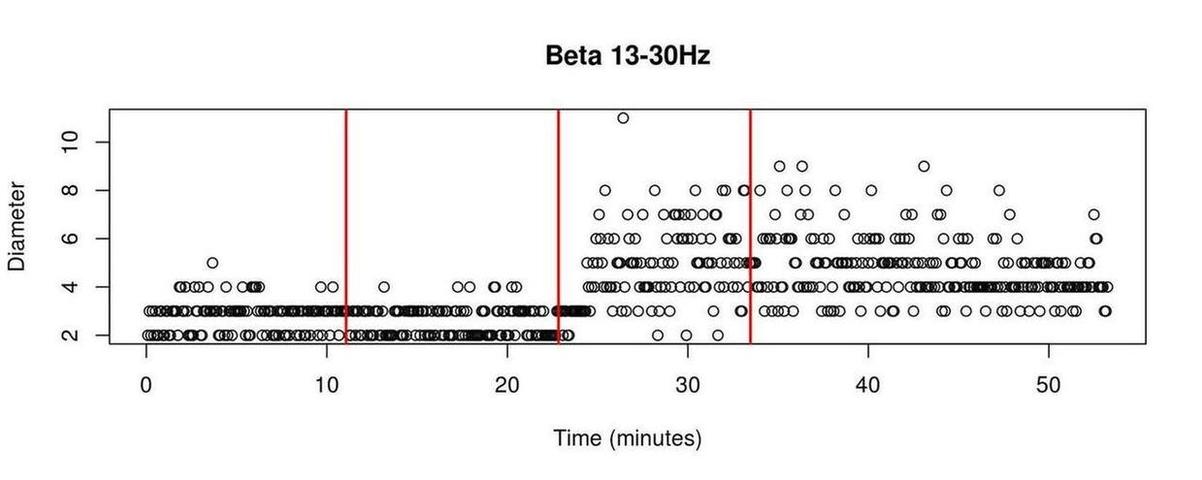}
 \caption{Parietal Lobe}
  \label{fig:sfig2}
\end{subfigure}\\
\centering
\begin{subfigure}{.5\textwidth}
\includegraphics[width=1\linewidth]{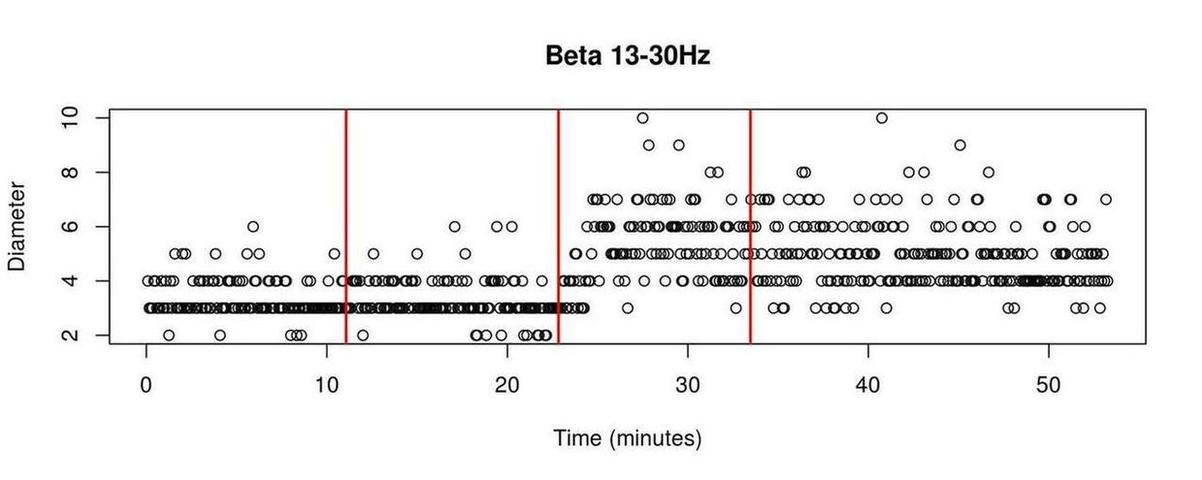}
  \caption{Temporal Lobe}
  \label{fig:sfig3}
\end{subfigure}%
\begin{subfigure}{.5\textwidth}
  \centering
  \includegraphics[width=1\linewidth]{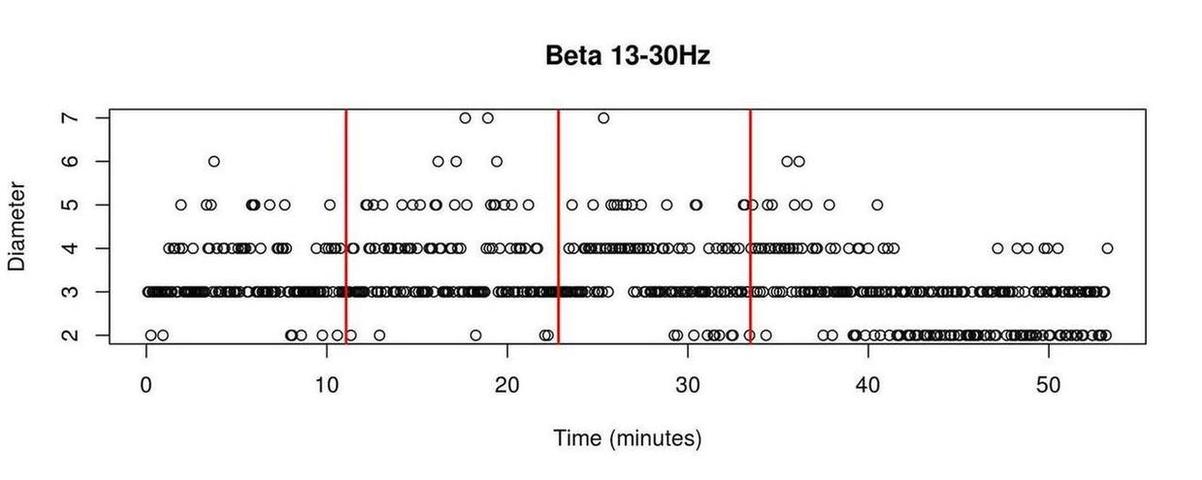}
  \caption{Occipital Lobe}
  \label{fig:sfig4}
\end{subfigure}\\
\caption{{\normalfont \textbf{Diameter.}} Diameter respective to the networks of the Theta frequency band (13-30Hz). Vertical axis diameter; Horizontal axis time (minutes). At t=11 minutes, the monkey was blindfolded, the first red line in each sub-figure represents the moment when a patch was placed over the eyes. At t=23 minutes the Ketamine-Medetomidine cocktail was injected, being represented by the second red line. The point of loss of consciousness (LOC) was registered at t=33 minutes, and is indicated by the third red line. Sub-figures: (\textbf{a}) Frontal Lobe; (\textbf{b}) Parietal Lobe; (\textbf{c}) Temporal Lobe; (\textbf{d}) Occipital Lobe.}
\label{fig:fig}
\end{figure}


\begin{table}[!h]
\centering
\caption{{\normalfont \textbf{Diameter.}} Mean, variance (Var), and standard deviation (SD) of the diameter of the networks respective to each one of the four cortical lobes, on the three different conditions in which the monkey was exposed during the experiment: awake with eyes open, awake with eyes closed and anesthesia (eyes closed). Frequency band Beta (13-30Hz).}
\vspace{0.5cm}
\begin{tabular}{l|lcr|lcr|lcr}
\hline 
\textbf{Beta  Band (13-30Hz)} & \multicolumn{3}{c}{Eyes Open} \vline &\multicolumn{3}{c}{Eyes Closed} \vline &\multicolumn{3}{c}{Anesthesia}\\
\hline
Corresponding Graph & Mean   & Var  & SD  & Mean  & Var  & SD & Mean   & Var  & SD\\ 
\hline                             

Frontal Lobe    & 2.43  &0.63  &0.80       &2.45  &0.76   &0.87      &7.51    & 3.75   & 1.94  \\

Parietal Lobe   &2.80  &0.45  &0.67      &2.54  &0.37   &0.61      &4.82    &1.82  &1.35   \\

Temporal Lobe   &3.34  & 0.46 &0.68      &3.27  &0.55   & 0.74     &5.08    &    1.61&  1.27  \\

Occipital Lobe  & 3.33 & 0.51 & 0.71     &3.61  &0.88   &0.94      & 2.99   &    0.66&  0.81  \\   

\end{tabular}
\end{table}

\clearpage

\subsubsection*{Gamma 25-100Hz}

\begin{figure}[!h]
\begin{subfigure}{.5\textwidth}
  \centering
  \includegraphics[width=1\linewidth]{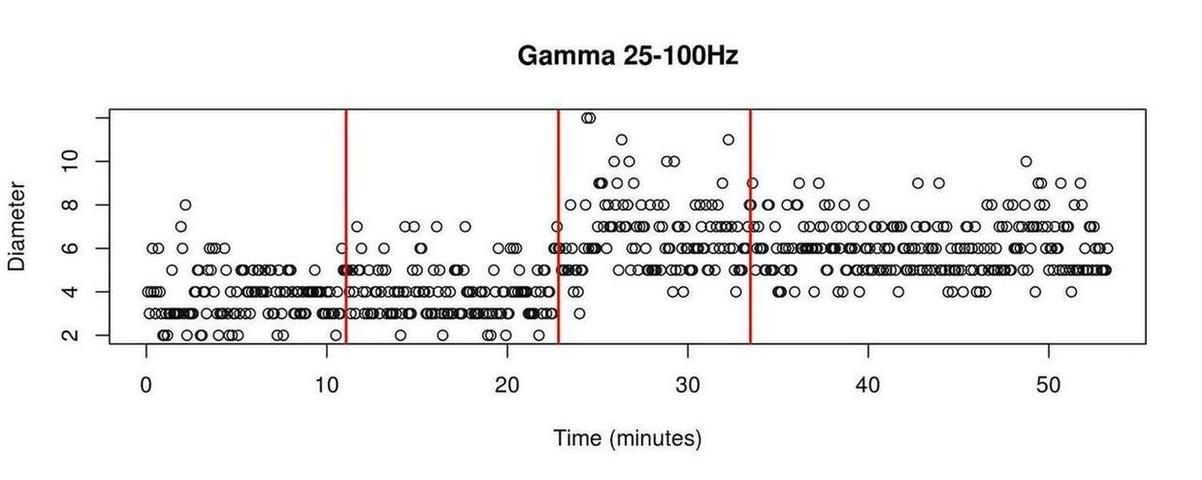}
  \caption{Frontal Lobe}
  \label{fig:sfig1}
\end{subfigure}%
\begin{subfigure}{.5\textwidth}
  \centering
  \includegraphics[width=1\linewidth]{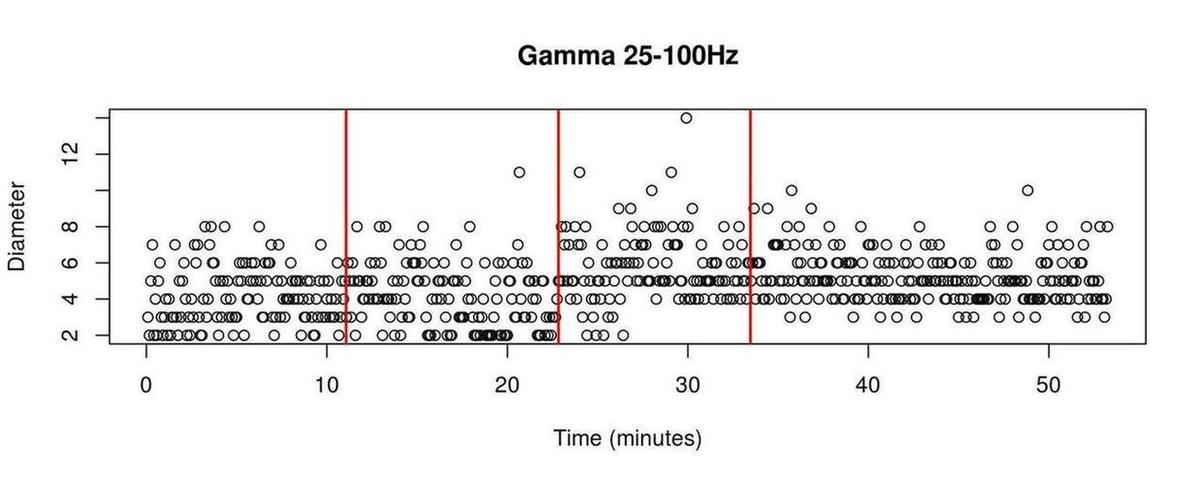}
 \caption{Parietal Lobe}
  \label{fig:sfig2}
\end{subfigure}\\
\centering
\begin{subfigure}{.5\textwidth}
\includegraphics[width=1\linewidth]{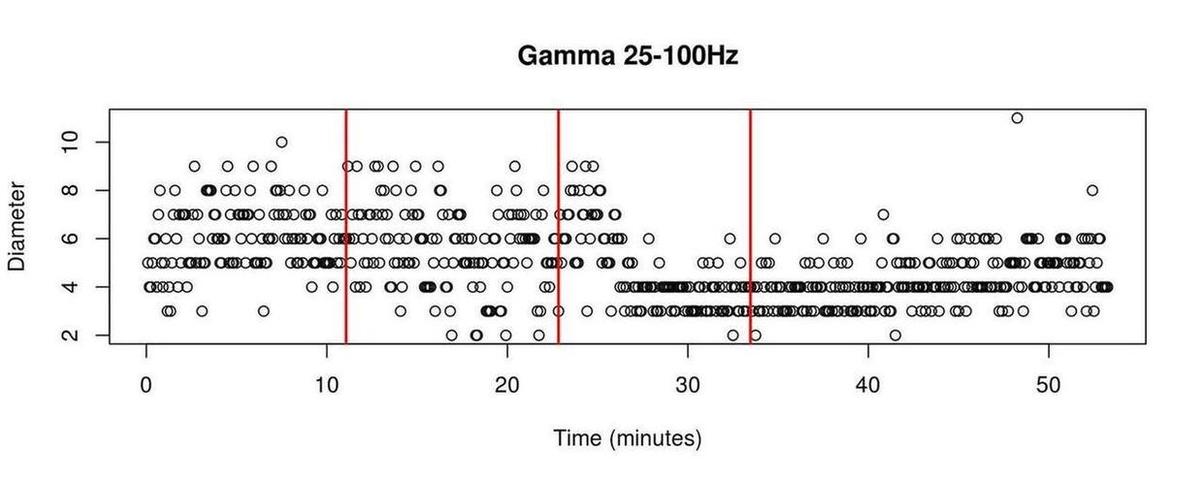}
  \caption{Temporal Lobe}
  \label{fig:sfig3}
\end{subfigure}%
\begin{subfigure}{.5\textwidth}
  \centering
  \includegraphics[width=1\linewidth]{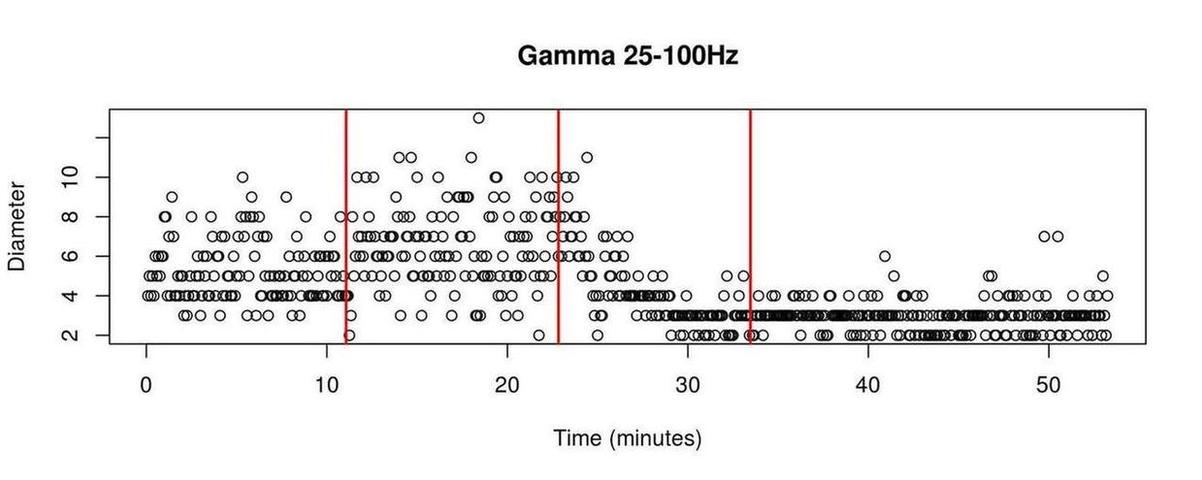}
  \caption{Occipital Lobe}
  \label{fig:sfig4}
\end{subfigure}\\
\caption{{\normalfont \textbf{Diameter.}} Diameter respective to the networks of the Gamma frequency band (25-100Hz). Vertical axis diameter; Horizontal axis time (minutes). At t=11 minutes, the monkey was blindfolded, the first red line in each sub-figure represents the moment when a patch was placed over the eyes. At t=23 minutes the Ketamine-Medetomidine cocktail was injected, being represented by the second red line. The point of loss of consciousness (LOC) was registered at t=33 minutes, and is indicated by the third red line. Sub-figures: (\textbf{a}) Frontal Lobe; (\textbf{b}) Parietal Lobe; (\textbf{c}) Temporal Lobe; (\textbf{d}) Occipital Lobe.}
\label{fig:fig}
\end{figure}


\begin{table}[!h]
\centering
\caption{{\normalfont \textbf{Diameter.}} Mean, variance (Var), and standard deviation (SD) of the diameter of the networks respective to each one of the four cortical lobes, on the three different conditions in which the monkey was exposed during the experiment: awake with eyes open, awake with eyes closed and anesthesia (eyes closed). Frequency band Gamma (25-100Hz). }
\vspace{0.5cm}
\begin{tabular}{l|lcr|lcr|lcr}
\hline 
\textbf{Gamma Band (25-100Hz)} & \multicolumn{3}{c}{Eyes Open} \vline &\multicolumn{3}{c}{Eyes Closed} \vline &\multicolumn{3}{c}{Anesthesia}\\
\hline
Corresponding Graph & Mean   & Var  & SD  & Mean  & Var  & SD & Mean   & Var  & SD\\ 
\hline                             

Frontal Lobe    &3.81   &1.32  &1.15       &3.93  &1.44   &1.20      &6.09    &1.66    & 1.29  \\

Parietal Lobe   &4.23  &2.62  &1.62      &4.02  &3.12   &1.77      &5.34    &2.34  &1.53   \\

Temporal Lobe   &6,.7  &2.03  &1.42      &5.46  &2.81   &1.68      &4.16    &    1.13&  1.06  \\

Occipital Lobe  &5.24  &2.47  &1.57      &6.57  &4.43   & 2.11     &2.98    &    0.63& 0.79   \\   

\end{tabular}
\end{table}

\clearpage

\subsection{Measures Related to Influence and Centrality}
\subsubsection{Average Betweenness Centrality Degree}

\subsubsection*{Delta 0-4Hz}

\begin{figure}[!h]
\begin{subfigure}{.5\textwidth}
  \centering
  \includegraphics[width=1\linewidth]{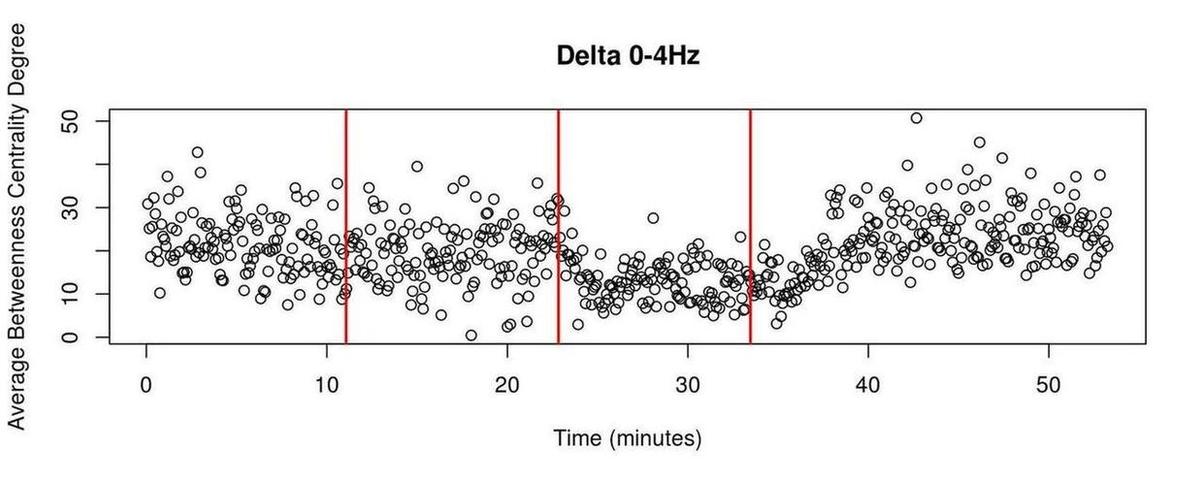}
  \caption{Frontal Lobe}
  \label{fig:sfig1}
\end{subfigure}%
\begin{subfigure}{.5\textwidth}
  \centering
  \includegraphics[width=1\linewidth]{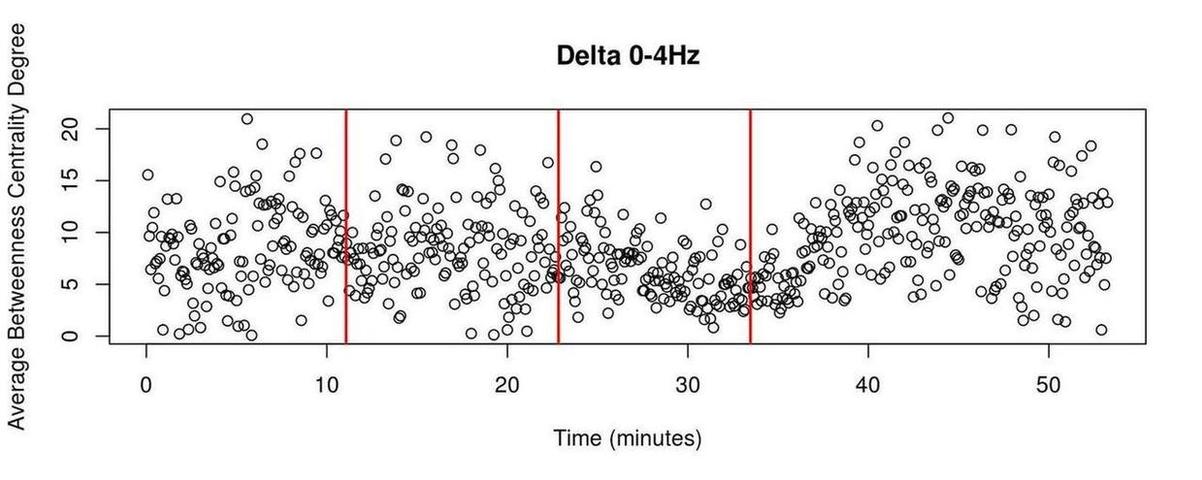}
 \caption{Parietal Lobe}
  \label{fig:sfig2}
\end{subfigure}\\
\centering
\begin{subfigure}{.5\textwidth}
\includegraphics[width=1\linewidth]{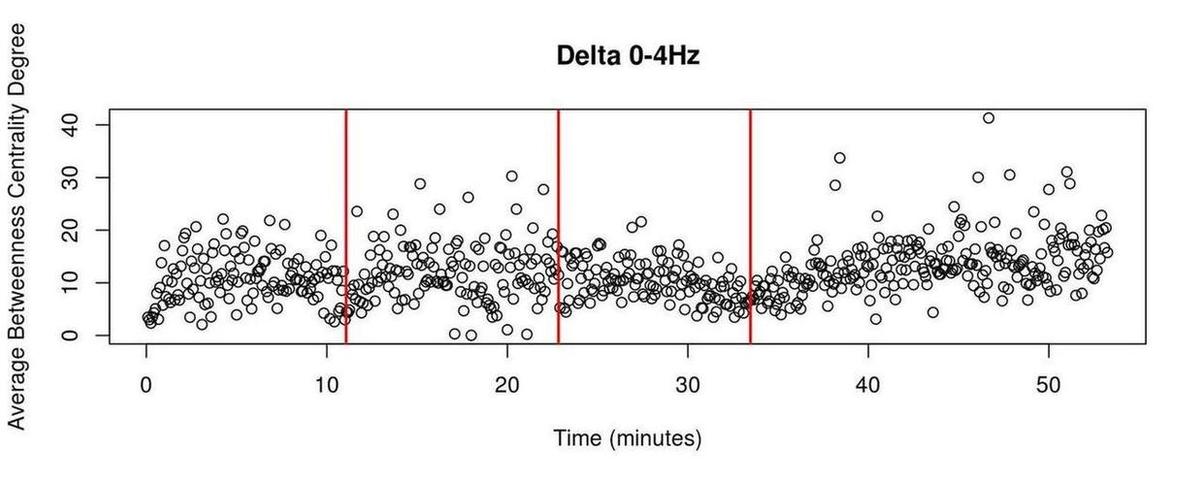}
  \caption{Temporal Lobe}
  \label{fig:sfig3}
\end{subfigure}%
\begin{subfigure}{.5\textwidth}
  \centering
  \includegraphics[width=1\linewidth]{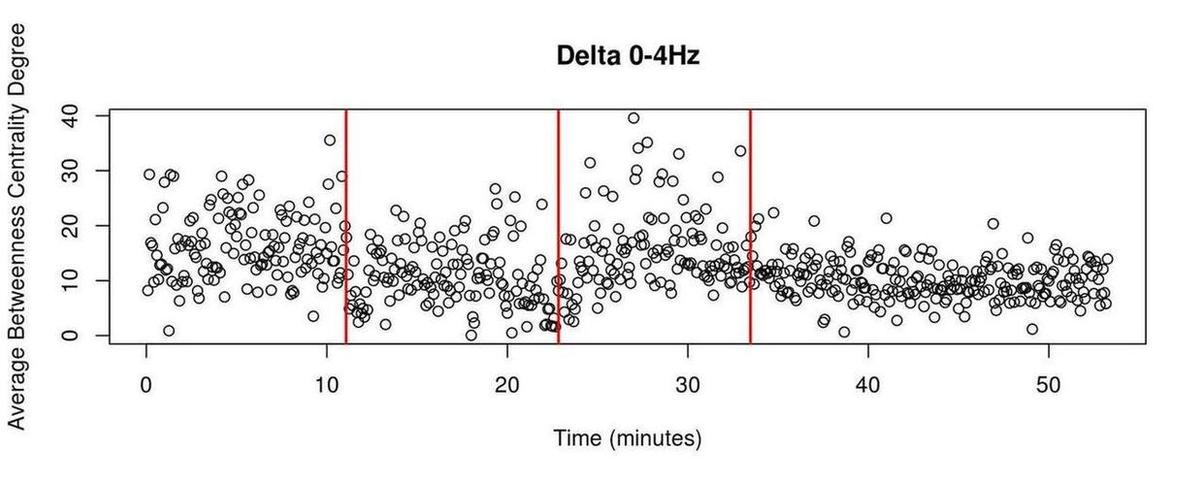}
  \caption{Occipital Lobe}
  \label{fig:sfig4}
\end{subfigure}\\
\caption{{\normalfont \textbf{Average Betweenness Centrality Degree.}}  Average betweenness centrality degree  respective to the networks of the Delta frequency band (0-4Hz). Vertical axis average betweenness centrality degree; Horizontal axis time (minutes). At t=11 minutes, the monkey was blindfolded, the first red line in each sub-figure represents the moment when a patch was placed over the eyes. At t=23 minutes the Ketamine-Medetomidine cocktail was injected, being represented by the second red line. The point of loss of consciousness (LOC) was registered at t=33 minutes, and is indicated by the third red line. Sub-figures: (\textbf{a}) Frontal Lobe; (\textbf{b}) Parietal Lobe; (\textbf{c}) Temporal Lobe; (\textbf{d}) Occipital Lobe.}
\label{fig:fig}
\end{figure}


\begin{table}[!h]
\centering
\caption{{\normalfont \textbf{Average Betweenness Centrality Degree.}}
 Mean, variance (Var), and standard deviation (SD) of the  average betweenness centrality degree of the networks respective to each one of the four cortical lobes, on the three different conditions in which the monkey was exposed during the experiment: awake with eyes open, awake with eyes closed and anesthesia (eyes closed). Frequency band Delta (0-4Hz). }
\vspace{0.5cm}
\begin{tabular}{l|lcr|lcr|lcr}
\hline 
\textbf{Delta Band (0-4Hz)} & \multicolumn{3}{c}{Eyes Open} \vline &\multicolumn{3}{c}{Eyes Closed} \vline &\multicolumn{3}{c}{Anesthesia}\\
\hline
Corresponding Graph & Mean   & Var  & SD  & Mean  & Var  & SD & Mean   & Var  & SD\\ 
\hline                             

Frontal Lobe    & 21.6  & 45.4 & 6.73      &19.8  & 54.1  & 7.35     & 20.0   & 63.4   &  7.96 \\

Parietal Lobe   &8.53  &17.9  &4.23      &8.52  &17.7   &4.21      &8.77    &19.9  &4.46   \\

Temporal Lobe   & 11.0 & 21.0 &  4.59    &12.4 &  3.62 &  5.62    &  12.5  &  27.7  &    5.26\\

Occipital Lobe  & 16.0 & 36.0 & 6.00     & 10.9 & 33.2  &  5.76    & 11.5   &  23.3  &  4.82  \\   

\end{tabular}
\end{table}

\clearpage

\subsubsection*{Theta 4-8Hz}

\begin{figure}[!h]
\begin{subfigure}{.5\textwidth}
  \centering
  \includegraphics[width=1\linewidth]{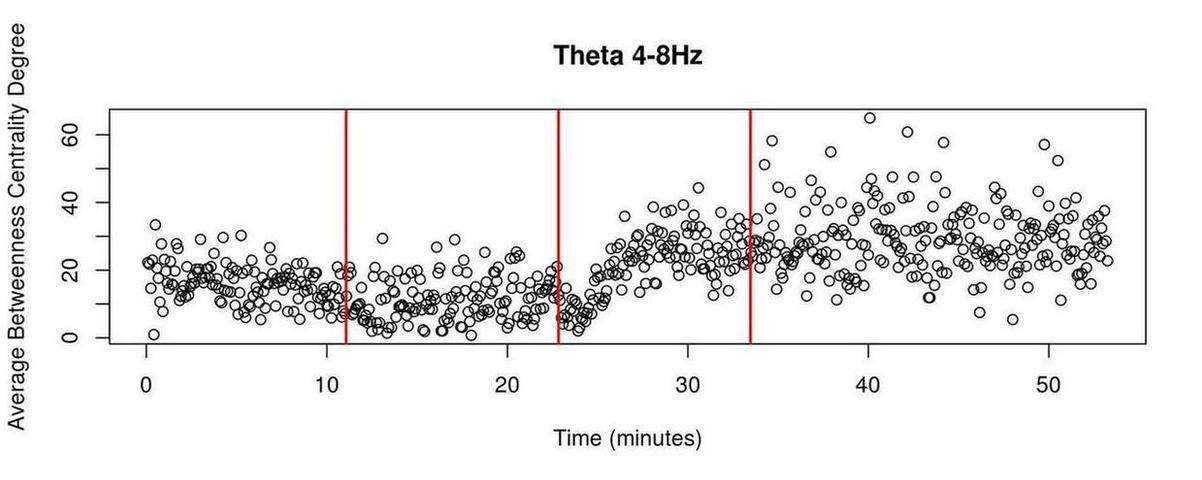}
  \caption{Frontal Lobe}
  \label{fig:sfig1}
\end{subfigure}%
\begin{subfigure}{.5\textwidth}
  \centering
  \includegraphics[width=1\linewidth]{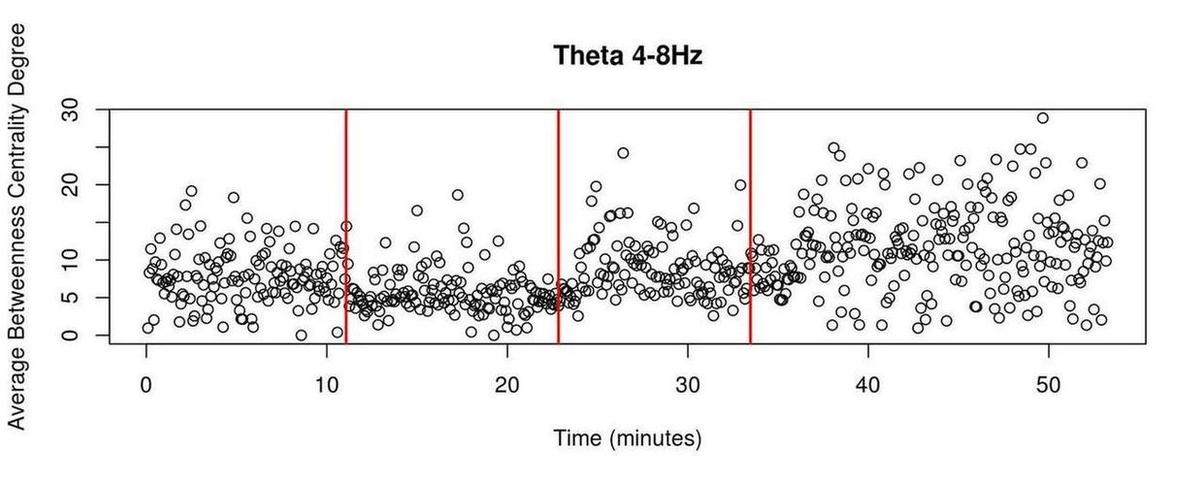}
 \caption{Parietal Lobe}
  \label{fig:sfig2}
\end{subfigure}\\
\centering
\begin{subfigure}{.5\textwidth}
\includegraphics[width=1\linewidth]{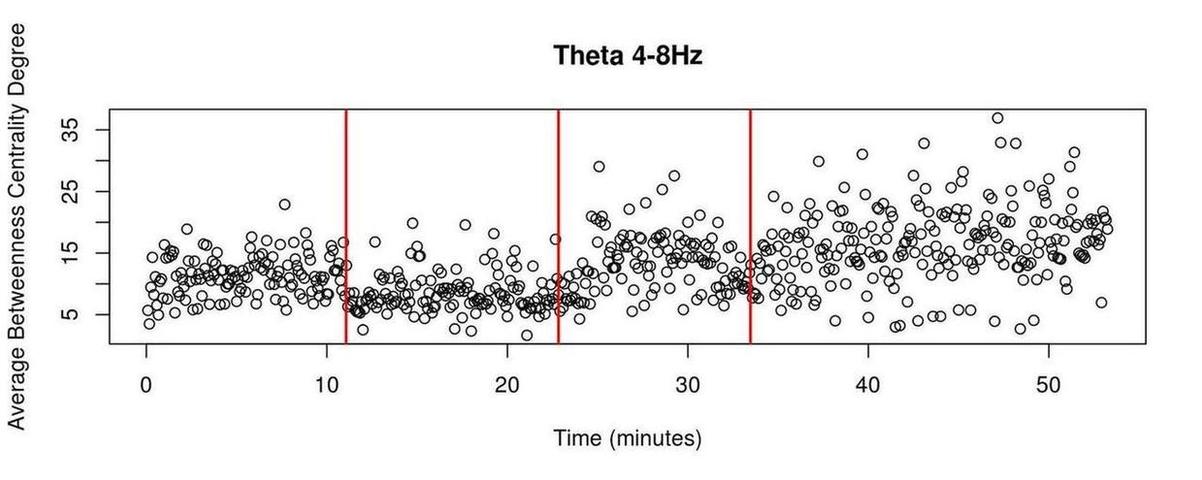}
  \caption{Temporal Lobe}
  \label{fig:sfig3}
\end{subfigure}%
\begin{subfigure}{.5\textwidth}
  \centering
  \includegraphics[width=1\linewidth]{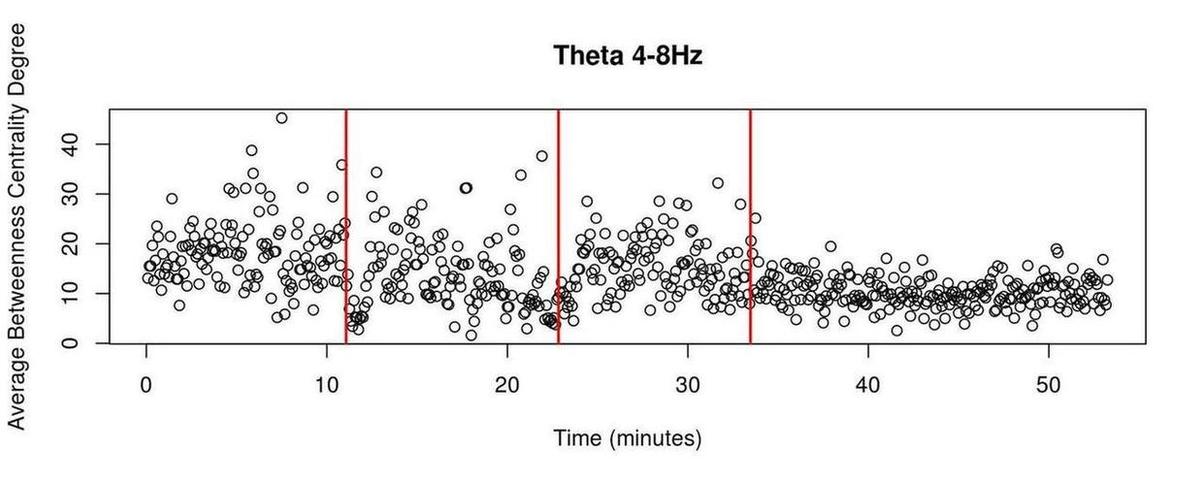}
  \caption{Occipital Lobe}
  \label{fig:sfig4}
\end{subfigure}\\
\caption{{\normalfont \textbf{Average Betweenness Centrality Degree.}} Average betweenness centrality degree  respective to the networks of the Theta frequency band (4-8Hz). Vertical axis average betweenness centrality degree; Horizontal axis time (minutes). At t=11 minutes, the monkey was blindfolded, the first red line in each sub-figure represents the moment when a patch was placed over the eyes. At t=23 minutes the Ketamine-Medetomidine cocktail was injected, being represented by the second red line. The point of loss of consciousness (LOC) was registered at t=33 minutes, and is indicated by the third red line. Sub-figures: (\textbf{a}) Frontal Lobe; (\textbf{b}) Parietal Lobe; (\textbf{c}) Temporal Lobe; (\textbf{d}) Occipital Lobe.}
\label{fig:fig}
\end{figure}


\begin{table}[!h]
\centering
\caption{ {\normalfont \textbf{Average Betweenness Centrality Degree.} }Mean, variance (Var), and standard deviation (SD) of the  average betweenness centrality degree of the networks respective to each one of the four cortical lobes, on the three different conditions in which the monkey was exposed during the experiment: awake with eyes open, awake with eyes closed and anesthesia (eyes closed). Frequency band Theta (4-8Hz). }
\vspace{0.5cm}
\begin{tabular}{l|lcr|lcr|lcr}
\hline 
\textbf{Theta Band (4-8Hz)} & \multicolumn{3}{c}{Eyes Open} \vline &\multicolumn{3}{c}{Eyes Closed} \vline &\multicolumn{3}{c}{Anesthesia}\\
\hline
Corresponding Graph & Mean   & Var  & SD  & Mean  & Var  & SD & Mean   & Var  & SD\\ 
\hline                             

Frontal Lobe    &16.4   &32.4 &  5.69  & 12,0      &41.8  & 6.46  &28.4      & 84.0   & 9.17      \\

Parietal Lobe   &7.76  &13.6  &3.69      &5.75  &8.79   &2.96      &10.8    &27.4  &5.23   \\

Temporal Lobe   &11.2  & 11.2 & 3.35  & 8.74     & 11.0 & 3.32  &15.8      & 33.9   &  5.82      \\

Occipital Lobe  &18.1  &42.0 & 6.48  & 13.5     & 47.3 &  6.88 & 11.5     & 19.7   &  4.44      \\   

\end{tabular}
\end{table}

\clearpage

\subsubsection*{Alpha 8-12Hz}

\begin{figure}[!h]
\begin{subfigure}{.5\textwidth}
  \centering
  \includegraphics[width=1\linewidth]{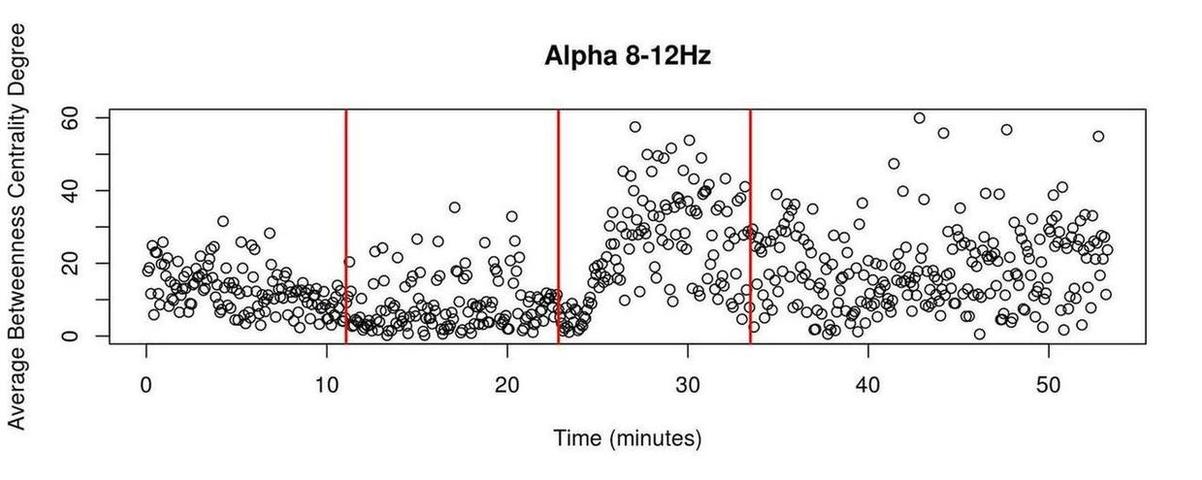}
  \caption{Frontal Lobe}
  \label{fig:sfig1}
\end{subfigure}%
\begin{subfigure}{.5\textwidth}
  \centering
  \includegraphics[width=1\linewidth]{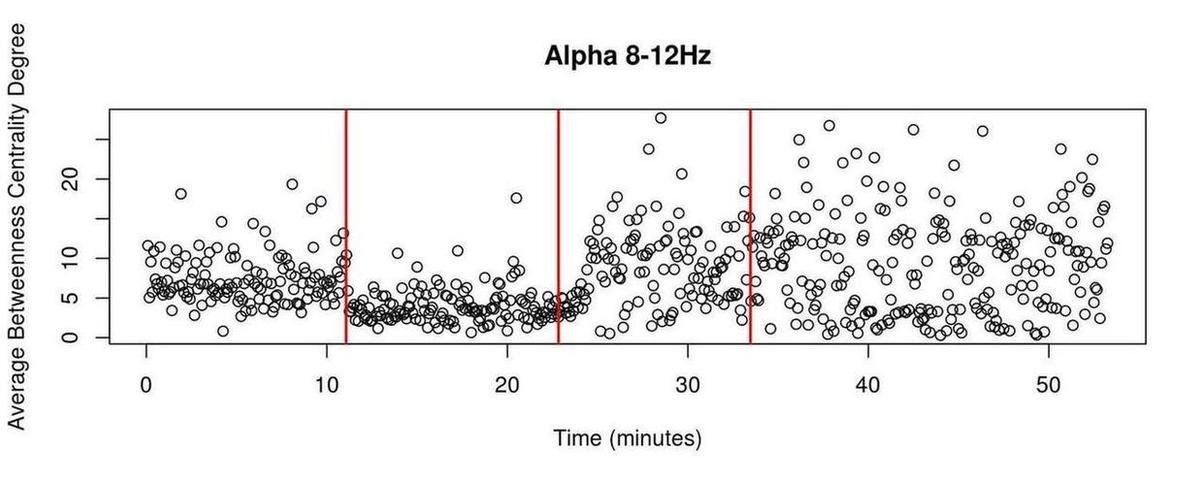}
 \caption{Parietal Lobe}
  \label{fig:sfig2}
\end{subfigure}\\
\centering
\begin{subfigure}{.5\textwidth}
\includegraphics[width=1\linewidth]{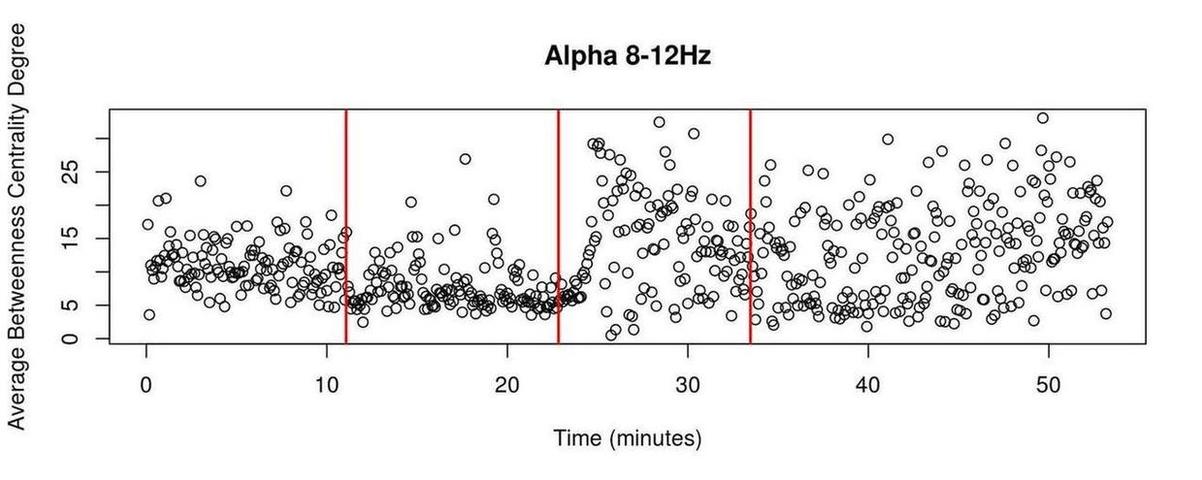}
  \caption{Temporal Lobe}
  \label{fig:sfig3}
\end{subfigure}%
\begin{subfigure}{.5\textwidth}
  \centering
  \includegraphics[width=1\linewidth]{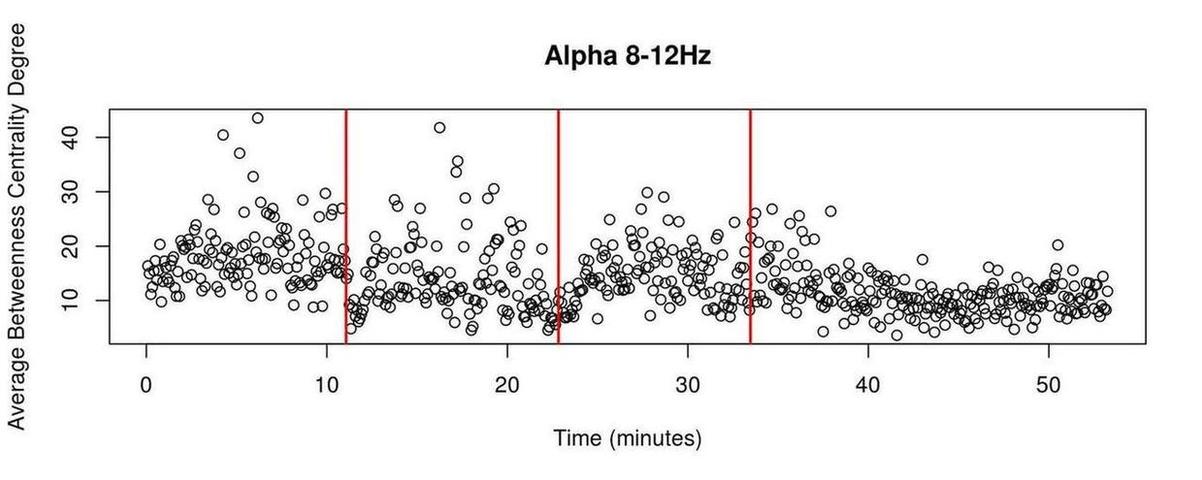}
  \caption{Occipital Lobe}
  \label{fig:sfig4}
\end{subfigure}\\
\caption{{\normalfont \textbf{Average Betweenness Centrality Degree.}} Average betweenness centrality degree  respective to the networks of the Alpha frequency band (8-12Hz). Vertical axis average betweenness centrality degree; Horizontal axis time (minutes). At t=11 minutes, the monkey was blindfolded, the first red line in each sub-figure represents the moment when a patch was placed over the eyes. At t=23 minutes the Ketamine-Medetomidine cocktail was injected, being represented by the second red line. The point of loss of consciousness (LOC) was registered at t=33 minutes, and is indicated by the third red line. Sub-figures: (\textbf{a}) Frontal Lobe; (\textbf{b}) Parietal Lobe; (\textbf{c}) Temporal Lobe; (\textbf{d}) Occipital Lobe.}
\label{fig:fig}
\end{figure}


\begin{table}[!h]
\centering
\caption{{\normalfont \textbf{Average Betweenness Centrality Degree.}} Mean, variance (Var), and standard deviation (SD) of the  average betweenness centrality degree of the networks respective to each one of the four cortical lobes, on the three different conditions in which the monkey was exposed during the experiment: awake with eyes open, awake with eyes closed and anesthesia (eyes closed). Frequency band Alpha (8-12Hz). }
\vspace{0.5cm}
\begin{tabular}{l|lcr|lcr|lcr}
\hline 
\textbf{Alpha Band (8-12Hz)} & \multicolumn{3}{c}{Eyes Open} \vline &\multicolumn{3}{c}{Eyes Closed} \vline &\multicolumn{3}{c}{Anesthesia}\\
\hline
Corresponding Graph & Mean   & Var  & SD  & Mean  & Var  & SD & Mean   & Var  & SD\\ 
\hline                             

Frontal Lobe  &13.3   &37.0  &6.09       &9.10  &52.8   & 7.26     &20.9   &145   & 12.1       \\

Parietal Lobe   &7.33  &10.3  &3.21      &4.09  &5.58   &2.36      &9.24    &34.8  &5.90   \\

Temporal Lobe  &11.1   &13.1  &3.61       &7.73  &14.4   &3.79      &13.2   &48.0   &6.93       \\

Occipital Lobe   &18.2   &35.5  &5.96       &14.1  &48.4   &6.96      &12.0   &20.5   &4.52     \\   

\end{tabular}
\end{table}

\clearpage

\subsubsection*{Beta 13-30Hz}

\begin{figure}[!h]
\begin{subfigure}{.5\textwidth}
  \centering
  \includegraphics[width=1\linewidth]{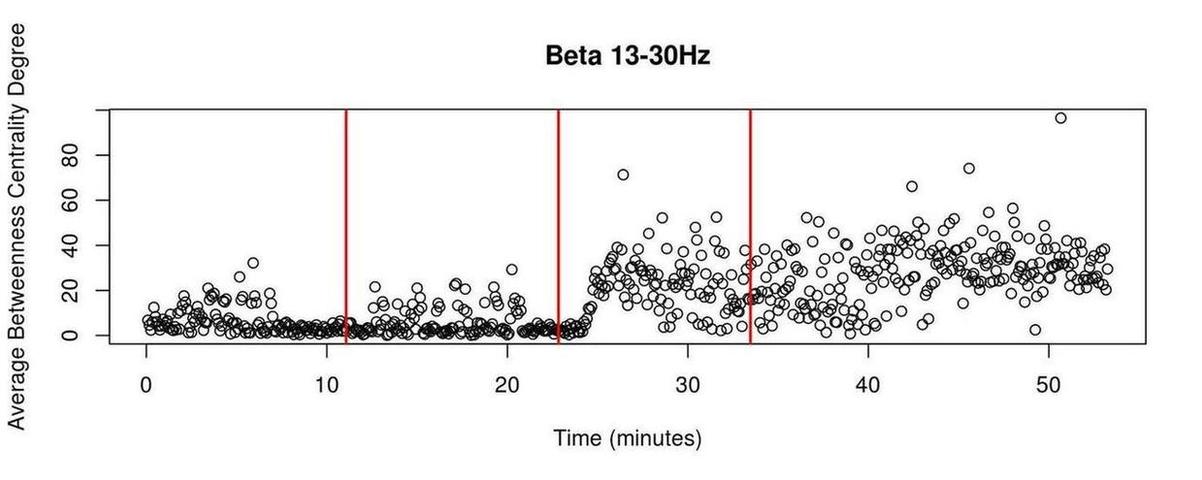}
  \caption{Frontal Lobe}
  \label{fig:sfig1}
\end{subfigure}%
\begin{subfigure}{.5\textwidth}
  \centering
  \includegraphics[width=1\linewidth]{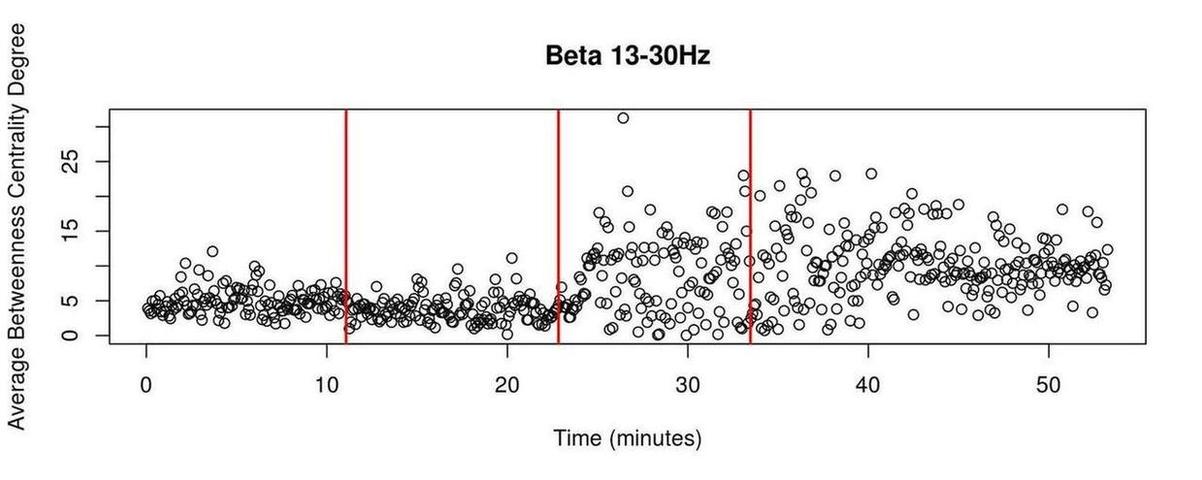}
 \caption{Parietal Lobe}
  \label{fig:sfig2}
\end{subfigure}\\
\centering
\begin{subfigure}{.5\textwidth}
\includegraphics[width=1\linewidth]{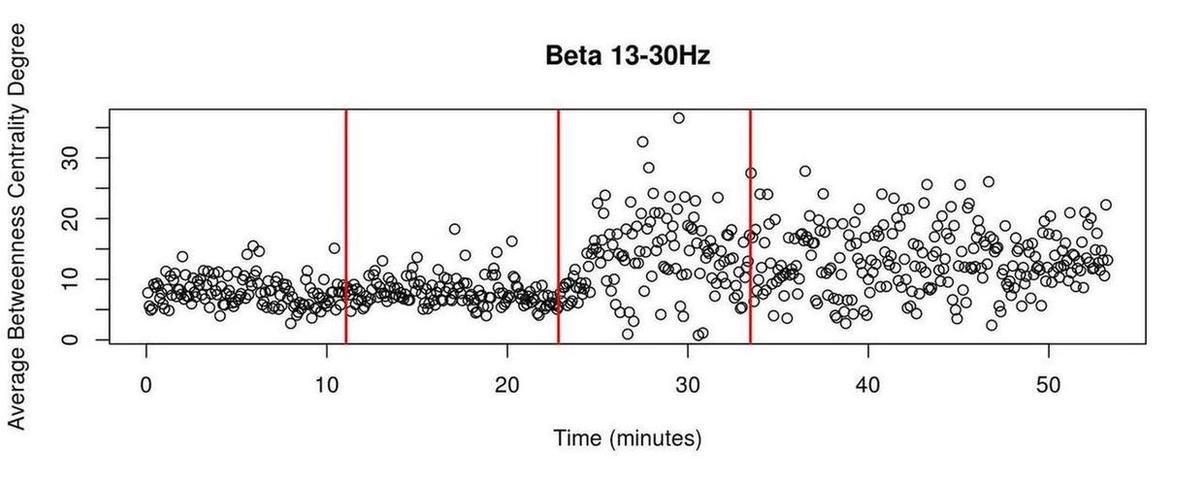}
  \caption{Temporal Lobe}
  \label{fig:sfig3}
\end{subfigure}%
\begin{subfigure}{.5\textwidth}
  \centering
  \includegraphics[width=1\linewidth]{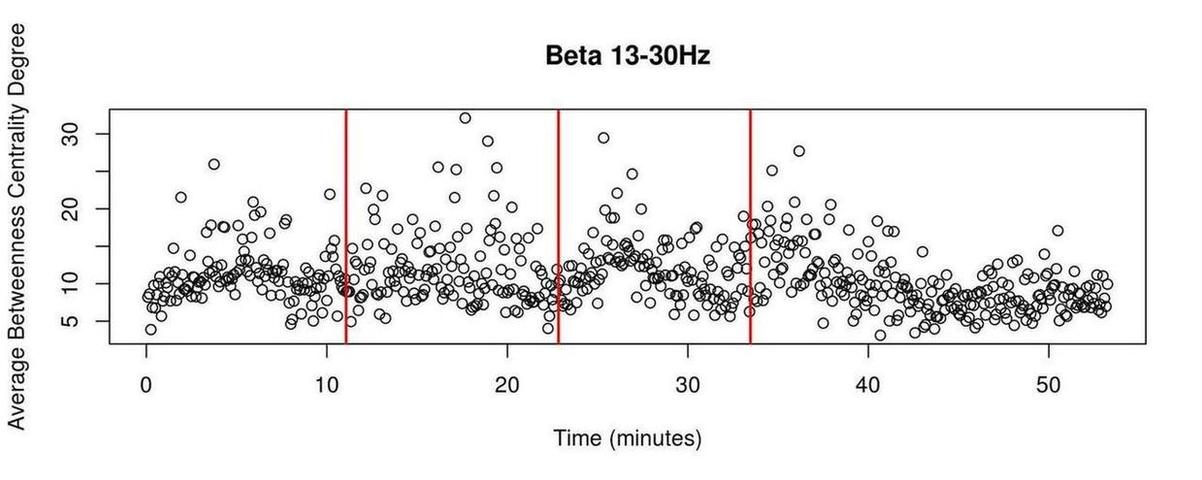}
  \caption{Occipital Lobe}
  \label{fig:sfig4}
\end{subfigure}\\
\caption{{\normalfont \textbf{Average Betweenness Centrality Degree.}} Average betweenness centrality degree  respective to the networks of the Beta frequency band (13-30Hz). Vertical axis average betweenness centrality degree; Horizontal axis time (minutes). At t=11 minutes, the monkey was blindfolded, the first red line in each sub-figure represents the moment when a patch was placed over the eyes. At t=23 minutes the Ketamine-Medetomidine cocktail was injected, being represented by the second red line. The point of loss of consciousness (LOC) was registered at t=33 minutes, and is indicated by the third red line. Sub-figures: (\textbf{a}) Frontal Lobe; (\textbf{b}) Parietal Lobe; (\textbf{c}) Temporal Lobe; (\textbf{d}) Occipital Lobe.}
\label{fig:fig}
\end{figure}


\begin{table}[!h]
\centering
\caption{{\normalfont \textbf{Average Betweenness Centrality Degree.}} Mean, variance (Var), and standard deviation (SD) of the  average betweenness centrality degree of the networks respective to each one of the four cortical lobes, on the three different conditions in which the monkey was exposed during the experiment: awake with eyes open, awake with eyes closed and anesthesia (eyes closed). Frequency band Beta (13-30Hz).  }
\vspace{0.5cm}
\begin{tabular}{l|lcr|lcr|lcr}
\hline 
\textbf{Beta Band (13-30Hz)} & \multicolumn{3}{c}{Eyes Open} \vline &\multicolumn{3}{c}{Eyes Closed} \vline &\multicolumn{3}{c}{Anesthesia}\\
\hline
Corresponding Graph & Mean   & Var  & SD  & Mean  & Var  & SD & Mean   & Var  & SD\\ 
\hline                             

Frontal Lobe    &6.52   &35.0  &5.91       &5.92  &37.1   & 6.09     &26.8    & 179   & 13.4  \\

Parietal Lobe   &4.89  &3.51  &1.87      &3.81  &3.48   &1.87      &9.82    &23.7  &4.87   \\

Temporal Lobe   &7.89  &5.53  &2.35      &7.87  & 5.73  &   2.39   & 13.5   & 30.1   &  5.49  \\

Occipital Lobe  &11.1  & 13.0 &3.60      & 11.9 & 24.5  &  4.95   & 10.0   & 14.6   &    3.82\\   

\end{tabular}
\end{table}

\clearpage

\subsubsection*{Gamma 25-100Hz}

\begin{figure}[!h]
\begin{subfigure}{.5\textwidth}
  \centering
  \includegraphics[width=1\linewidth]{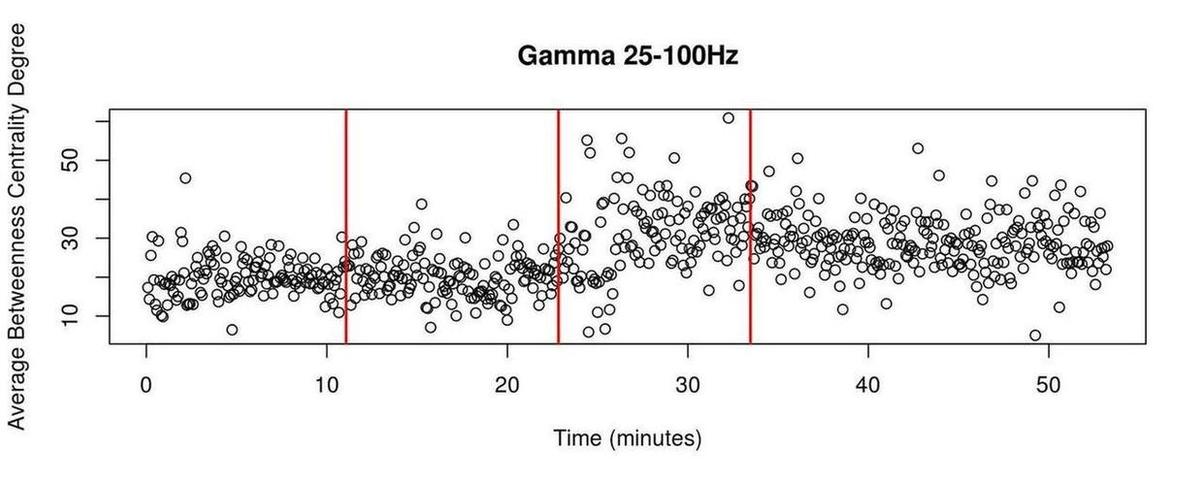}
  \caption{Frontal Lobe}
  \label{fig:sfig1}
\end{subfigure}%
\begin{subfigure}{.5\textwidth}
  \centering
  \includegraphics[width=1\linewidth]{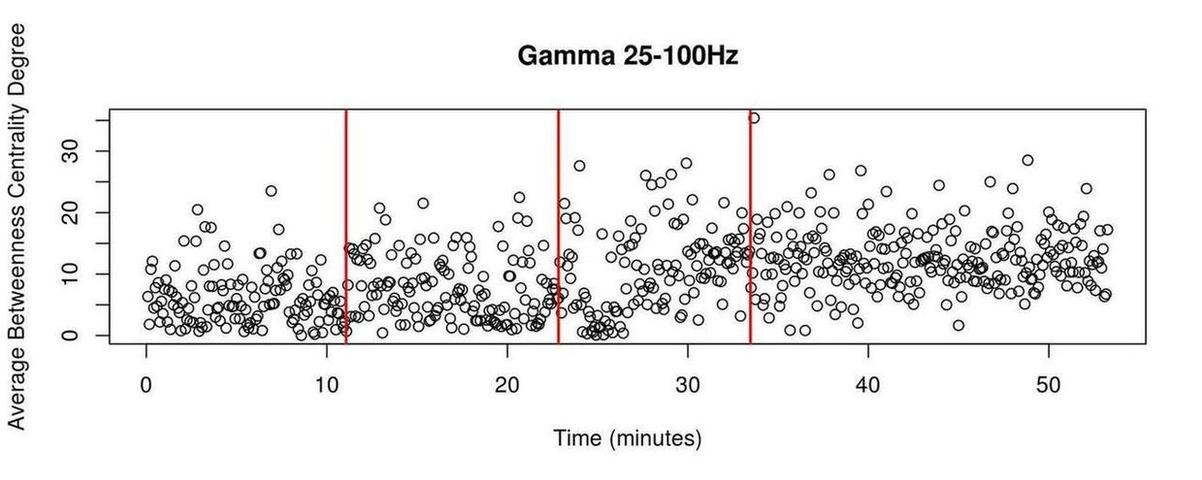}
 \caption{Parietal Lobe}
  \label{fig:sfig2}
\end{subfigure}\\
\centering
\begin{subfigure}{.5\textwidth}
\includegraphics[width=1\linewidth]{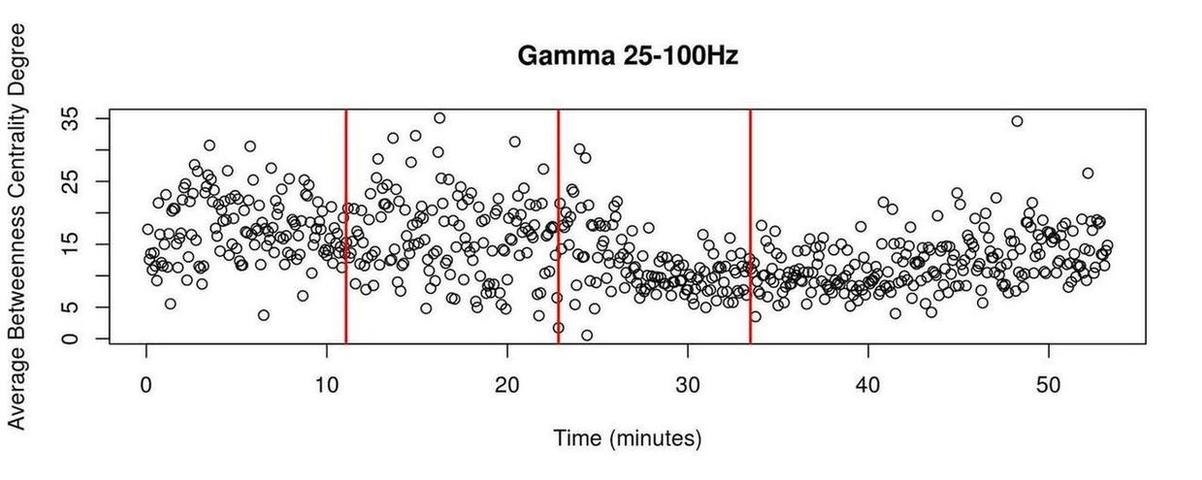}
  \caption{Temporal Lobe}
  \label{fig:sfig3}
\end{subfigure}%
\begin{subfigure}{.5\textwidth}
  \centering
  \includegraphics[width=1\linewidth]{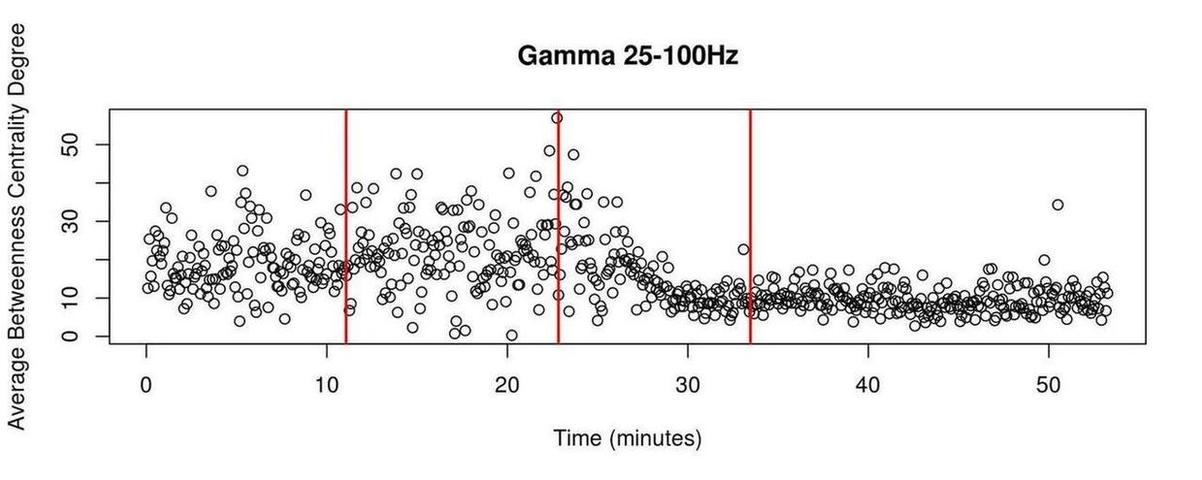}
  \caption{Occipital Lobe}
  \label{fig:sfig4}
\end{subfigure}\\
\caption{{\normalfont \textbf{Average Betweenness Centrality Degree.}} Average betweenness centrality degree  respective to the networks of the Gamma frequency band (25-100Hz). Vertical axis average betweenness centrality degree; Horizontal axis time (minutes). At t=11 minutes, the monkey was blindfolded, the first red line in each sub-figure represents the moment when a patch was placed over the eyes. At t=23 minutes the Ketamine-Medetomidine cocktail was injected, being represented by the second red line. The point of loss of consciousness (LOC) was registered at t=33 minutes, and is indicated by the third red line. Sub-figures: (\textbf{a}) Frontal Lobe; (\textbf{b}) Parietal Lobe; (\textbf{c}) Temporal Lobe; (\textbf{d}) Occipital Lobe.}
\label{fig:fig}
\end{figure}


\begin{table}[!h]
\centering
\caption{{\normalfont \textbf{Average Betweenness Centrality Degree.}} Mean, variance (Var), and standard deviation (SD) of the  average betweenness centrality degree of the networks respective to each one of the four cortical lobes, on the three different conditions in which the monkey was exposed during the experiment: awake with eyes open, awake with eyes closed and anesthesia (eyes closed). Frequency band Gamma (25-100Hz). }
\vspace{0.5cm}
\begin{tabular}{l|lcr|lcr|lcr}
\hline 
\textbf{Gamma Band 25-100Hz)} & \multicolumn{3}{c}{Eyes Open} \vline &\multicolumn{3}{c}{Eyes Closed} \vline &\multicolumn{3}{c}{Anesthesia}\\
\hline
Corresponding Graph & Mean   & Var  & SD  & Mean  & Var  & SD & Mean   & Var  & SD\\ 
\hline                             

Frontal Lobe    &20.1   & 27.2 &5.21       & 20.0 & 28.1  & 5.3     & 29.7   & 51.0   & 7.14  \\

Parietal Lobe   &6.42  &21.3  &4.62      &7.30  &25.3   &5.03      &12.5    &26.9  &5.19   \\

Temporal Lobe   & 17.6 & 27.5 & 5.25     & 16.1 & 44.3  &  6.66    &  11.6  & 16.4   &  4.05  \\

Occipital Lobe  & 18.9 & 54.5 & 7.38     & 22.0 & 100  &  10.0    &  9.84  &  13.0  &    3.60\\   

\end{tabular}
\end{table}

\clearpage

\subsection*{Measures Related to Local Integration}

\subsubsection{Transitivity}

\subsubsection*{Delta 0-4Hz}

\begin{figure}[!h]
\begin{subfigure}{.5\textwidth}
  \centering
  \includegraphics[width=1\linewidth]{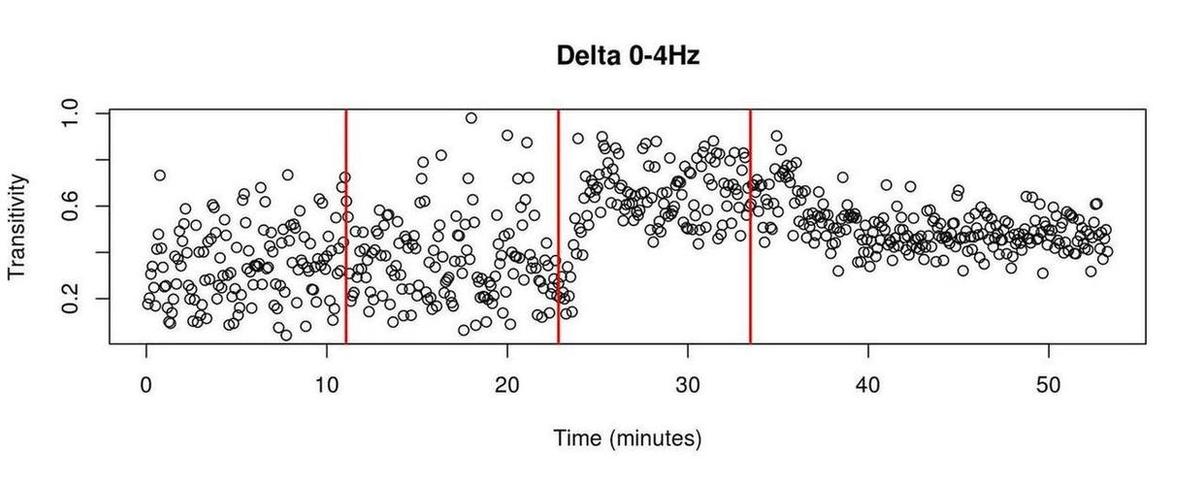}
  \caption{Frontal Lobe}
  \label{fig:sfig1}
\end{subfigure}%
\begin{subfigure}{.5\textwidth}
  \centering
  \includegraphics[width=1\linewidth]{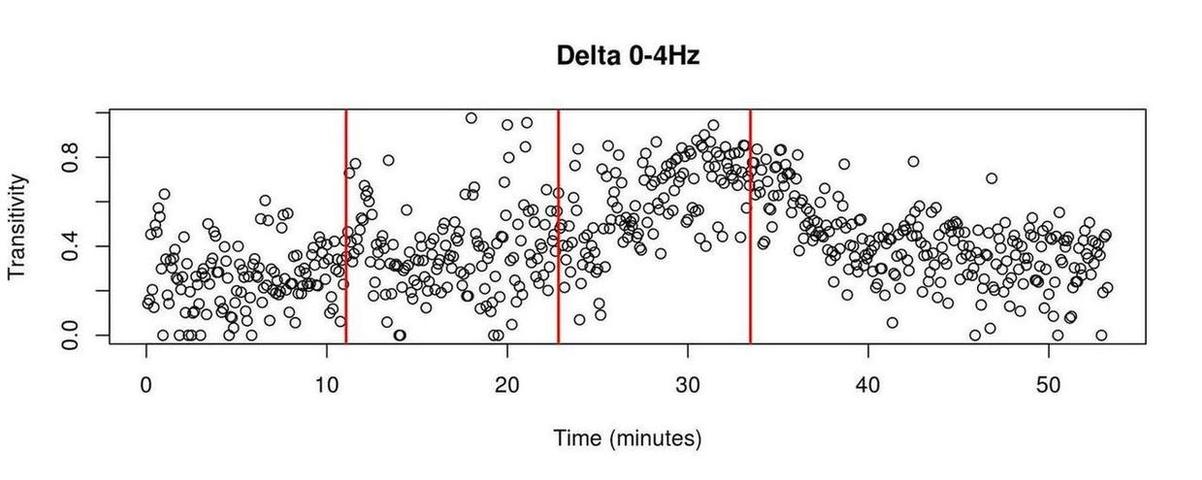}
 \caption{Parietal Lobe}
  \label{fig:sfig2}
\end{subfigure}\\
\centering
\begin{subfigure}{.5\textwidth}
\includegraphics[width=1\linewidth]{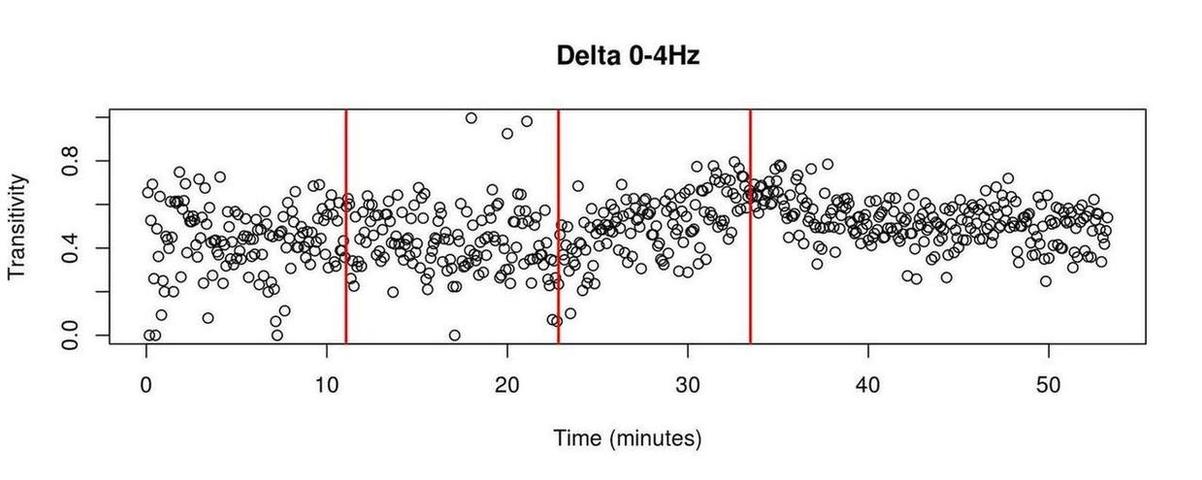}
  \caption{Temporal Lobe}
  \label{fig:sfig3}
\end{subfigure}%
\begin{subfigure}{.5\textwidth}
  \centering
  \includegraphics[width=1\linewidth]{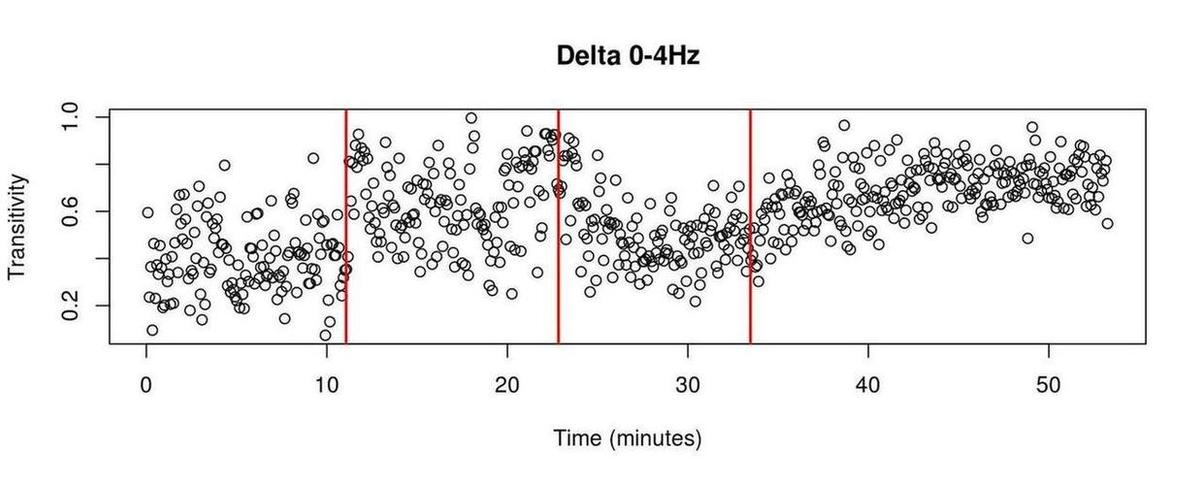}
  \caption{Occipital Lobe}
  \label{fig:sfig4}
\end{subfigure}\\
\caption{{\normalfont \textbf{Transitivity.}} Transitivity coefficient respective to the networks of the Delta frequency band (0-4Hz). Vertical axis transitivity coefficient; Horizontal axis time (minutes). At t=11 minutes, the monkey was blindfolded, the first red line in each sub-figure represents the moment when a patch was placed over the eyes. At t=23 minutes the Ketamine-Medetomidine cocktail was injected, being represented by the second red line. The point of loss of consciousness (LOC) was registered at t=33 minutes, and is indicated by the third red line. Sub-figures: (\textbf{a}) Frontal Lobe; (\textbf{b}) Parietal Lobe; (\textbf{c}) Temporal Lobe; (\textbf{d}) Occipital Lobe.}
\label{fig:fig}
\end{figure}


\begin{table}[!h]
\centering
\caption{{\normalfont \textbf{Transitivity.}}  Mean, variance (Var), and standard deviation (SD) of the  transitivity coefficient of the networks respective to each one of the four cortical lobes, on the three different conditions in which the monkey was exposed during the experiment: awake with eyes open, awake with eyes closed and anesthesia (eyes closed). Frequency band Delta (0-4Hz).  }
\vspace{0.5cm}
\begin{tabular}{l|lcr|lcr|lcr}
\hline 
\textbf{Delta Band (0-4Hz)} & \multicolumn{3}{c}{Eyes Open} \vline &\multicolumn{3}{c}{Eyes Closed} \vline &\multicolumn{3}{c}{Anesthesia}\\
\hline
Corresponding Graph & Mean   & Var  & SD  & Mean  & Var  & SD & Mean   & Var  & SD\\ 
\hline                             

Frontal Lobe    & 0.34  & 0.02 & 0.16      & 0.37 & 0.03  & 0.19     & 0.54   & 0.02   &  0.12 \\

Parietal Lobe   &0.26  &0.02  &0.14      &0.36  &0.04   &0.19      &0.48    &0.04  &0.20   \\

Temporal Lobe   & 0.44 & 0.02 & 0.16     & 0.43 & 0.03  &  0.16    & 0.53   &  0.01  &  0.11  \\

Occipital Lobe  & 0.40 & 0.02 & 0.15     & 0.63 &  0.03 & 0.18     &  0.64  & 0.02   &  0.15  \\   

\end{tabular}
\end{table}

\clearpage

\subsubsection*{Theta 4-8Hz}

\begin{figure}[!h]
\begin{subfigure}{.5\textwidth}
  \centering
  \includegraphics[width=1\linewidth]{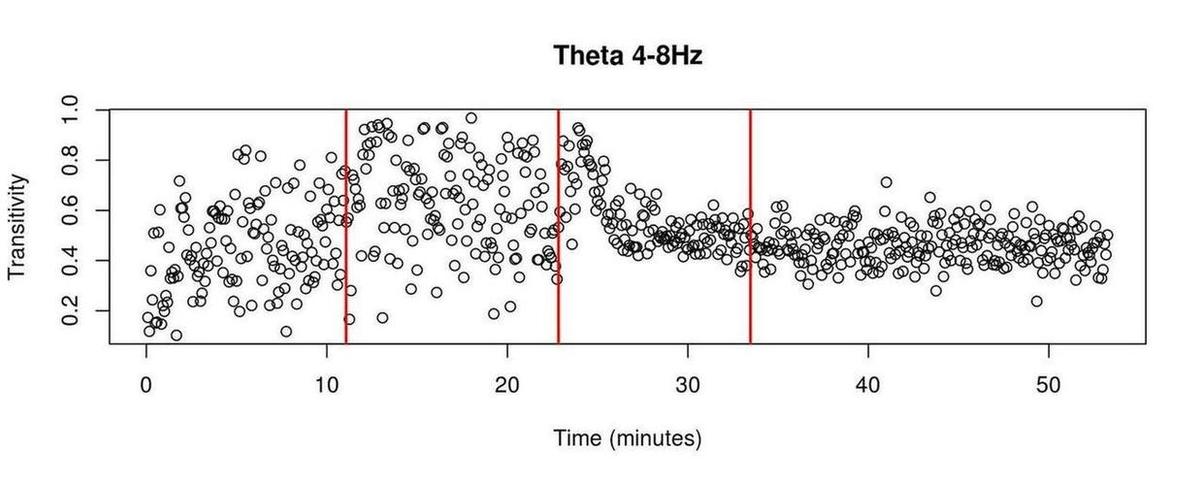}
  \caption{Frontal Lobe}
  \label{fig:sfig1}
\end{subfigure}%
\begin{subfigure}{.5\textwidth}
  \centering
  \includegraphics[width=1\linewidth]{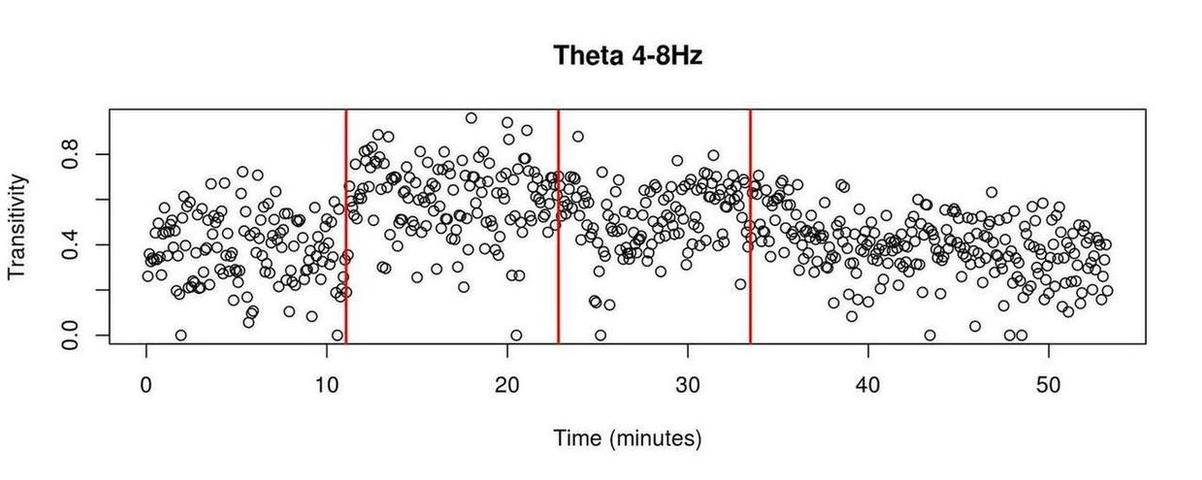}
 \caption{Parietal Lobe}
  \label{fig:sfig2}
\end{subfigure}\\
\centering
\begin{subfigure}{.5\textwidth}
\includegraphics[width=1\linewidth]{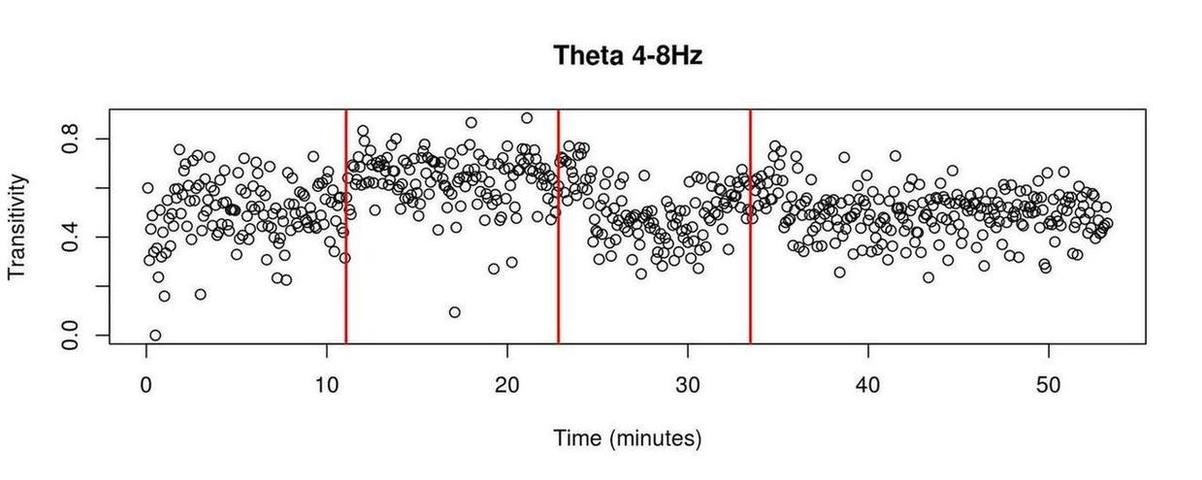}
  \caption{Temporal Lobe}
  \label{fig:sfig3}
\end{subfigure}%
\begin{subfigure}{.5\textwidth}
  \centering
  \includegraphics[width=1\linewidth]{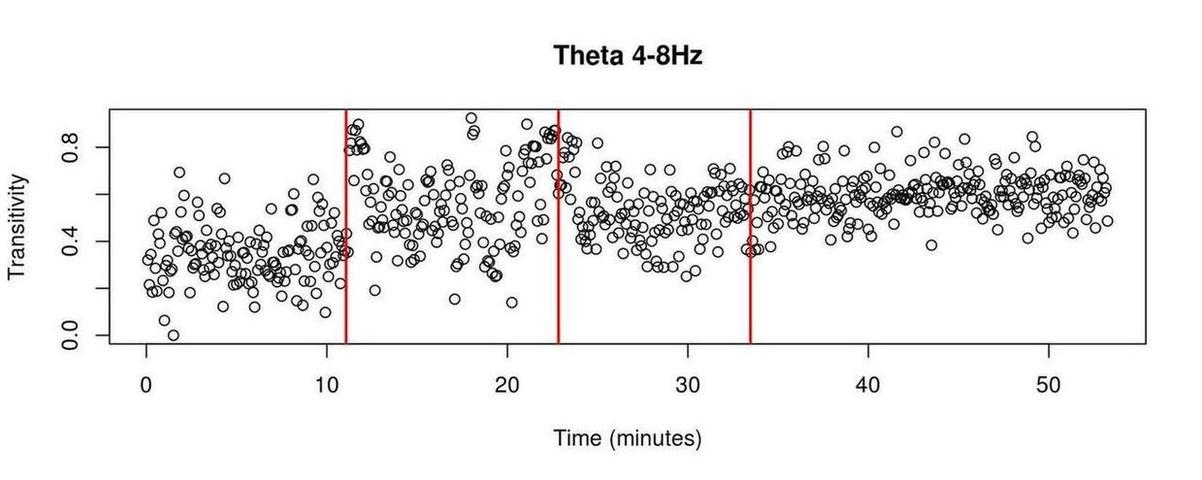}
  \caption{Occipital Lobe}
  \label{fig:sfig4}
\end{subfigure}\\
\caption{{\normalfont \textbf{Transitivity.}} Transitivity coefficient respective to the networks of the Theta frequency band (4-8Hz). Vertical axis transitivity coefficient; Horizontal axis time (minutes). At t=11 minutes, the monkey was blindfolded, the first red line in each sub-figure represents the moment when a patch was placed over the eyes. At t=23 minutes the Ketamine-Medetomidine cocktail was injected, being represented by the second red line. The point of loss of consciousness (LOC) was registered at t=33 minutes, and is indicated by the third red line. Sub-figures: (\textbf{a}) Frontal Lobe; (\textbf{b}) Parietal Lobe; (\textbf{c}) Temporal Lobe; (\textbf{d}) Occipital Lobe.}
\label{fig:fig}
\end{figure}


\begin{table}[!h]
\centering
\caption{{\normalfont \textbf{Transitivity.}}  Mean, variance (Var), and standard deviation (SD) of the  transitivity coefficient of the networks respective to each one of the four cortical lobes, on the three different conditions in which the monkey was exposed during the experiment: awake with eyes open, awake with eyes closed and anesthesia (eyes closed). Frequency band Theta (4-8Hz). }
\vspace{0.5cm}
\begin{tabular}{l|lcr|lcr|lcr}
\hline 
\textbf{Theta Band (4-8Hz)} & \multicolumn{3}{c}{Eyes Open} \vline &\multicolumn{3}{c}{Eyes Closed} \vline &\multicolumn{3}{c}{Anesthesia}\\
\hline
Corresponding Graph & Mean   & Var  & SD  & Mean  & Var  & SD & Mean   & Var  & SD\\ 
\hline                             

Frontal Lobe    &0.44   &0.03  &0.17       &0.62  & 0.03  &0.19      &0.47    & 0.01   & 0.07 \\

Parietal Lobe   &0.38  &0.02  &0.15      &0.59  &0.03   &0.16      &0.43    &0.02  &0.15   \\

Temporal Lobe   & 0.50 & 0.02 & 0.13     & 0.63 & 0.01  &0.11     & 0.50   & 0.01   &    0.10\\

Occipital Lobe  &0.34 & 0.02 &  0.13    &0.56  & 0.03  & 0.18     & 0.58   & 0.01   &    0.10\\   

\end{tabular}
\end{table}

\clearpage

\subsubsection*{Alpha 8-12Hz}

\begin{figure}[!h]
\begin{subfigure}{.5\textwidth}
  \centering
  \includegraphics[width=1\linewidth]{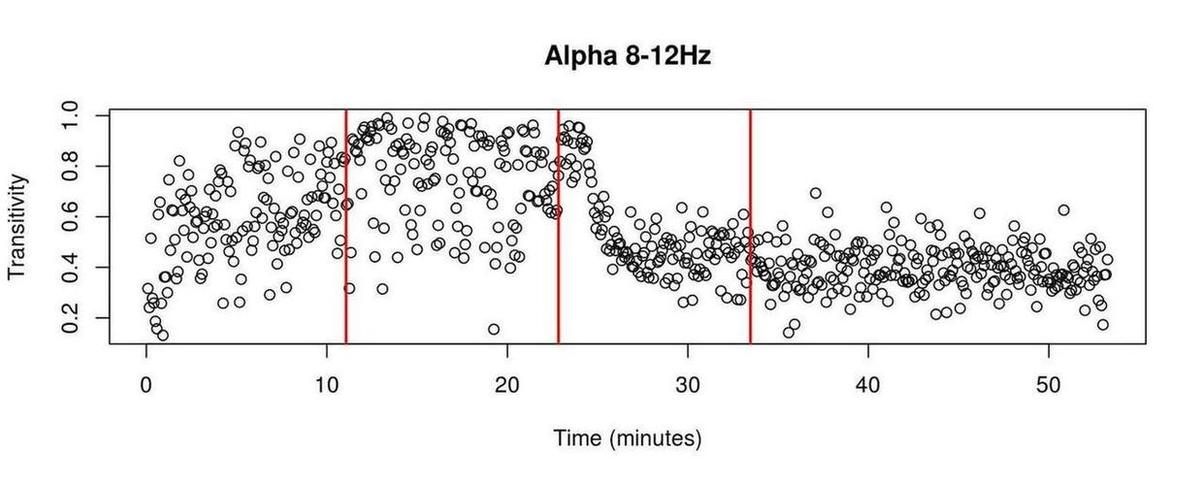}
  \caption{Frontal Lobe}
  \label{fig:sfig1}
\end{subfigure}%
\begin{subfigure}{.5\textwidth}
  \centering
  \includegraphics[width=1\linewidth]{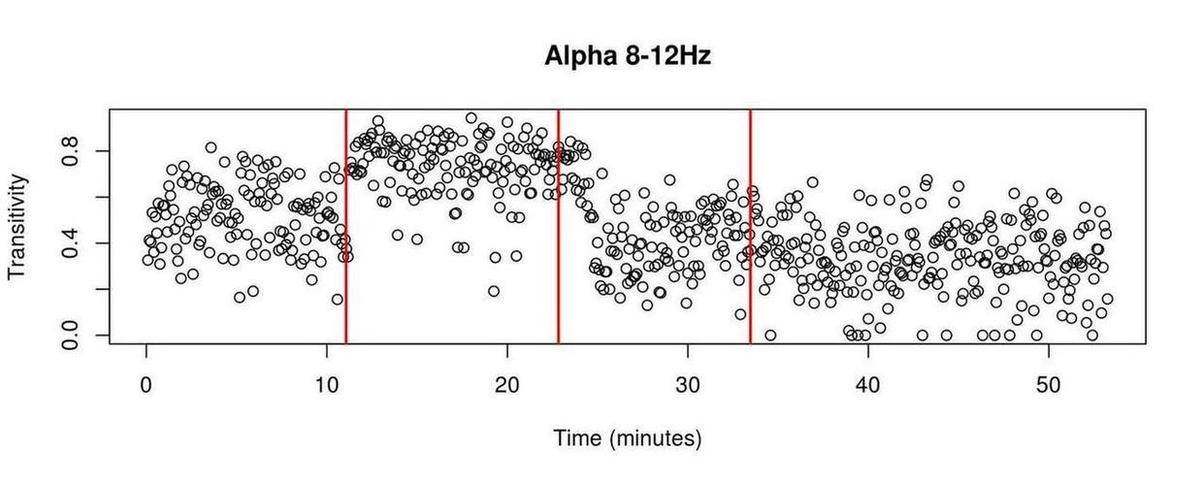}
 \caption{Parietal Lobe}
  \label{fig:sfig2}
\end{subfigure}\\
\centering
\begin{subfigure}{.5\textwidth}
\includegraphics[width=1\linewidth]{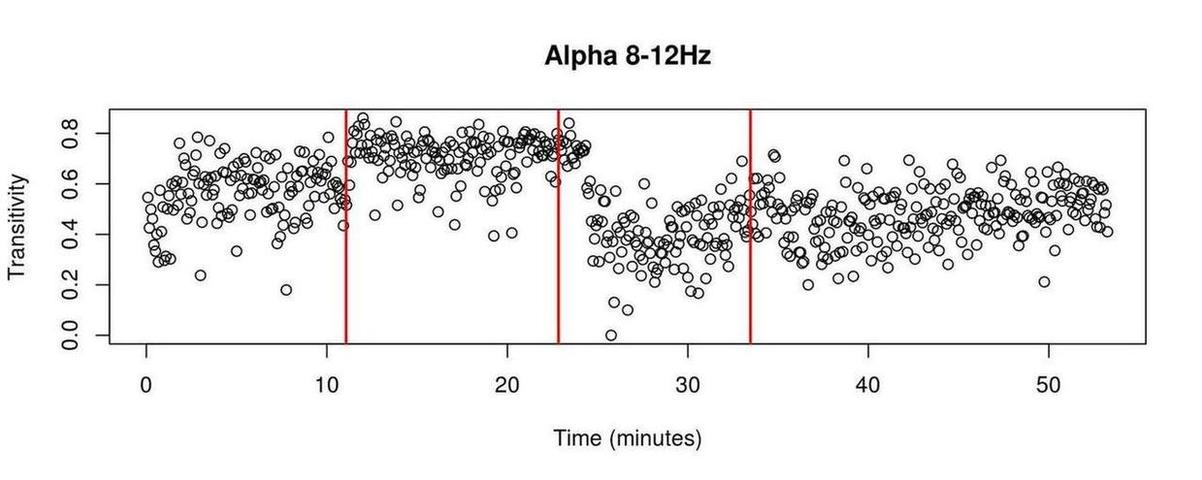}
  \caption{Temporal Lobe}
  \label{fig:sfig3}
\end{subfigure}%
\begin{subfigure}{.5\textwidth}
  \centering
  \includegraphics[width=1\linewidth]{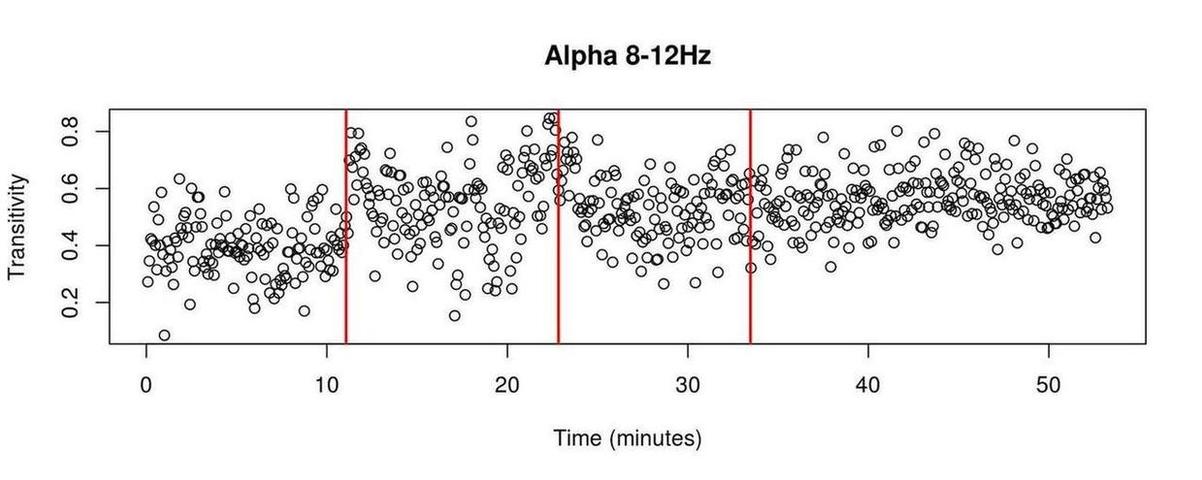}
  \caption{Occipital Lobe}
  \label{fig:sfig4}
\end{subfigure}\\
\caption{{\normalfont \textbf{Transitivity.}} Transitivity coefficient respective to the networks of the Alpha frequency band (8-12Hz). Vertical axis transitivity coefficient; Horizontal axis time (minutes). At t=11 minutes, the monkey was blindfolded, the first red line in each sub-figure represents the moment when a patch was placed over the eyes. At t=23 minutes the Ketamine-Medetomidine cocktail was injected, being represented by the second red line. The point of loss of consciousness (LOC) was registered at t=33 minutes, and is indicated by the third red line. Sub-figures: (\textbf{a}) Frontal Lobe; (\textbf{b}) Parietal Lobe; (\textbf{c}) Temporal Lobe; (\textbf{d}) Occipital Lobe.}
\label{fig:fig}
\end{figure}


\begin{table}[!h]
\centering
\caption{{\normalfont \textbf{Transitivity.}}  Mean, variance (Var), and standard deviation (SD) of the  transitivity coefficient of the networks respective to each one of the four cortical lobes, on the three different conditions in which the monkey was exposed during the experiment: awake with eyes open, awake with eyes closed and anesthesia (eyes closed). Frequency band Alpha (8-12Hz). }
\vspace{0.5cm}
\begin{tabular}{l|lcr|lcr|lcr}
\hline 
\textbf{Alpha Band (8-12Hz)} & \multicolumn{3}{c}{Eyes Open} \vline &\multicolumn{3}{c}{Eyes Closed} \vline &\multicolumn{3}{c}{Anesthesia}\\
\hline
Corresponding Graph & Mean   & Var  & SD  & Mean  & Var  & SD & Mean   & Var  & SD\\ 
\hline                             

Frontal Lobe  &0.58   &0.03  &0.18       &0.75  &0.03   &0.17      &0.40   &0.01   & 0.09     \\

Parietal Lobe   &0.52  &0.02  &0.14      &0.72  &0.02   &0.13      &0.36    & 0.02 & 0.15  \\

Temporal Lobe   &0.56   &0.01  &0.12       &0.71  &0.01   &0.09      &0.46   &0.01   &0.11     \\

Occipital Lobe  &0.39   &0.01  &0.10       &0.54  &0.02   &0.15      &0.55   &0.01   &0.10    \\   

\end{tabular}
\end{table}

\clearpage

\subsubsection*{Beta 13-30Hz}

\begin{figure}[!h]
\begin{subfigure}{.5\textwidth}
  \centering
  \includegraphics[width=1\linewidth]{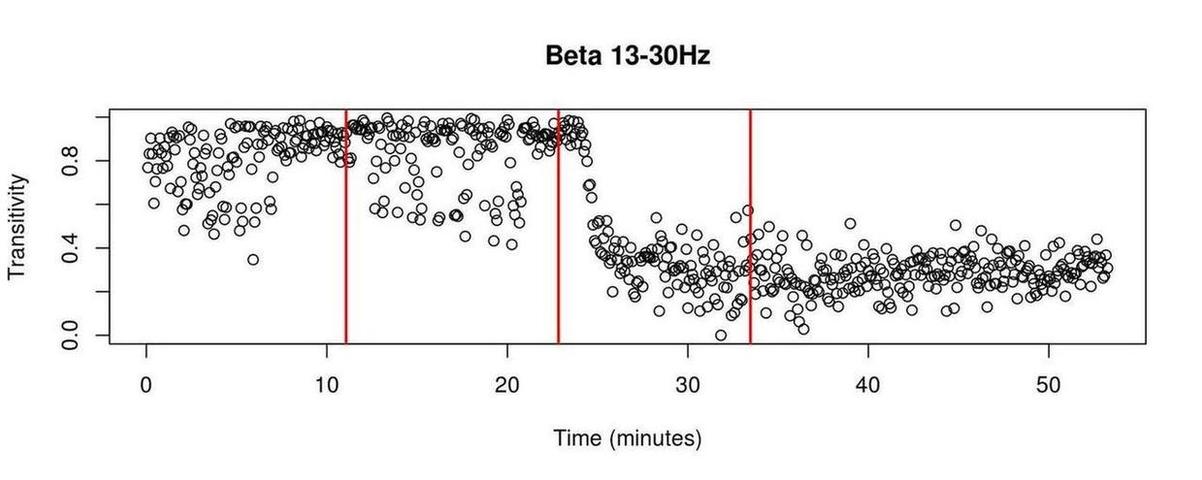}
  \caption{Frontal Lobe}
  \label{fig:sfig1}
\end{subfigure}%
\begin{subfigure}{.5\textwidth}
  \centering
  \includegraphics[width=1\linewidth]{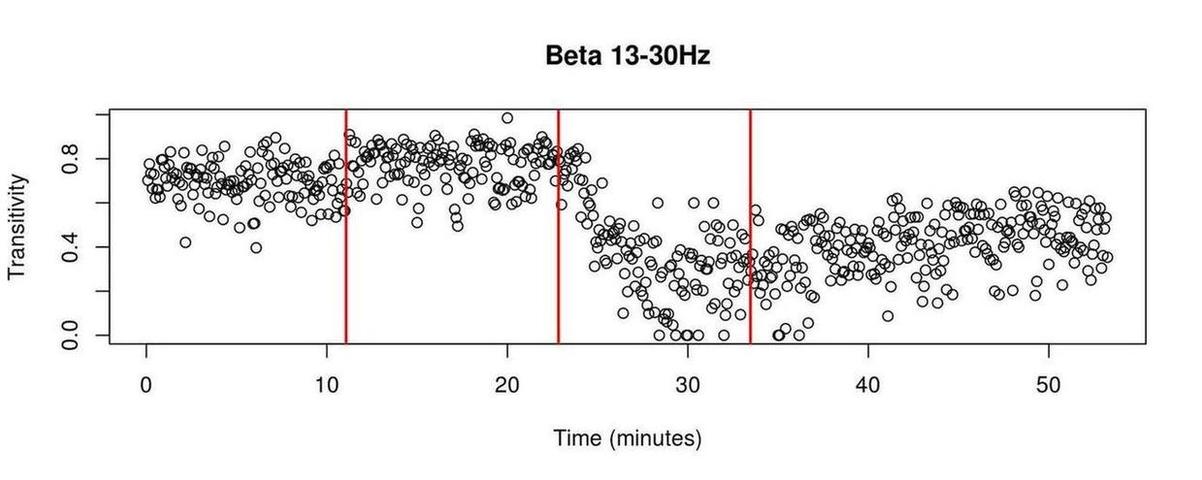}
 \caption{Parietal Lobe}
  \label{fig:sfig2}
\end{subfigure}\\
\centering
\begin{subfigure}{.5\textwidth}
\includegraphics[width=1\linewidth]{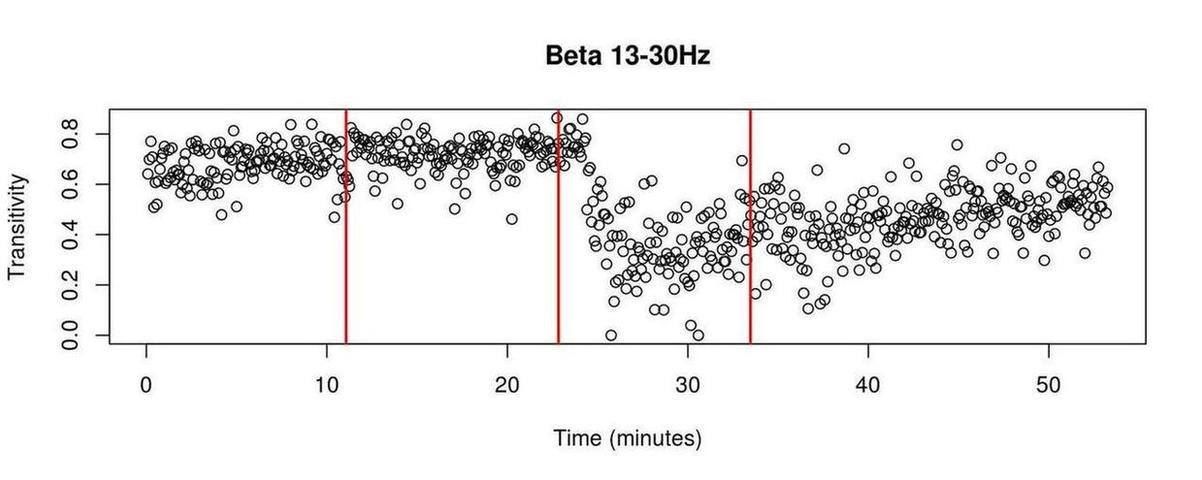}
  \caption{Temporal Lobe}
  \label{fig:sfig3}
\end{subfigure}%
\begin{subfigure}{.5\textwidth}
  \centering
  \includegraphics[width=1\linewidth]{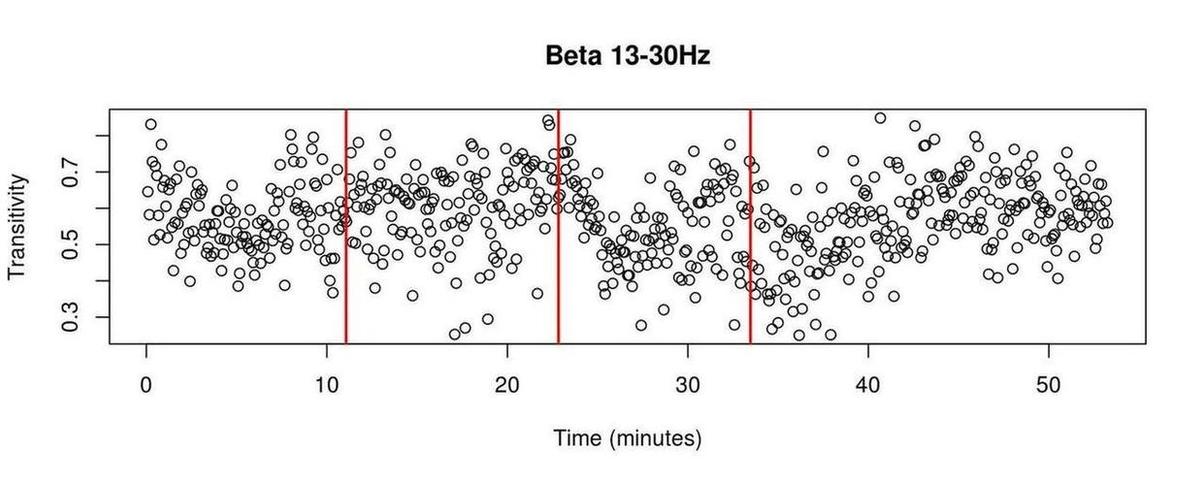}
  \caption{Occipital Lobe}
  \label{fig:sfig4}
\end{subfigure}\\
\caption{{\normalfont \textbf{Transitivity.}} Transitivity coefficient respective to the networks of the Beta frequency band (13-30Hz). Vertical axis transitivity coefficient; Horizontal axis time (minutes). At t=11 minutes, the monkey was blindfolded, the first red line in each sub-figure represents the moment when a patch was placed over the eyes. At t=23 minutes the Ketamine-Medetomidine cocktail was injected, being represented by the second red line. The point of loss of consciousness (LOC) was registered at t=33 minutes, and is indicated by the third red line. Sub-figures: (\textbf{a}) Frontal Lobe; (\textbf{b}) Parietal Lobe; (\textbf{c}) Temporal Lobe; (\textbf{d}) Occipital Lobe.}
\label{fig:fig}
\end{figure}


\begin{table}[!h]
\centering
\caption{{\normalfont \textbf{Transitivity.}}  Mean, variance (Var), and standard deviation (SD) of the  transitivity coefficient of the networks respective to each one of the four cortical lobes, on the three different conditions in which the monkey was exposed during the experiment: awake with eyes open, awake with eyes closed and anesthesia (eyes closed). Frequency band Beta (13-30Hz).  }
\vspace{0.5cm}
\begin{tabular}{l|lcr|lcr|lcr}
\hline 
\textbf{Beta Band (13-30Hz)} & \multicolumn{3}{c}{Eyes Open} \vline &\multicolumn{3}{c}{Eyes Closed} \vline &\multicolumn{3}{c}{Anesthesia}\\
\hline
Corresponding Graph & Mean   & Var  & SD  & Mean  & Var  & SD & Mean   & Var  & SD\\ 
\hline                             

Frontal Lobe    &0.81   &0.02  &0.14       &0.83  & 0.02  & 0.16     & 0.29   & 0.01   & 0.09  \\

Parietal Lobe   &0.70  &0.01  &0.09      &0.77  &0.01   &0.10      &0.38    &0.02  &0.15   \\

Temporal Lobe   &0.68  &0.0  &0.07      &0.72 & 0.0  &  0.06    & 0.44   & 0.02   &    0.12\\

Occipital Lobe  &0.58  &0.01  &0.09      & 0.61 & 0.01  &  0.11    & 0.56   & 0.01   & 0.11   \\   

\end{tabular}
\end{table}

\clearpage

\subsection*{Gamma 25-100Hz}

\begin{figure}[!h]
\begin{subfigure}{.5\textwidth}
  \centering
  \includegraphics[width=1\linewidth]{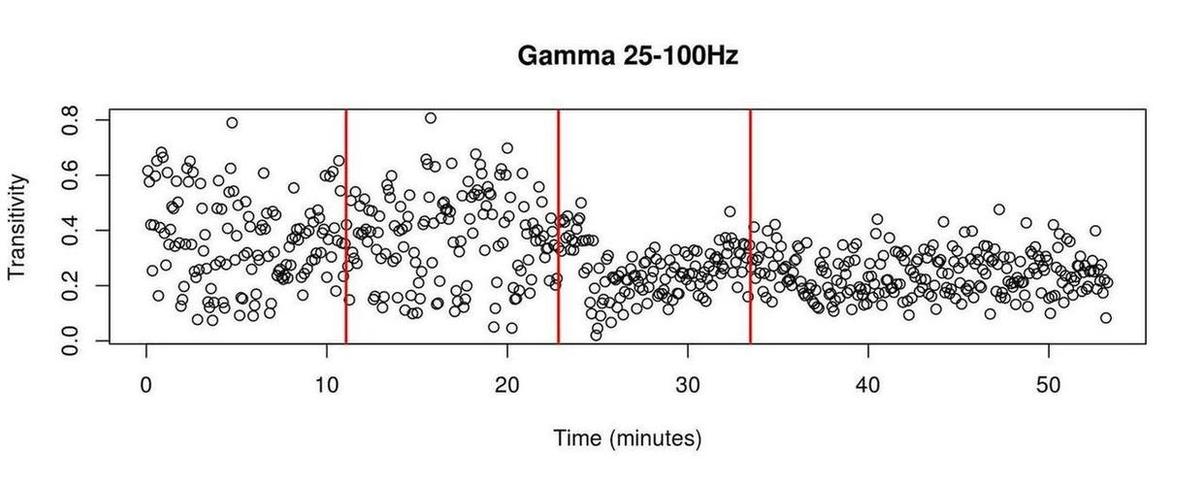}
  \caption{Frontal Lobe}
  \label{fig:sfig1}
\end{subfigure}%
\begin{subfigure}{.5\textwidth}
  \centering
  \includegraphics[width=1\linewidth]{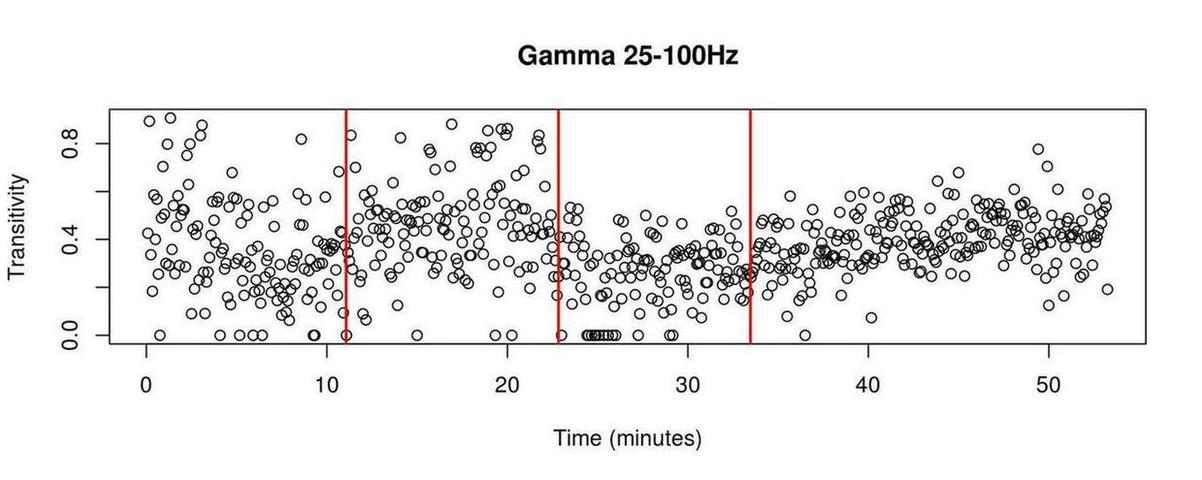}
 \caption{Parietal Lobe}
  \label{fig:sfig2}
\end{subfigure}\\
\centering
\begin{subfigure}{.5\textwidth}
\includegraphics[width=1\linewidth]{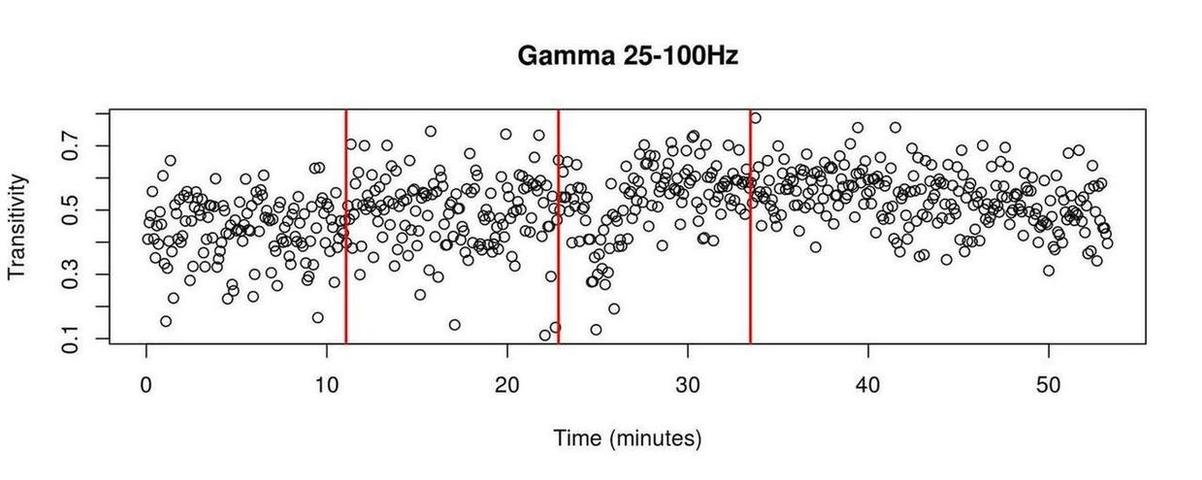}
  \caption{Temporal Lobe}
  \label{fig:sfig3}
\end{subfigure}%
\begin{subfigure}{.5\textwidth}
  \centering
  \includegraphics[width=1\linewidth]{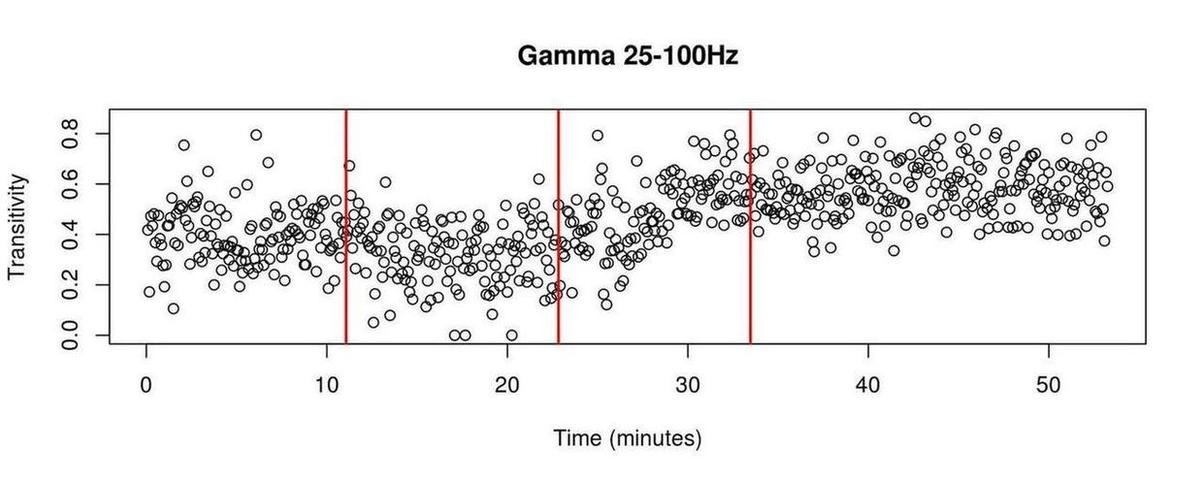}
  \caption{Occipital Lobe}
  \label{fig:sfig4}
\end{subfigure}\\
\caption{{\normalfont \textbf{Transitivity.}} Transitivity coefficient respective to the networks of the Gamma frequency band (25-100Hz). Vertical axis transitivity coefficient; Horizontal axis time (minutes). At t=11 minutes, the monkey was blindfolded, the first red line in each sub-figure represents the moment when a patch was placed over the eyes. At t=23 minutes the Ketamine-Medetomidine cocktail was injected, being represented by the second red line. The point of loss of consciousness (LOC) was registered at t=33 minutes, and is indicated by the third red line. Sub-figures: (\textbf{a}) Frontal Lobe; (\textbf{b}) Parietal Lobe; (\textbf{c}) Temporal Lobe; (\textbf{d}) Occipital Lobe.}
\label{fig:fig}
\end{figure}


\begin{table}[!h]
\centering
\caption{{\normalfont \textbf{Transitivity.}}  Mean, variance (Var), and standard deviation (SD) of the  transitivity coefficient of the networks respective to each one of the four cortical lobes, on the three different conditions in which the monkey was exposed during the experiment: awake with eyes open, awake with eyes closed and anesthesia (eyes closed). Frequency band Gamma (25-100Hz). }
\vspace{0.5cm}
\begin{tabular}{l|lcr|lcr|lcr}
\hline 
\textbf{Gamma Band (25-100Hz)} & \multicolumn{3}{c}{Eyes Open} \vline &\multicolumn{3}{c}{Eyes Closed} \vline &\multicolumn{3}{c}{Anesthesia}\\
\hline
Corresponding Graph & Mean   & Var  & SD  & Mean  & Var  & SD & Mean   & Var  & SD\\ 
\hline                             

Frontal Lobe    & 0.36  & 0.03 & 0.16      & 0.38 & 0.03  & 0.17     & 0.25   & 0.01   & 0.08  \\

Parietal Lobe   &0.36  &0.04  &0.21      &0.47  &0.04   &0.19      &0.37    &0.02  &0.12   \\

Temporal Lobe   &0.44 & 0.01 &  0.10    & 0.49 &  0.01 & 0.11     &  0.55  & 0.01   &    0.08\\

Occipital Lobe  & 0.40 & 0.01 &  0.12    & 0.32 &  0.02  &  0.12   & 0.57   & 0.01   &   0.10 \\   

\end{tabular}
\end{table}

\clearpage



\clearpage

\newpage

\begin{multicols}{2}

\section{Discussion}

In this research, administering the anesthetics has led to alterations in distinct properties of functional brain networks in the four brain lobes and across the five frequency bands analyzed. Those results constitute experimental evidence that reveals distinctions in how functional interactions are established among cortical areas within each brain lobe (frontal, parietal, temporal, and occipital) during awake and general anesthesia conditions.
 
Within less than two minutes after administering the anesthetic cocktail, remarkable changes were concurrently observed in several network properties. The manner in which these alterations occurred strongly suggests a phase transition in which the functional brain networks underwent changes and adopted a distinct architecture. The fact that the changes were observed following the anesthetic injection indicates that these alterations were due to the pharmacological effects of the anesthetics, which involved a significant reduction in awareness and level of consciousness in the animal model.

As it is believed that functional brain networks constitute a substrate that underlies different types of physical dynamics and emergent properties \citep{bassett2006small}, alterations in many network properties at the onset of the transition to unconsciousness suggest that the structure of the functional networks presented during general anesthesia no longer supports neural activities necessary for awareness and conscious experiences. Thus, the results of this research indicate potential links between the structural properties of functional brain networks and neural correlates of consciousness.

\subsection{Average Degree}

Significant alterations in the average degree of the network's vertices were observed during the experiment (see Figs. 1-5; Tables 1-5). In addition, the average degree of networks respective to different brain lobes presented specific and characteristic dynamic behavior in response to the experimental conditions (see Figs. 1-5; Tables 1-5).

On lower frequency bands (0-4 Hz), the anesthetics induced an increase in the average degree on the frontal, parietal, and temporal lobes (compare 0-20 min to 23 -33 min on Fig. 1 $ \lceil $a, b, c$ \rceil $; Table 1). However, after establishing LOC in the same brain regions, the average degree decreased and presented a smaller variation (see Fig. 1 $ \lceil$a, b, c$ \rceil $).

In the experiment, the most remarkable alterations in the average degree of the network's vertices were observed at medium frequencies (4-30Hz). In Theta, Alpha, and Beta frequency bands, the anesthetics promoted an expressive reduction in functional connectivity on the frontal, parietal, and temporal lobes (see Figs. 2,3,4 $ \lceil $a, b, c$ \rceil $; Tables 2,3,4). A different response to the administration of the drug cocktail occurred at the occipital lobe, where the mean connectivity substantially increased (see Figs. 2,3,4 $ \lceil $d$ \rceil $ ).
 
In the Gamma frequency band (25-100Hz), the anesthetic induction led to a decrease in the average degree and its variation over time on the frontal and parietal lobes (see Fig. 5 $\lceil$a, b$ \rceil$; Table 5). In the same frequency band, an increase in the connectivity of the networks was verified after the drug injection on the occipital and temporal lobes (see Fig. 5 $ \lceil $c, d$ \rceil $; Table 5).

The average degree reflects the vertice's average number of connections \citep{rubinov2010complex} bringing information related to the graph's global connectivity. Thus, this network measure indicates how interactive the elements of the system are. Regarding the functional brain networks of the present study, this property provides an estimative of the functional connectivity established locally on the respective brain lobe and frequency band.
The results obtained revealed that the administration of the anesthetics led to an expressive reduction in the average degree on networks of the frontal, parietal, and temporal lobes (Theta, Alpha, and Beta bands) (see Figs. 2,3,4; Table 2,3,4).
From this experimental evidence, it can be concluded that the Ketamine-Medetomidine anesthetic induction led to a drastic reduction in functional connectivity in frontal, parietal, and temporal areas.

\subsection{Assortativity}

Expressive alterations in the assortativity were observed during the experiment (see Figs. 6-10; Tables 6-10). The networks of distinct brain lobes presented characteristic assortativity; the distinct controlled experimental conditions to which the animal model was exposed during the experiment promoted different behaviors on this network property.
The most remarkable alterations in the network's assortativity occurred in the frontal lobe. During awake conditions (eyes open and closed) on the frequency bands Delta, Theta, Alpha, and Beta, the functional networks of the frontal lobe were disassortative. Within less than two minutes after the administration of the anesthetics, a clear transition on this network structural property was verified, the networks once disassortative turned to be assortative\footnote{Networks were assortative most of the time, prevailing this character, but there were also some instants that networks were disassortative (see Figs. 6-9 $ \lceil $a$ \rceil $).} (see Fig 6-9 $ \lceil $a$ \rceil $; Tables 6-9). 

In awake conditions, between (13-30Hz), most of the networks of the temporal and occipital lobes were assortative. However, after the administration of the Ketamine-Medetomidine cocktail, most of the time, the networks prevailed disassortative (see Fig 9 $ \lceil $c, d$ \rceil $; Table 9).

The assortativity \citep{boccaletti2006complex} is related to preferential
attachment between network vertices concerning their connectivity degree \citep{boccaletti2006complex}. Alterations in the assortative character reveal structural changes on the graphs in the manner in which the connections are established among the nodes. As reported by Costa, the assortativity character may greatly influence the dynamic processes supported by the system \citep{di2005synchronization,
brede2005assortative,costa2007characterization}. From this observation, it is possible to hypothesize that the expressive and consistent alterations observed on the frontal lobe (see Fig 6-9 $ \lceil $a$ \rceil $; Tables 6-9) may have impacted the networks in such a way that they no longer supported certain types of neural processes and functional dynamics necessary for awareness and conscious experiences. Thus, these experimental findings indicate potential associations between the assortativity of the functional networks of the frontal lobe and the neural correlates of consciousness.

\subsection{Average Path Length}

Noticeable alterations in the average path length of the networks were observed within about one and a half minutes after administering the anesthetics (see Figs.11-15; Tables 11-15). 
The most remarkable alterations due to the anesthetic induction were verified on the frontal, parietal, and temporal lobes. In the frequency bands Theta, Alpha, and Beta, a substantial increase in the average path length occurred (see Figs. 12-14 $ \lceil $a, b, c$ \rceil $, Tables 12-14). The networks of the occipital lobe presented a somewhat different response to the same experimental conditions, once during general anesthesia (mainly after established LOC), a tendency of the networks to present shorter geodesic paths was verified (see Figs. 12-14 $ \lceil $d$ \rceil $).

On the Gamma frequency band (25-100Hz), administering the anesthetics led to a decrease in the average path length on the temporal and occipital lobe's networks (see Fig. 15 $ \lceil $c, d$ \rceil $; Table 15).
According to Latora and Marchiori, the capacity and global efficiency of a network in transmitting information are directly related to the average of its minimum paths, being the most
efficient, those networks having the shortest paths \citep{latora2001efficient}.
The experimental results of this research demonstrate that the administration of the anesthetics led to a substantial increase in the average path length on the frontal, parietal, and temporal lobes. Thus, the results indicate that in these brain lobes, the capacity and efficiency of information transmission are reduced during general anesthesia.

\subsection{Diameter}

In the experiment, alterations in the diameter were observed in response to the different experimental conditions to which the animal model was exposed (see Figs. 16-20; Tables 16-20).

 In the frequency bands Delta, Theta, Alpha, and Beta, the diameter of the networks of the frontal, parietal, and temporal lobes increased after the administration of the anesthetics (see Figs. 16-19 $ \lceil $a, b, c$ \rceil $, Tables 16-19). However, a different phenomenon occurred in the occipital lobe, where the diameter decreased during general anesthesia (see Figs. 16-19 $ \lceil $d$ \rceil $, Tables 16-19).

In the Gamma frequency band (25-100Hz), the diameter of the networks respective to the temporal and occipital lobes was shorter during general anesthesia (see Figs. 16-19 $ \lceil $c, d$ \rceil $, Tables 16-19).

Compared to the average path length, the diameter of a network \citep{costa2007characterization}  is a less informative measure once it is related only to the larger geodesic paths of the graph. In contrast, the average path length considers all the minimum paths of the network.

The diameter reflects the length of the largest minimum path, representing the largest distance existent on the network\footnote{Assuming that the graph is connected.}.
In the present study, the main alteration was an increase in the length of the diameter of the networks of the frontal, parietal, and temporal lobes on Delta, Theta, Alpha, and Beta frequency bands (see Figs. 16-19 $ \lceil $a, b, c$ \rceil $, Tables 16-19). Such a result supports the conclusions obtained by analyzing the average path length of the networks that, under general anesthesia, the global transmission of information is reduced in the frontal, parietal, and temporal lobes.

\subsection{Average Betweenness  Centrality Degree}

Alterations in the vertex's mean betweenness centrality degree were observed during the experiment on all frequency bands, and brain lobes were analyzed (see Figs. 21-25; Tables 21-25).

The most remarkable changes occurred on the frontal, parietal, and temporal lobes; on the Alpha and Beta frequency bands, the administration of the anesthetics led to a substantial increase and also a considerably higher variation on the vertices average betweenness centrality degree (see Figs. 23, 24 $ \lceil $a, b, c$ \rceil $; Tables 23, 24).

\subsection{Transitivity}

In this study, the administration of the anesthetics has led to alterations in the transitivity coefficient of the networks on the distinct brain lobes and frequency bands analyzed (see Figs. 26-30; Tables 26-30).

The most prominent changes occurred in the frontal, parietal, and temporal lobes on the Alpha and Beta frequency bands. In these brain regions, within less than two minutes after administering the anesthetics, the transitivity coefficient of the networks substantially decreased (see Figs. 28, 29 $ \lceil $a, b, c$ \rceil $; Tables 28, 29). 

According to Latora and Marchiori, the local efficiency of the transmission of information in a network is directly related to its transitivity coefficient; the higher the coefficient, the greater
the local efficiency of the network \citep{latora2003economic}.
The results obtained in this study reveal that administering the anesthetics
led to a reduction in the transitivity coefficient (see Figs. 28, 29 $ \lceil $a, b, c$ \rceil $; Tables 28, 29). Such a decrease observed in the frontal, parietal, and temporal lobes indicates that during the induced state of anesthesia, information transmission efficiency is diminished in these cortical lobes.
\section{Conclusions}

The experimental results of this research revealed that the anesthetic agents altered the functional brain networks. Furthermore, concomitant alterations in distinct network properties, observed within less than two minutes after the administration of the drugs, strongly indicate that the loss of consciousness experienced by the animal model during the experiment was associated with a phase transition in the network's architecture. Such results also support the conclusion that awareness and conscious experiences may depend on specifically structured functional brain networks.
The networks of the four brain lobes on the five frequency bands analyzed had characteristic properties, and each lobe also presented a distinct response due to the administration of the anesthetics. This observation reveals that functional brain networks are structured in a specific way over distinct cortical areas.
It was observed that functional neural activities are dynamic, as significant changes were found in the network measures over short time intervals. Furthermore, this dynamic nature was present in both awake (eyes open and closed) and general anesthesia conditions. Those results also reveal that the anesthetic induction, far from ``shutting down'' brain activity, led the brain into a specific, complex, and dynamic state.
 
Regarding the network measures, the most remarkable results were: An accentuated decrease in the average degree on the frontal, parietal, and temporal lobes \footnote{On frequency bands Theta, Alpha, and Beta.} which indicates that the anesthetic agents compromised the functional connectivity significantly in these anatomical areas; a prominent alteration in the assortativity character of the functional networks of the frontal lobe that changed from disassortative (awake) to assortative (general anesthesia). This experimental evidence potentially indicates the assortativity character of the frontal lobe functional networks as a neural correlate of consciousness.


\bibliography{biblio}


\end{multicols}

\end{document}